\documentclass[iop,revtex4]{emulateapj}
\usepackage[flushleft]{threeparttable}
\usepackage{graphics,graphicx}
\usepackage{epstopdf}
\usepackage{epsfig}
\usepackage{helvet}
\usepackage[utf8]{inputenc}
\usepackage[T1]{fontenc}
\usepackage{longtable}
\usepackage{soul}
\usepackage{amsmath, amssymb, gensymb} 
\usepackage{natbib}
\usepackage{microtype}
\usepackage{multirow}
\usepackage{morefloats}
\usepackage{lscape}
\usepackage{aas_macros}
\def\eg {e.g.,} 

\def\jybm {Jy\,bm$^{-1}$}

\newcommand{\mir}[1]{\textrm{\fontfamily{lmvtt}\selectfont #1}} 
\newcommand\miriad{\mir{MIRIAD}} 
\newcommand\XYphase{\mir{XYphase}} 

\begin{document}

\slugcomment{\textit{Accepted by the Journal of Astronomical Instrumentation on 17 May 2015}}

\title{The 1.3\,mm Full-Stokes Polarization System at CARMA}
\shorttitle{TADPOL Data Release}

\author{Charles~L.~H.~Hull\altaffilmark{1,2} and Richard~L.~Plambeck,\altaffilmark{3}}

\shortauthors{Hull \& Plambeck}
\email{chat.hull@cfa.harvard.edu}

\altaffiltext{1}{Harvard-Smithsonian Center for Astrophysics, 60 Garden St., Cambridge, MA 02138, USA}
\altaffiltext{2}{C.L.H.H. is a Jansky Fellow of the National Radio Astronomy Observatory, which is a facility of the National Science Foundation operated under cooperative agreement by Associated Universities, Inc.}
\altaffiltext{3}{Astronomy Department \& Radio Astronomy Laboratory, University of California, Berkeley, CA 94720-3411, USA}

\begin{abstract}
\normalsize

The CARMA 1.3\,mm polarization system consists of dual-polarization receivers
that are sensitive to right- ($R$) and left-circular ($L$) polarization, and a
spectral-line correlator that measures all four cross polarizations ($RR$,
$LL$, $LR$, $RL$) on each of the 105 baselines connecting the 15 telescopes.
Each receiver comprises a single feed horn, a waveguide circular polarizer, an
orthomode transducer (OMT), two heterodyne mixers, and two low-noise amplifiers
(LNAs), all mounted in a cryogenically cooled dewar.  Here we review the basics
of polarization observations, describe the construction and performance of key
receiver components (circular polarizer, OMT, and mixers---but not the
correlator), and discuss in detail the calibration of the system, particularly
the calibration of the $R$--$L$ phase offsets and the polarization leakage
corrections.  The absolute accuracy of polarization position angle measurements
was checked by mapping the radial polarization pattern across the disk of Mars.
Transferring the Mars calibration to the well known polarization calibrator
3C286, we find a polarization position angle of $\chi = 39.2 \pm 1\degree$ for
3C286 at 225\,GHz, consistent with other observations at millimeter wavelengths.
Finally, we consider what limitations in accuracy are expected due to
the signal-to-noise ratio, dynamic range, and primary beam polarization. 

\end{abstract}


\clearpage
\section{Introduction}
\label{sec:intro} 

Millimeter-wave radiation from astronomical objects may be polarized by a
number of processes, including many that involve magnetic fields (e.g.,
synchrotron radiation, Zeeman splitting, dust emission, etc.).  This paper
describes the design, calibration, and performance of the 1.3\,mm (210--270\,GHz) 
receiver system used for polarization measurements with the Combined Array
for Research in Millimeter Astronomy (CARMA).  CARMA is an 23-antenna aperture
synthesis telescope; the 1.3\,mm receivers are installed on the six 10\,m-diameter 
and nine 6\,m-diameter antennas.  

The CARMA system makes it possible to observe the polarized thermal emission
from dust in protostellar cores with angular resolutions of 1--2$''$, allowing
one to study the magnetic field morphologies on scales of $\sim$\,1000\,AU in the
nearest star-forming regions \citep{Hull2013, Krumholz2013, Stephens2013, Hull2014,
Davidson2014, Wright2014, Stephens2014, SeguraCox2015} or toward evolved stars
\citep{Sabin2015}.  The CARMA polarization system also has been used to measure
Faraday rotation toward the active galactic nucleus in 3C84
\citep{Plambeck2014}, constraining the mode of accretion onto the black hole in
this source, and in Very Long Baseline Interferometer experiments that promise
to probe the event horizons of nearby black holes \citep{Johnson2014}.

The plan of this paper is as follows.  We first review the measurement of
polarization via the Stokes parameters, and we motivate our choice of crossed circular
feeds for the CARMA system.  We then describe the design and performance of the
key receiver components---broadband waveguide circular polarizers, orthomode
transducers, and mixers---and discuss in detail the calibration of the $R$--$L$
phase offsets and polarization leakage terms.  We describe measurements of the
absolute polarization position angle (PA) using Mars, and transfer this calibration
to source 3C286.  Finally, we consider what limitations in accuracy are
expected due to the signal-to-noise ratio (SNR), dynamic range, and primary beam polarization.

\subsection{Stokes parameters}

It is conventional to characterize the polarization
properties of the incoming radiation field in terms of Stokes parameters.  The
Stokes parameters fully describe the characteristics of radiation that is fully
polarized, partially polarized, or unpolarized.  In an $X$--$Y$ coordinate system
where $+x$ points North, $+y$ points East,\footnote{\,Note that, as in Figure 4.8 of \citet{TMS}
the $X$ and $Y$ axes are rotated 90$^{\circ}$ counterclockwise relative to
standard Cartesian axes.} 
and radiation propagates toward us
along the $+z$ axis, the Stokes parameters are given in terms of the
time-averaged products of the complex voltages $E_X$ and $E_Y$
(Equations 2.47a--2.47d in \citealt{RybickiLightman1979}, 
Equations 1 in \citealt{Hamaker1996b},
and Equations 4.19 in \citealt{TMS}):
\begin{align}
\label{eqn:stokes_xy1}
I &= \langle E_X E_X^* \rangle + \langle E_Y E_Y^* \rangle \\
\label{eqn:stokes_xy2}
Q &= \langle E_X E_X^* \rangle - \langle E_Y E_Y^* \rangle \\
\label{eqn:stokes_xy3}
U &= \langle E_X E_Y^* \rangle + \langle E_X^* E_Y \rangle \\
\label{eqn:stokes_xy4}
V &= -i \left( \langle E_X E_Y^* \rangle - \langle E_X^* E_Y \rangle \right) \,\,,
\end{align}
\noindent
where $E_X^*$ denotes the complex conjugate of $E_X$.  

Alternatively, as described in more detail in the Appendix, the Stokes
parameters may be expressed in a circular polarization basis:
\begin{align}
\label{eqn:stokes_rl1}
I &= \langle E_R E_R^* \rangle + \langle E_L E_L^* \rangle  \\
\label{eqn:stokes_rl2}
Q &= \langle E_R E_L^* \rangle + \langle E_R^* E_L \rangle  \\
\label{eqn:stokes_rl3}
U &= -i\, \left( \langle E_R E_L^* \rangle - \langle E_R^* E_L \rangle \right)  \\
\label{eqn:stokes_rl4}
V &= \langle E_R E_R^* \rangle - \langle E_L E_L^* \rangle \,\,.
\end{align}

Note that Equations \ref{eqn:stokes_xy1}--\ref{eqn:stokes_xy4} and
\ref{eqn:stokes_rl1}--\ref{eqn:stokes_rl4} are appropriate for a single dish
telescope.  As described in \citet{TMS}, an interferometer like CARMA actually
measures Stokes {\it visibilities} from cross-correlations between antennas.
For example, for telescopes $m$ and $n$, Stokes $Q_{mn} = \langle E_{Xm} E^*_{Xn}
\rangle - \langle E_{Ym} E_{Yn}^* \rangle$.  These visibilities are Fourier
transformed to produce images of Stokes $I$, $Q$, $U$, and $V$ toward the
astronomical source.  These are maps of radio brightness in units of
Jansky/beam.  Stokes $I$ is always positive, but Stokes $Q$, $U$, and $V$ can
be negative or positive (but not complex).

From the Stokes images, one can derive the fractional linear polarization
$\Pi_{l}$ and position angle $\chi$:
\begin{align}
\Pi_{l} &= \frac{\sqrt{Q^2 + U^2}}{I} \\  
\chi &= \frac{1}{2} \arctan{\frac{U}{Q}}\,\,,\,\,\,\,0 < \chi < \pi.
\end{align}

\subsection{Choice of crossed circular feeds.} 

For polarization measurements with an aperture synthesis array, the receivers
may be configured either with crossed linear ($E_X$ and $E_Y$) or crossed
circular ($E_R$ and $E_L$) feeds.  For observations at 1.3\,mm wavelength, the
science goals are mostly to measure weak linear polarization; few sources have
appreciable circular polarization.  Linear polarization is derived from Stokes
$Q$ and $U$.  With crossed linear feeds, Stokes $Q$ (Equation
\ref{eqn:stokes_xy2}) is derived from the difference in power measured by the
$X$ and $Y$ receivers.  One must therefore take the difference of two large
numbers ($E_X E_X^*$ and $E_Y E_Y^*$) in order to measure a small number ($Q$),
which demands that the $X$ and $Y$ gains be extremely stable and well
calibrated.  The gain stability requirement is relaxed considerably if crossed
circular feeds are used, since then both $Q$ and $U$ are derived from
cross-products of $E_R$ and $E_L$. The magnitudes of these cross products are
nearly zero if the source is weakly polarized, hence gain fluctuations are much
less of a problem.  For this reason, it is generally considered better to
observe weak linear polarization with crossed circular feeds, and this is the
approach that CARMA uses.  For a more detailed discussion of the pros and cons
of linear and circular feeds see \citet{Cotton1998}, as well as the summary 
(Section \ref{sec:summary}).

\subsection{Observing modes}

The 1.3\,mm receiver system operates in three modes: single-polarization mode
($LL$ or $RR$) dual-polarization ($LL$ and $RR$ simultaneously), and
full-Stokes ($LL, RR, LR, RL$).  Depending on the correlator setup, 
all three modes can use up to the full 8\,GHz bandwidth
of the CARMA correlator (4\,GHz per sideband), but with different combinations
of correlator bands and polarizations.  See Table \ref{table:obs_modes} for the properties
of the different observing modes in the wide-band (500\,MHz/band) setup, which we used 
for observations of dust polarization.

There is no benefit to using dual-polarization mode for wide-band (continuum)
observing.  Typically the RCP receivers have higher noise temperatures than the
LCP receivers, so it is better to use the available correlator bandwidth to
sample a wider frequency range in LCP, rather than a narrower range in both LCP
and RCP.  The dual-polarization mode offers higher sensitivity only for
spectral line observations, where it is beneficial to get independent
information from the second polarization, even if the receiver temperature for
this polarization is a little higher.

\begin{table*}
\centerline{Table \ref{table:obs_modes}: CARMA 1.3\,mm wide-band Observing Modes \vspace*{0.1in}}
\begin{center}
{\normalsize
\begin{tabular}{lccccccc}
\hline
Mode & Bands & BW/band & Pol. & SB & Total BW & Cross-corr. & Channels \\
                                & (\#) & (MHz) & (\#) & (\#) & (GHz) &    & (\# per band) \\
\hline
Single-pol & 8 & 500 & 1 & 2 & 8 & RR \textit{or} LL & 95 \\
Dual-pol & 4 & 500 & 2 & 2 & 8 & RR \textit{and} LL & 95 \\
Full-Stokes & 4 & 500 & 2 & 2 & 8 & RR, LL, RL, LR & 47 \\
\hline
\end{tabular}

\bigskip
\caption{\small 1.3\,mm CARMA wide-band (500\,MHz/band) observing modes. The total bandwidth of each mode is 8\,GHz, and
is equal to the number of correlator bands $\times$ the bandwidth per correlator band (BW/band) $\times$ the
number of polarizations (Pol.) $\times$ the number of sidebands (SB).  Observations with narrower bandwidths also
are possible, and provide a greater number of channels.}
\label{table:obs_modes}
}
\end{center}
\end{table*}

\begin{figure} [hbt!]
\centering
\includegraphics[scale=0.34]{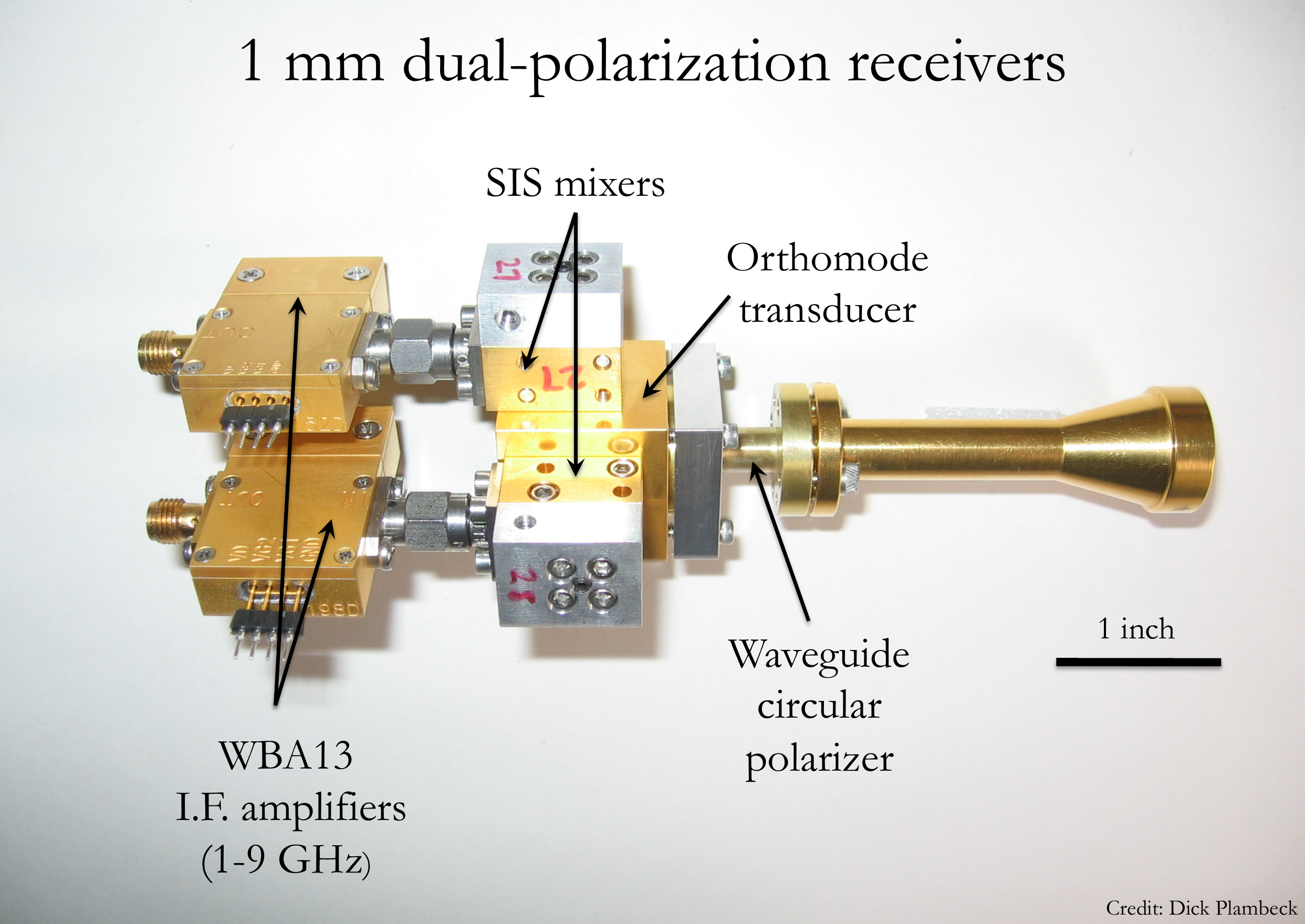}

\caption{\small The 1.3\,mm dual-polarization receiver module, installed inside
the dewars on all the 10\,m and 6\,m antennas at CARMA.  On the 10\,m
telescopes all these parts are cooled to 4\,K. On the 6\,m telescopes, to reduce
the heat load on the 4\,K cryocooler stages, the two WBA13 amplifiers are
connected to the 15\,K cryocooler stages, while all other parts are cooled to
4\,K; the heat load from 15\,K to 4\,K through the stainless steel SMA connectors
is only 1--2\,mW.  } 
\label{fig:rx} 
\end{figure}

\section{Hardware}

Figure~\ref{fig:rx} is a photo of the 1.3\,mm dual-polarization receiver module that is mounted
in the cryogenically cooled dewar on each telescope.  It includes a single feed
horn, a waveguide circular polarizer (Section \ref{sec:polarizer}), 
an orthomode transducer (OMT; Section \ref{sec:OMT}), two
heterodyne mixers (Section \ref{sec:mixer}), and two low-noise amplifiers (LNAs).  The local oscillator
(LO) and sky signals are combined using a mylar beamsplitter in front of the
dewar window.

\vspace{0.3in}
\subsection{Waveguide Polarizer}
\label{sec:polarizer}

\begin{figure*} [hbt!]
\centering
\includegraphics[scale=0.6, clip, trim=0cm 4cm 0cm 4cm]{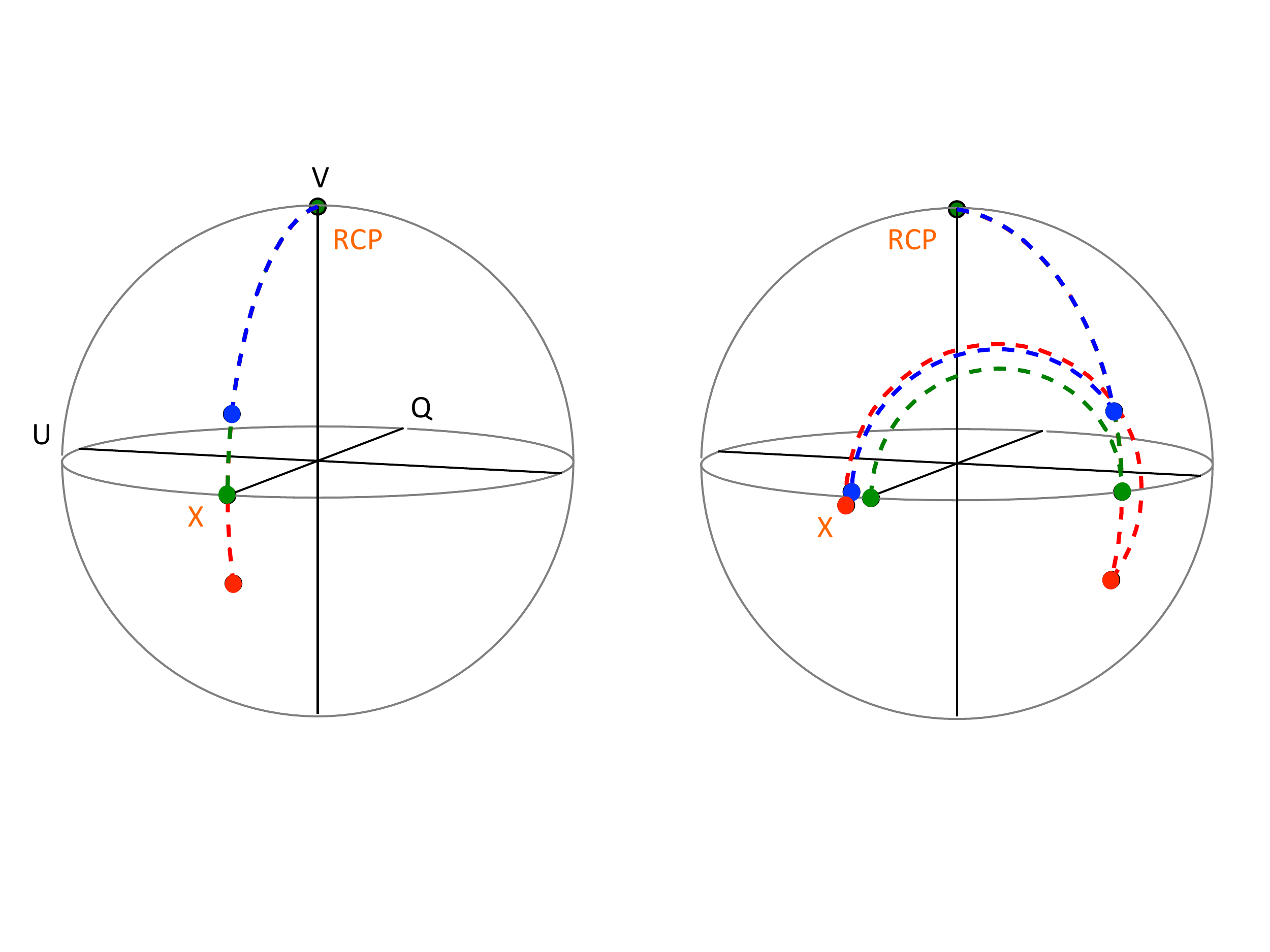}
\caption{\small Action of single- and 2-element circular polarizers visualized
on a Poincar\'{e} sphere.  The principal axes of the sphere correspond to
Stokes $Q$, $U$, and $V$.
The goal is to transform $R$ and $L$ circularly polarized radiation, at the north and south
poles, to $X$ and $Y$ linear polarization, at opposite ends of the $Q$-axis on the
equator.  \textit{Left:} Passing RCP radiation through a single quarter-wave
retarder with its principal axes aligned at 45$^\circ$ to the $X$-direction
corresponds to a rotation on the sphere that places the central frequency
(230~GHz, green) exactly on the equator, but spreads neighboring frequencies
(210~GHz, red; 260~GHz, blue) along an arc, leading to substantial ellipticity
in their polarizations.  \textit{Right:} In a 2-stage polarizer the quarter wave
section is followed by a half-wave retarder that rotates the center frequency
back to the equator; the dispersion of this half-wave retarder nearly cancels
chromatic errors produced by the quarter-wave section, thus broadening the
bandwidth.}
\label{fig:poincare}
\end{figure*}

The CARMA dual-polarization receivers use waveguide polarizers, operating at
temperatures of 4\,K, to convert incoming $R$ and $L$ polarized signals into
$X$ and $Y$ linearly polarized signals.  These polarizers have lower loss and
are more compact than quarter waveplates mounted outside the dewars.  Since the
polarizers are permanently installed inside the dewars, they must operate over
the full 210--270\,GHz tuning range of the receivers.  A polarizer consisting
of a single quarter-wave retarder section would have much too narrow a
bandwidth for this.  However, it is possible to stack one or more half-wave
retarder sections, rotated axially with respect to the quarter-wave retarder,
to achieve broader bandwidth.  Such multi-section polarizers were first
described by \citet{Pancharatnam1955} in the context of birefringent wave
plates.  The action of the multi-section design is most easily visualized on a
Poincar\'e sphere (Figure~\ref{fig:poincare}).  

\begin{figure*} [hbt!]
\begin{minipage}[b]{0.5\linewidth}
\centering
\includegraphics[width=0.95\textwidth,clip,trim=12cm 17cm 16cm 12cm]{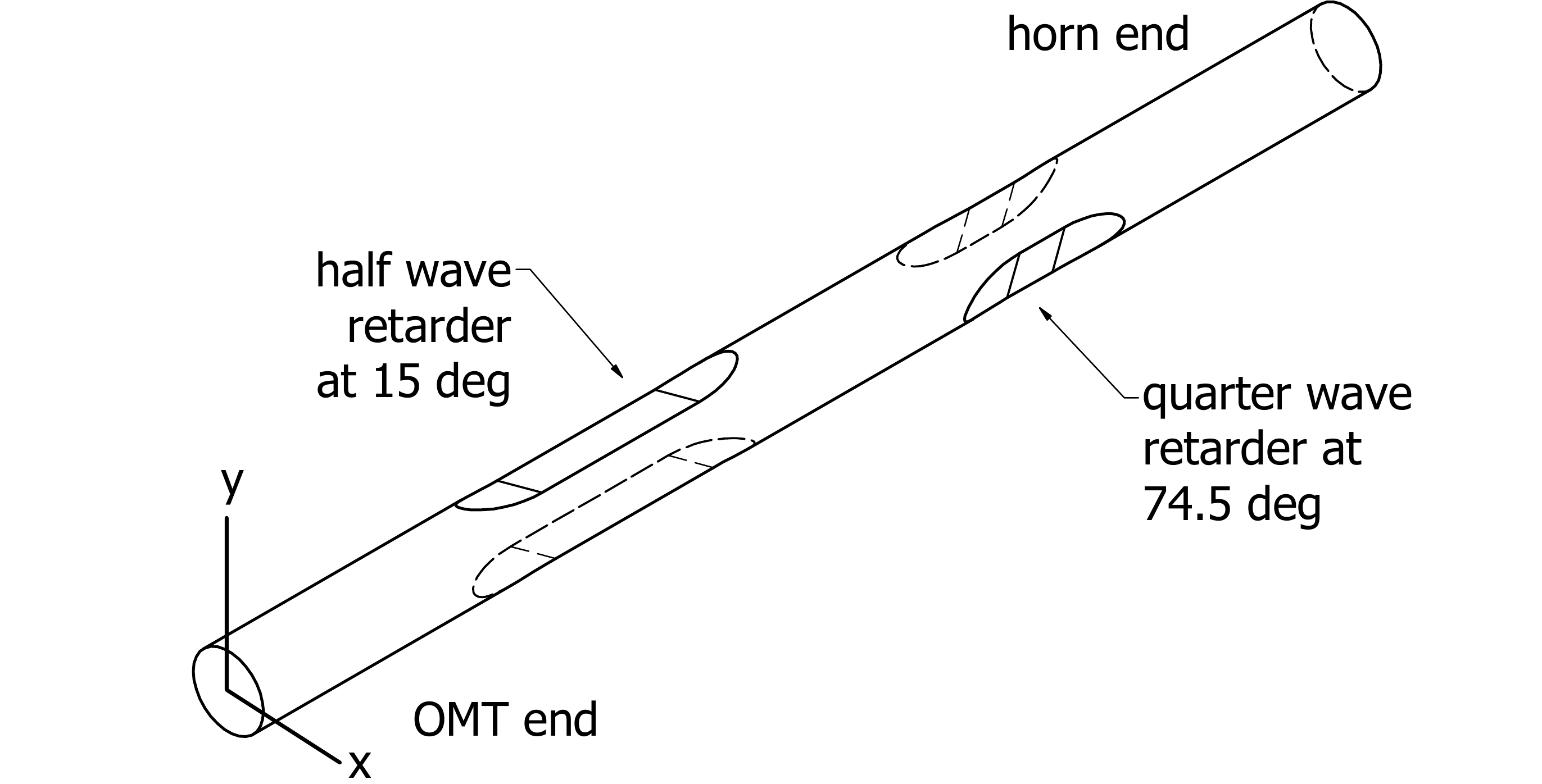}
\end{minipage}
\hspace{0.02\linewidth}
\begin{minipage}[b]{0.46\linewidth}
\centering
{\includegraphics[width=0.95\textwidth,clip,trim=5cm 0.5cm 5cm 0.5cm]{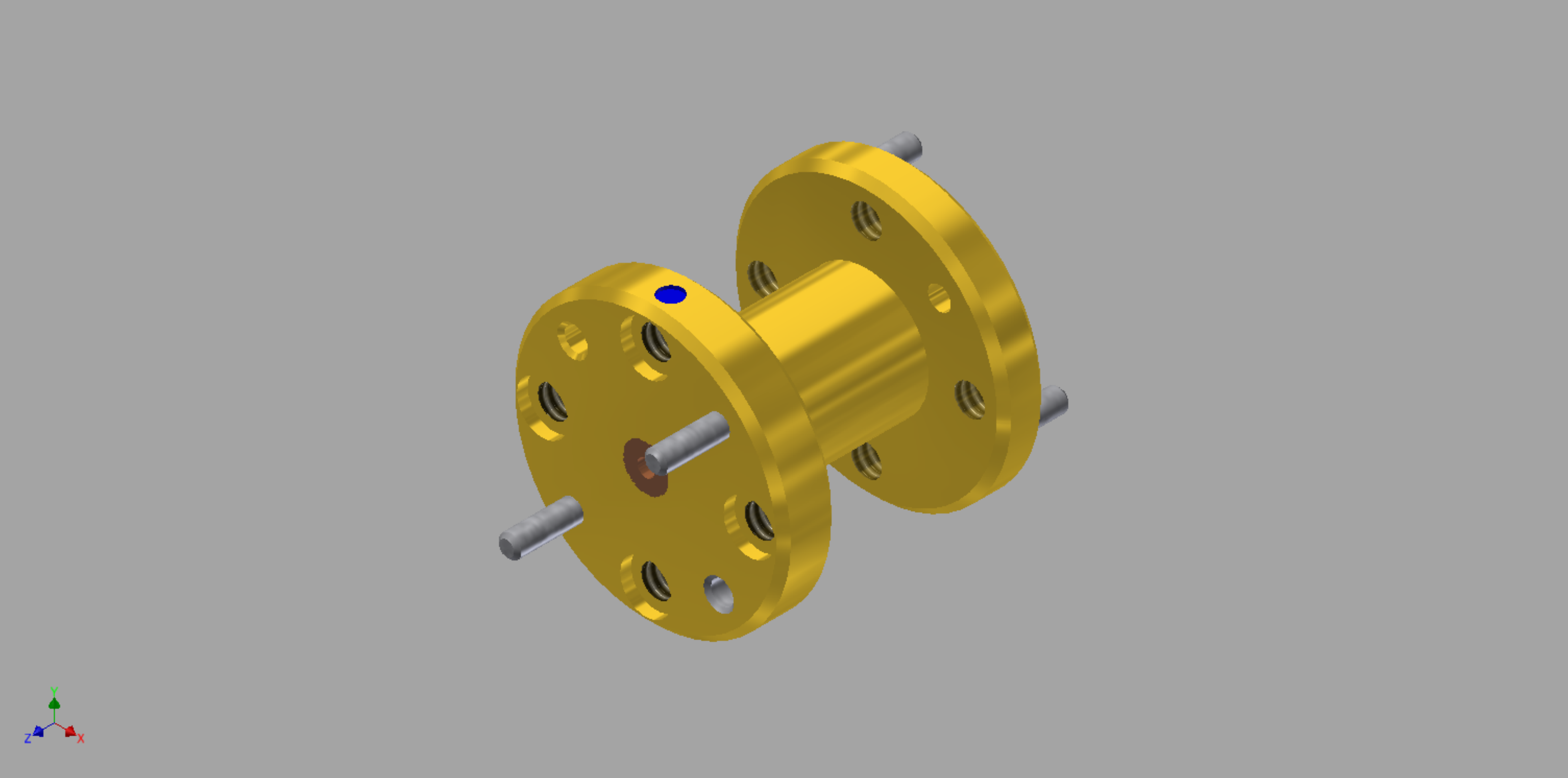}}
\end{minipage}
\caption{\small \textit{Left:} Waveguide layout.  The polarizer converts an $L$-polarized
signal entering the horn end to a $Y$-polarized signal exiting the OMT end, while
$R$ is converted to $X$. \textit{Right:} Outer view of the polarizer. A thick flange
with a dot indicates the OMT end.  The dot is aligned with the polarizer's
$Y$-axis for serial numbers CP01--CP19, but (due to an unfortunate mixup) with the
$X$-axis for CP20--CP36. \\}
\label{fig:polOuter}
\end{figure*}

\begin{figure*} [hbt!]
\centering
\includegraphics[width=0.9\textwidth, clip, trim=3cm 1cm 3cm 7cm]{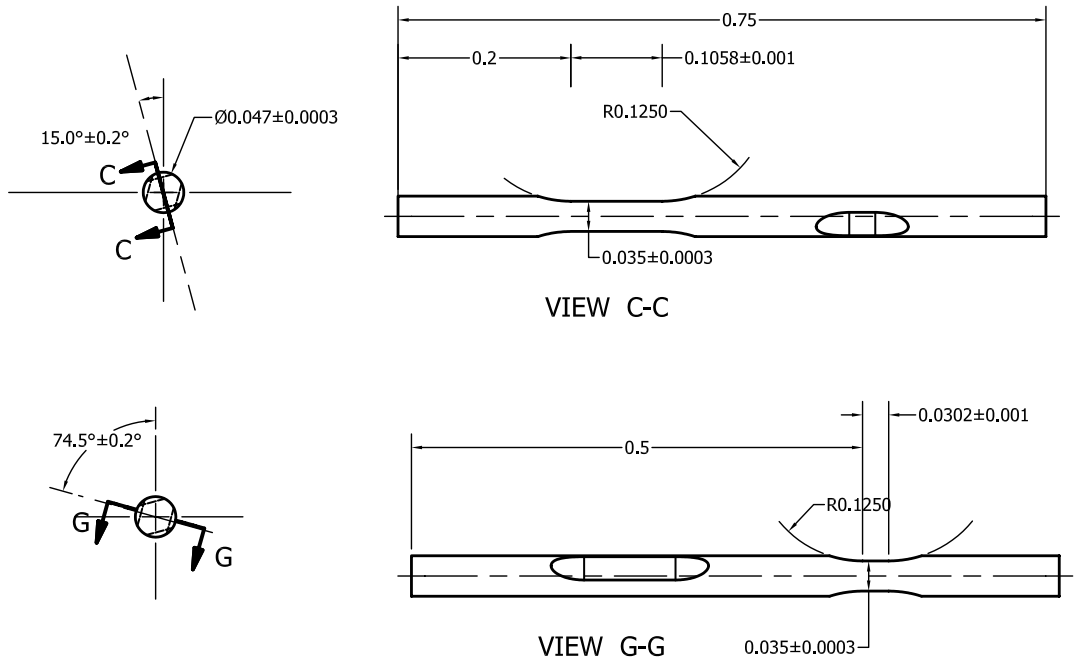}
\caption{\small Polarizer dimensions (in inches). \\}
\label{fig:PolDimensions}
\end{figure*}

\begin{figure*} [hbt!]
\centering
\includegraphics[scale=0.6, clip, trim=1cm 4.5cm 2cm 9cm]{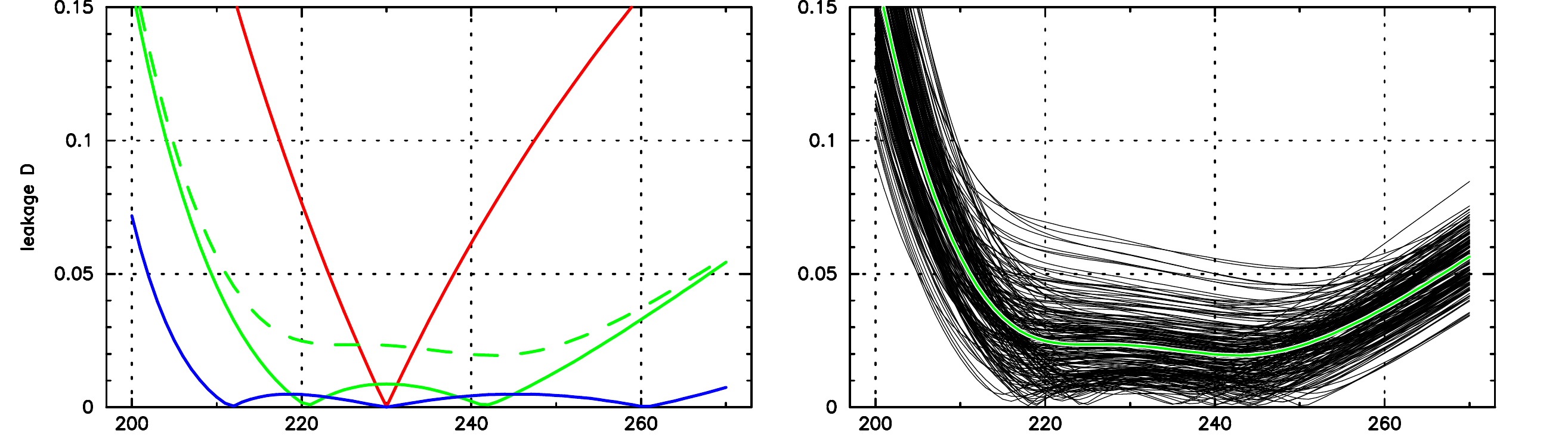}
\caption{\small \textit{Left:} Theoretical polarization leakage for 1-, 2-, and
3-section waveguide polarizers (red, green, blue curves) using faceted circular
retarder sections.  The dashed green curve shows the expected performance of a
2-section polarizer when machining tolerances are taken into account.
\textit{Right:} Simulated leakages for 200 polarizers with machining tolerances
of $\pm\,0.0003''$ on the thickness of the faceted sections, and
$\pm\,0.2^\circ$ on the angles of the flats.  The green curve shows the mean
result that is reproduced (dashed) in the left-hand panel. \\}
\label{fig:PolLeak}
\end{figure*}

The CARMA polarizer is similar to the 2-section waveguide polarizer developed
by Kovac for the DASI experiment \citep{Leitch2002,Kovac2004}.  Whereas the
(30\,GHz) DASI polarizer used dielectric vanes as retarder sections, the much
higher frequency (230\,GHz) CARMA polarizer uses sections of faceted circular
waveguide as retarders.  The guide wavelengths differ parallel and
perpendicular to the facets, causing a phase shift.  The design of the
polarizer is described in detail in CARMA Memo 54 \citep{Plambeck2010}.
Figure~\ref{fig:polOuter} shows the waveguide layout, and 
Figure~\ref{fig:PolDimensions} gives the polarizer dimensions.  The performance
of the polarizer is characterized by the polarization leakages (see the
discussion in Section \ref{sec:leakage}).  The leakages for ideal 1-, 2- and
3-section polarizers are shown in the left-hand panel of
Figure~\ref{fig:PolLeak}; the leakages expected for a 2-section polarizer
when machining tolerances
are taken into account are shown in the right-hand panel.  When one allows for
machining tolerances, the performance of a 3-stage polarizer is no better than
that of a 2-stage polarizer.

Note that a multi-section polarizer cannot be flipped end to end---this is because
rotations do not commute on the surface of a sphere.  On the CARMA polarizers,
a thicker flange with a dot marks the OMT end.

\begin{figure} [hbt!]
\centering
\includegraphics[scale=0.34, clip, trim=0cm 0cm 0cm 1cm]{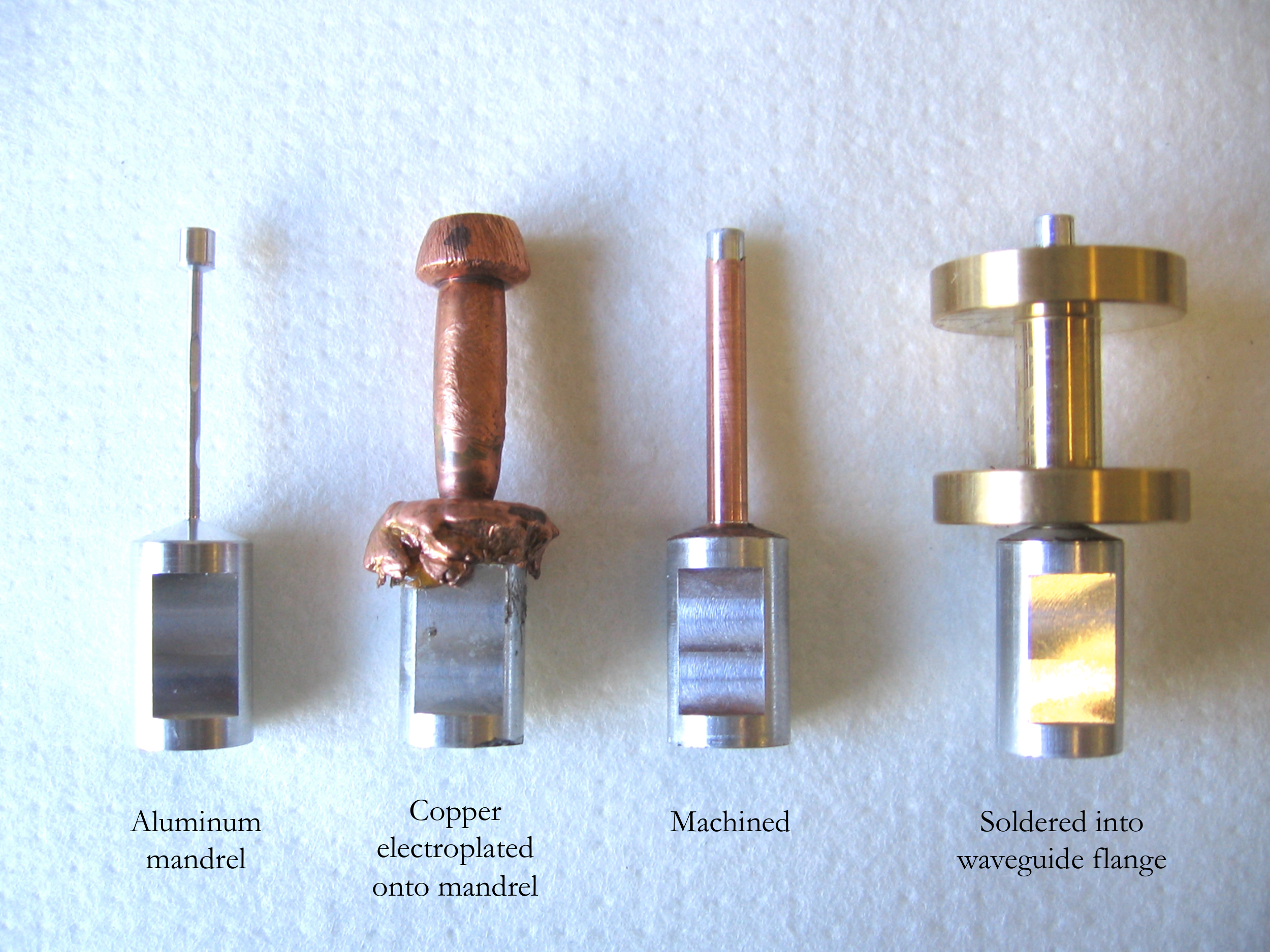}
\caption{\small Photo of the different stages of polarizer construction. \\}
\label{fig:PolConstruction}
\end{figure}

\begin{figure*} [hbt!] \centering
\includegraphics[width=1.07\textwidth, clip, trim=1cm 1.8cm 0cm 1.5cm]{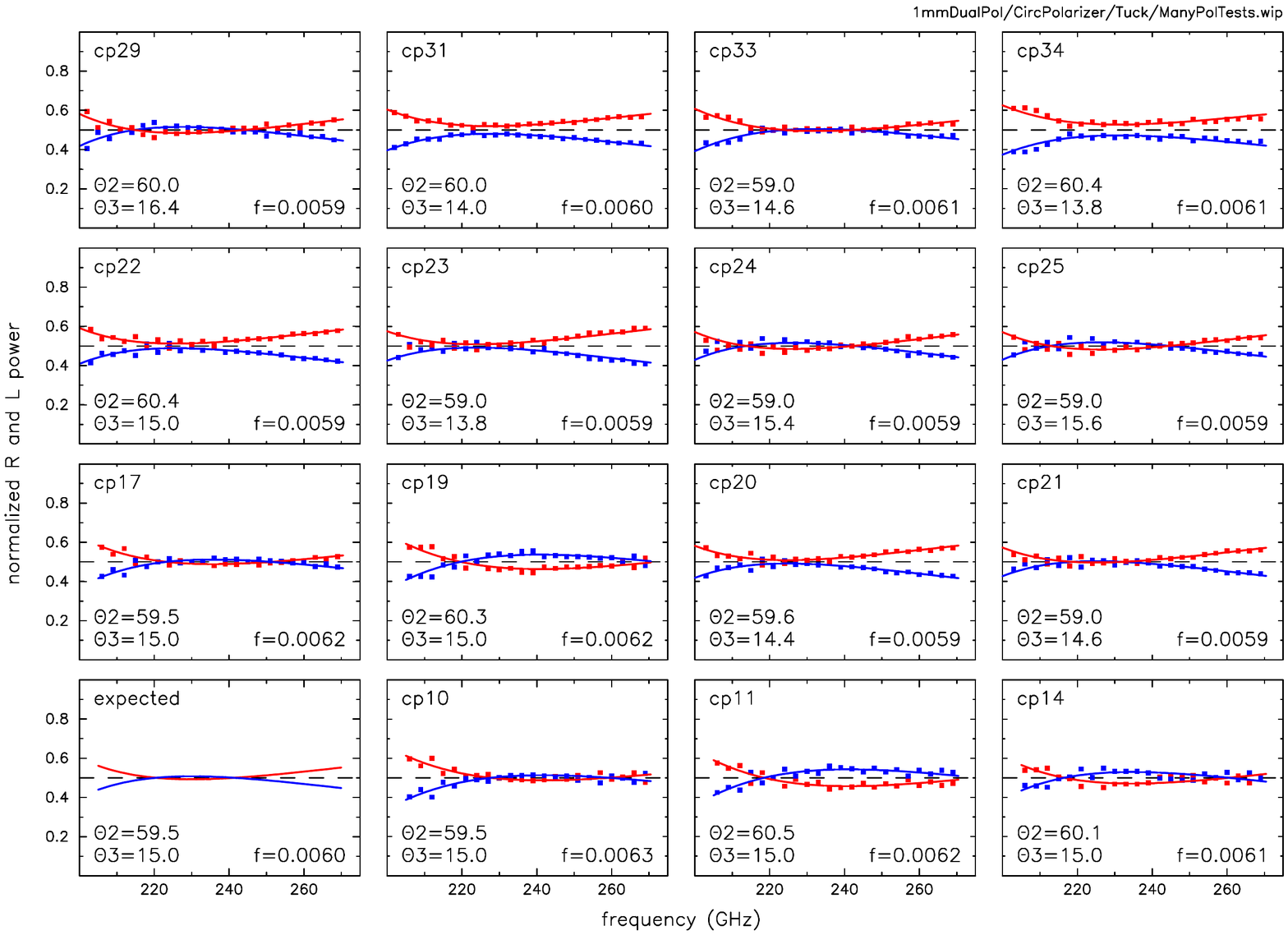}
\caption{\small Test results for all the circular polarizers used on the
telescopes.  A linearly polarized signal (which can be represented as a
superposition of LCP and RCP with equal amplitude) was injected into the
``horn'' end of the polarizer, and the normalized power emerging from the $R$
(red) and $L$ (blue) outputs of the OMT are plotted as a function of frequency.
The power levels would be perfectly equal for an ideal polarizer; the expected
power split for a 2-section polarizer of our design in shown in the lower left
panel. 
The data for each polarizer were fitted to an analytical model that allowed for
errors in the angles and thicknesses of the faceted sections.  \textit{f} is
the depth of the facets (the target dimension is $0.0060''$), $\Theta3$ is the
angle of the half-wave section relative to the $Y$ axis of the OMT (the target
value is 15.0$\degree$), and $\Theta2$ is the angle between the quarter-wave and
half-wave sections (the target value is 59.5$\degree$). \\
}
\label{fig:AllPolTests}
\end{figure*}

As shown in Figure~\ref{fig:PolConstruction}, the polarizers are manufactured
by electroplating gold and copper onto an aluminum mandrel.  The copper
electroform is machined to the correct diameter and soldered into a cylindrical
hole in a brass shell.  After machining is finished, the part is electroplated with
gold and the aluminum mandrel is dissolved with sodium hydroxide to leave the
finished waveguide.  All electroplating steps were done by A.J. Tuck Co.~(Brookfield, CT).

We checked the performance of the polarizers by injecting a linearly polarized signal 
into the the horn end of each polarizer.  By definition, this linearly
polarized wave is the superposition of equal amplitude $R$- and $L$-polarized
waves.  Thus, for an ideal polarizer one would measure equal powers out
the $X$- and $Y$-ports of an OMT attached to the output port of the polarizer.
Figure~\ref{fig:AllPolTests} shows the results of these tests. 
The input signal was tuned from 205--270\,GHz.
Two sets of power measurements were made, with the OMT and two power meters
rotated by 90$^{\circ}$ for the second test in order to average out differences
in the OMT loss and the power meter calibrations.  The theoretically expected
power split for our polarizer design is shown in the lower left panel of the
figure.  

The measured data for each polarizer were fitted to an analytic model (based on
the one described in CARMA Memo 54) that allowed for errors in the angles and
thicknesses of the faceted waveguide sections.  Smooth curves in each panel
show the $R$ and $L$ powers predicted by the model for that polarizer.  The model
is a simplification in that it assumes that the facet depths of each section
are uniform and equal for both retarder sections.  However, while the fits suggest
that the thicknesses of the faceted retarder sections normally were within the
$\pm$\,$0.0003''$ allowed tolerance, the angles of these retarder sections
often were outside the $\pm$\,$0.2^\circ$ tolerance.  Although it was difficult to
inspect the inside of the waveguide on the completed polarizers, careful
measurements with an optical comparator found angular offsets that were roughly
consistent with the fitted results.  The expected polarization leakages may be
derived from the fitted dimensions; for most of the polarizers the predicted
leakage amplitudes are less than 5\% across the 210--270\,GHz band.

\subsection{Orthomode transducer}
\label{sec:OMT}

\begin{figure*} [bt!]
\centering
\includegraphics[scale=0.6, clip, trim=0cm 2.5cm 0cm 4cm]{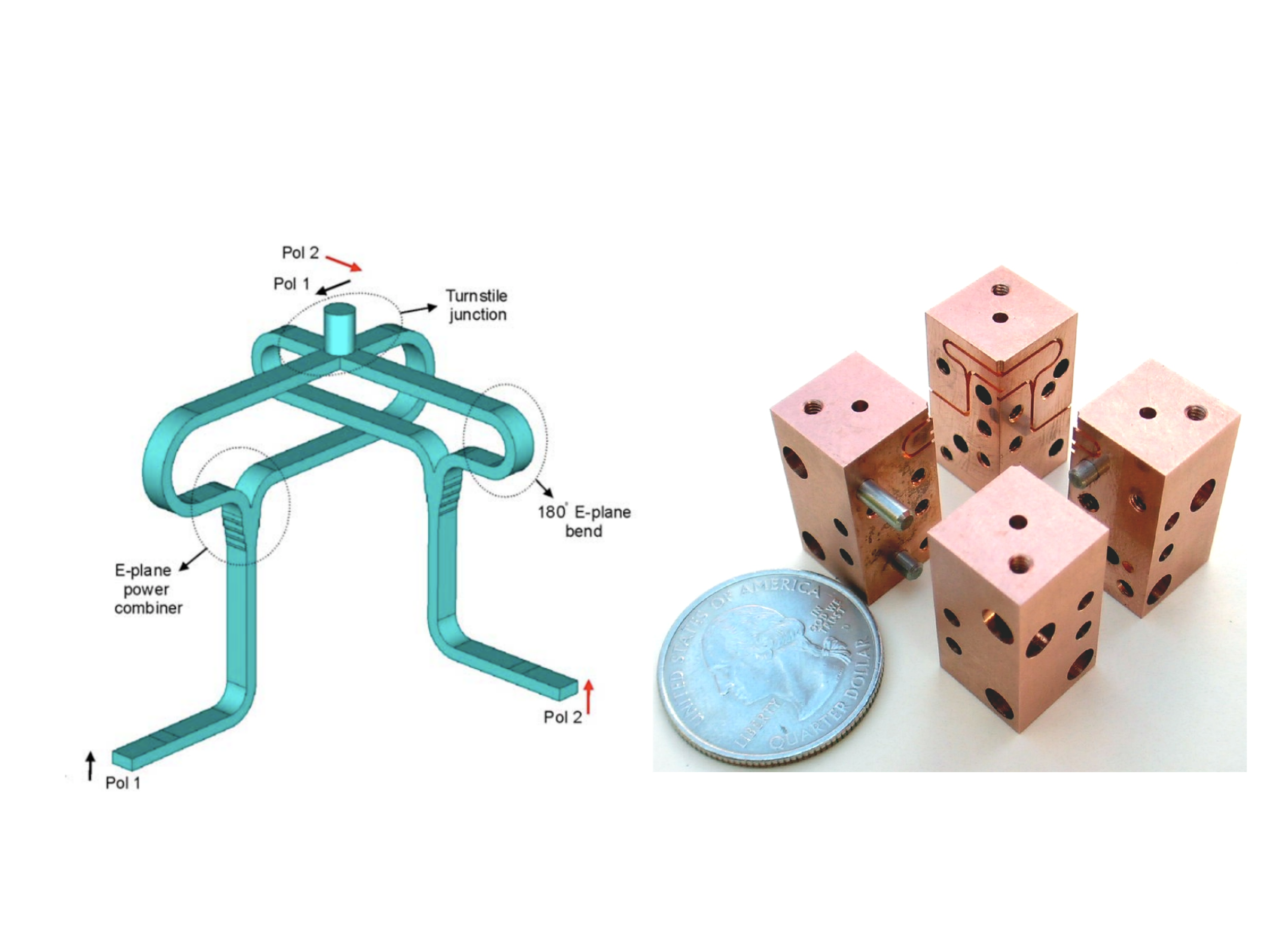}
\caption{\small \textit{Left:} OMT design.  A turnstile junction splits
incoming signals of each polarization into two rectangular arms that
recombine in a waveguide $E$-plane junction. \textit{Right:} The polarizer
is constructed from 4 blocks that meet along a common edge. \\}
\label{fig:OMTdesign}
\end{figure*}

\begin{figure*} [bt!]
\centering
\includegraphics[scale=0.72, clip, trim=1.5cm 1.6cm 1cm 1.4cm]{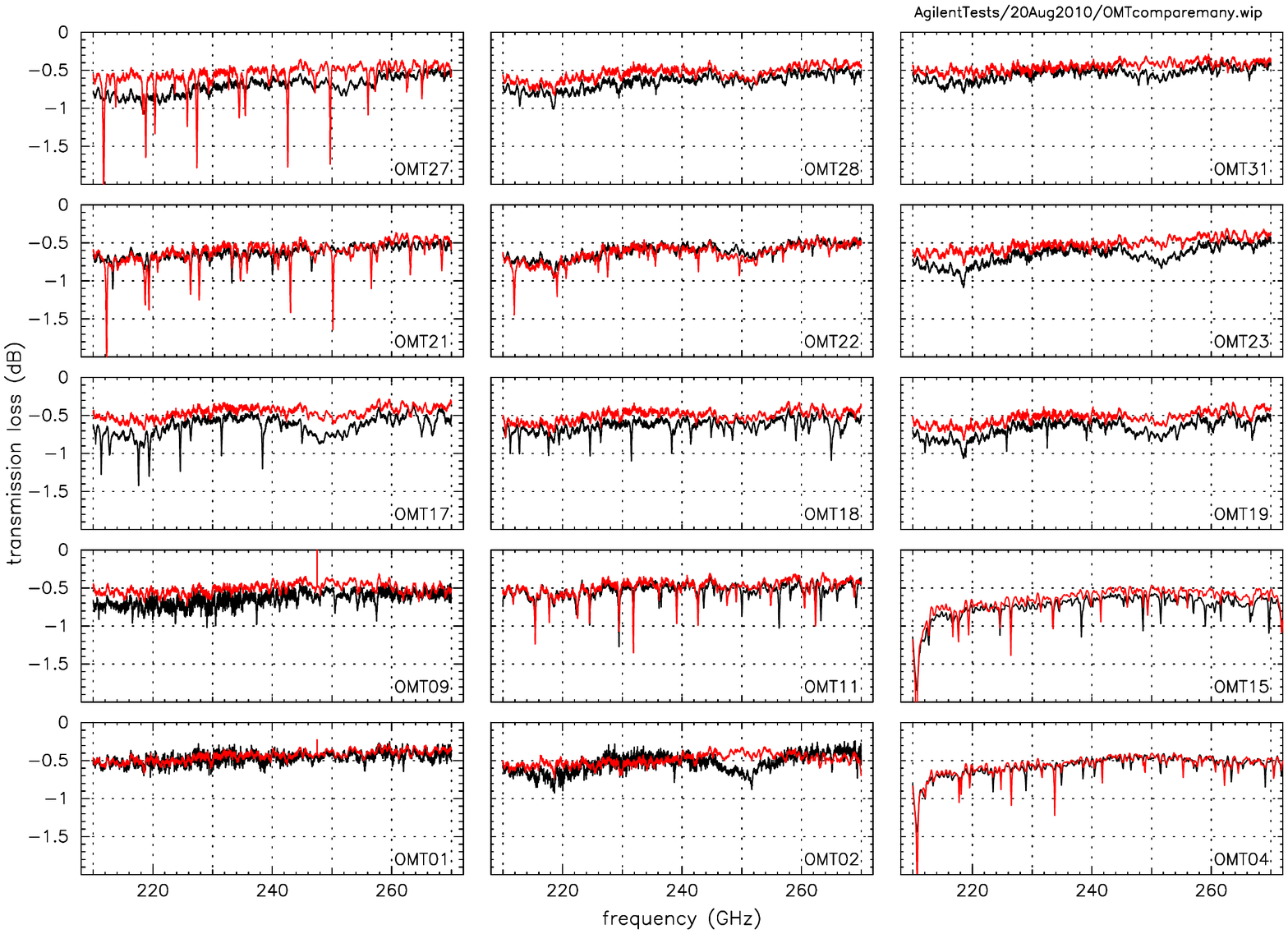}
\caption{\small Transmission loss (S21) measurements of the OMTs used on the telescopes, made with
an Agilent N4252A network analyzer.  Losses in waveguide transitions to the OMT
have been calibrated out.  Red curves are the transmission from the circular
input to the ``short'' arm of the OMT (rectangular waveguide port closer to the
input); black curves are transmission to the ``long'' arm.  The step size is
18.75\,MHz.  The lossy resonances are roughly 150\,MHz wide at the half-power
points. \\ \\}
\label{fig:OMTcomparemany} 
\end{figure*}

After the waveguide polarizer converts incoming $R$ and $L$ polarized signals
into $X$ and $Y$ in the circular waveguide, these linearly polarized signals
are coupled into separate rectangular waveguides by an orthomode transducer
(OMT).  As shown in Figure~\ref{fig:OMTdesign}, the OMT employs a turnstile
junction to split each of the two incoming polarizations into two opposite
waveguide arms.  These arms recombine in $E$-plane Y-junctions.  

The OMT is fabricated from four blocks that meet along a common edge. Initial
tests of the design were made with a 20\,GHz scale model
\citep{Navarrini2005b,Navarrini2006a}.  To test the manufacturability at 1\,mm,
OMTs were then made by 4 independent machine shops (RAL, Custom Microwave,
Protofab Inc., U of Arizona).  These were tested by Alessandro Navarrini and Alberto
Bolatto using a vector network analyzer at NRAO in Charlottesville, VA.
Details of these tests are reported in CARMA Memo 32 \citep{Navarrini2006b},
and are summarized in \citet{Navarrini2006a}.  The test results showed
excellent input return loss (better than --15\,dB) and polarization isolation
(better than --40\,dB).  The transmission losses were initially 1--2\,dB, a
little higher than expected, but improved by several tenths of a dB when gaps
in the tuning stub at the base of the tunstile were filled with indium.
Unfortunately, the tuning stub is split between the 4 blocks, and getting the
blocks to join tightly at the tuning stub is difficult.  

Some of the prototype OMTs showed narrow resonances in their transmission
curves.  Simulations showed that these could be caused by small differences in
the lengths of the opposite waveguide arms between the turnstile and the power
combiner, perhaps due to misalignments of the 4 blocks (see Figure 19 in
\citealt{Navarrini2006b}).  Energy coming from the turnstile junction reflects
back from the power combiner if the two signals reaching it are not exactly
180$^\circ$ out of phase.  Thus, standing waves can be set up between the
turnstile and power combiner at frequencies where an integral number of half
wavelengths fit within the waveguide arm.

Subsequently, a set of 30 OMTs were machined by Protofab Inc. (Petaluma, CA).
These brass parts were gold plated.  All but one of the OMTs were tested on a
demonstration N4252A network analyzer at Agilent, Inc. (Santa Rosa, CA), in
2010 January and August.  We gratefully acknowledge Suren Singh for making these tests
possible.  Transmission measurements for the 15 polarizers ultimately used
on the telescopes are shown in
Figure~\ref{fig:OMTcomparemany}.  
Average transmission losses
are in the range 0.4 to 0.7\,dB.  Narrow ($\sim$150\,MHz FWHM [full width at half maximum]) resonances in the
transmission, with losses of up to 2\,dB, were often a problem, however.  For
the handful of OMTs measured both in January and August, the frequencies of the
strongest resonances were identical. Often the resonances disappeared if we
loosened the screws holding the 4 quadrants of the OMT together, and in some
cases it was possible to retighten them in a sequence that maintained better
performance.  Loosening and tightening the screws changes the waveguide
widths, hence the guide wavelengths, hence the electrical lengths between the
turnstile junction and the Y-junction power combiners. 
The arms are about an inch long, so one expects resonances spaced by about
6\,GHz, as observed.

\begin{figure} [bt!]
\centering
\includegraphics[scale=0.45, clip, trim=1.2cm 1cm 0cm 7cm]{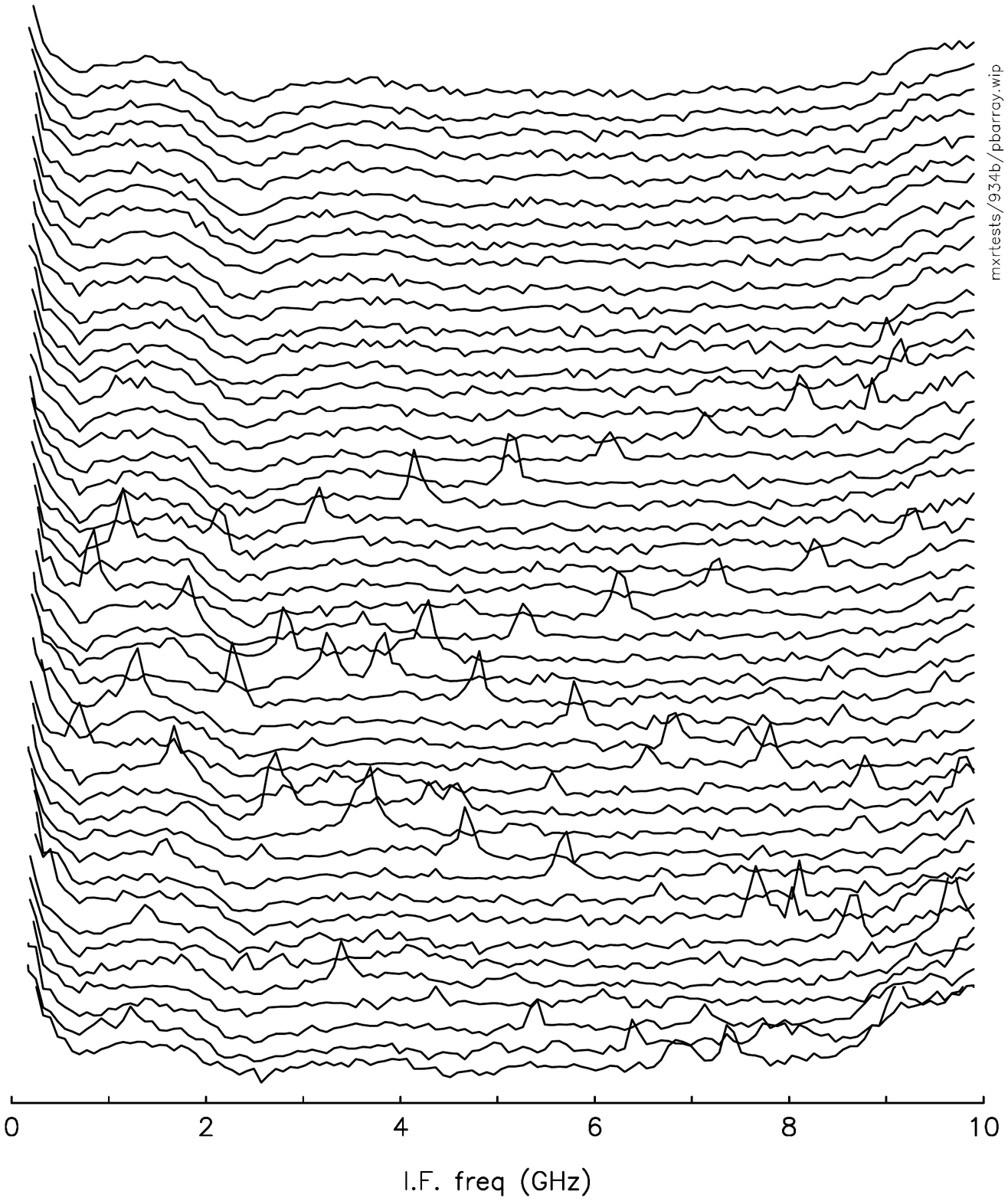}
\caption{\small Receiver noise temperature vs. IF frequency, measured at 46 different
LO frequencies, from 210\,GHz (bottom) to 255\,GHz (top) in 1\,GHz steps.  In this
plot resonances in the OMT show up as sharp peaks that march across the IF
band, forming ``boat wakes.''  In this example mixer 54-22 was mounted on the
short arm of OMT10.  The handful of OMTs that showed this
behavior were disqualified for use on the array.}
\label{fig:boat-wakes} 
\end{figure}

Occasionally an OMT that worked well at room temperature developed resonances
when cooled to 4\,K, probably because of differential thermal contraction
between the screws and the 4 OMT quadrants.  We were able to detect these
resonances by measuring the receiver noise temperature across the 1--10\,GHz IF
passband over a range of LO frequencies.  A lossy resonance in the OMT shows up
as a narrow spike in the noise temperature, and this spike moves through the IF
as the LO is stepped to different frequencies, forming what we termed ``boat
wakes'' in the noise temperature plots.  A good example is shown in
Figure~\ref{fig:boat-wakes}.  Since it was not possible to adjust any of the
screws when the OMT was at 4\,K, we chose simply to replace any OMT showing
such behavior with a spare.

\subsection{Mixers}
\label{sec:mixer} 

The mixers use ALMA (Atacama Large Millimeter-submillimeter Array) Band 6 
SIS (superconductor-insulator-superconductor) tunnel
junctions fabricated at the University of Virginia by Arthur Lichtenberger.
These are arrays of 4 SIS devices in series, offering greater dynamic
range and higher saturation power than a single SIS device.  Most of our
mixers were constructed using ``C12,L56'' or ``C14,L56'' devices from wafer 1489.

\begin{figure} [bt!]
\centering
\includegraphics[scale=0.49, clip, trim=1.1cm 0cm 0cm 1cm]{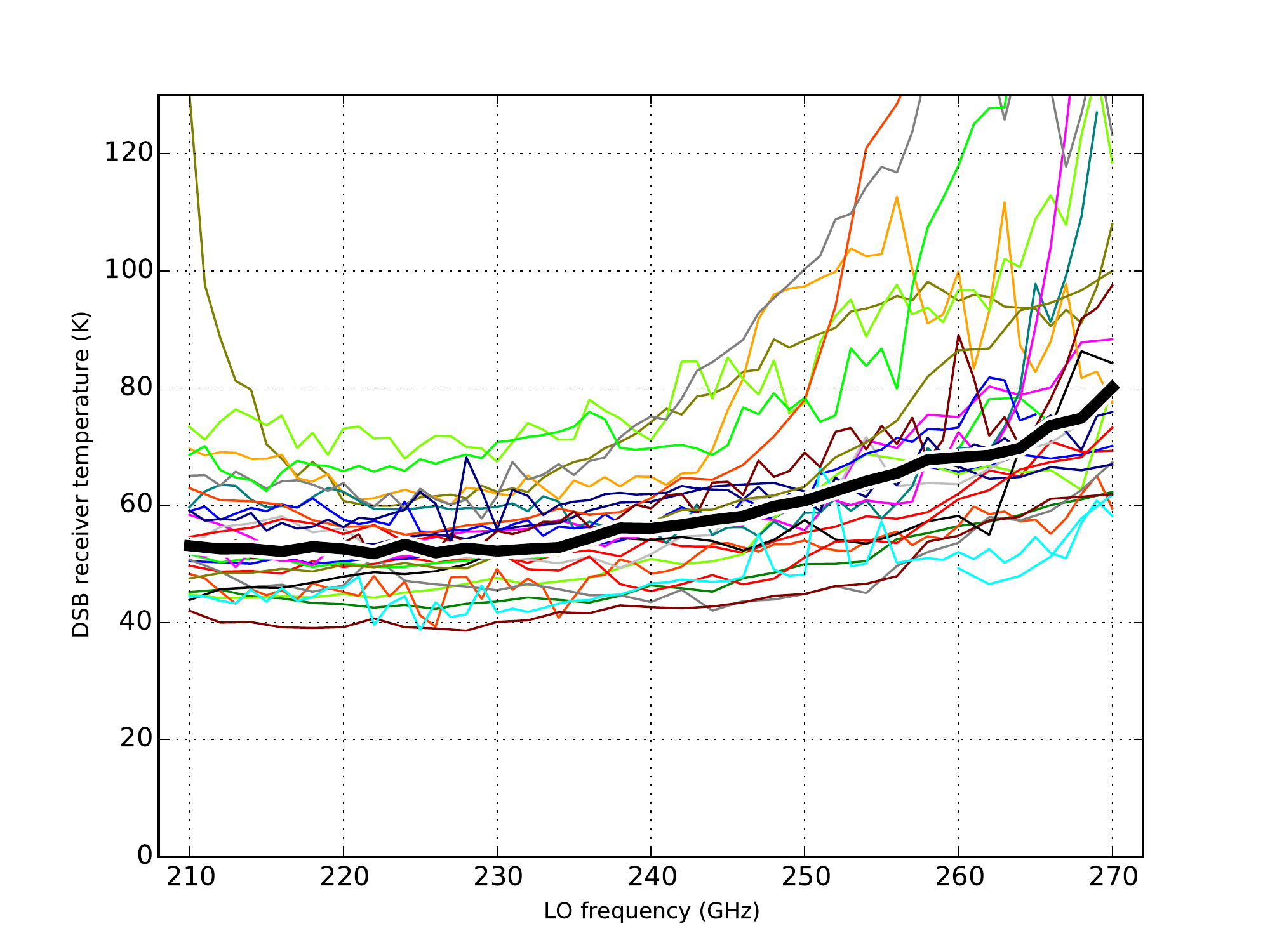}
\caption{\small DSB receiver temperatures as a function of LO frequency for the mixers used on
the array (except 4 that are not shown because their data were lost in a disk crash).  These
are based on total power measurements of the entire 1--9\,GHz IF passband with a broadband
power meter, with no band limiting filters.
The median receiver temperature is shown by the thick black curve.}
\label{fig:mixerscompare} 
\end{figure}

\begin{figure} [bt!]
\centering
\includegraphics[scale=0.36, clip, trim=2.6cm 1.5cm 0cm 1cm]{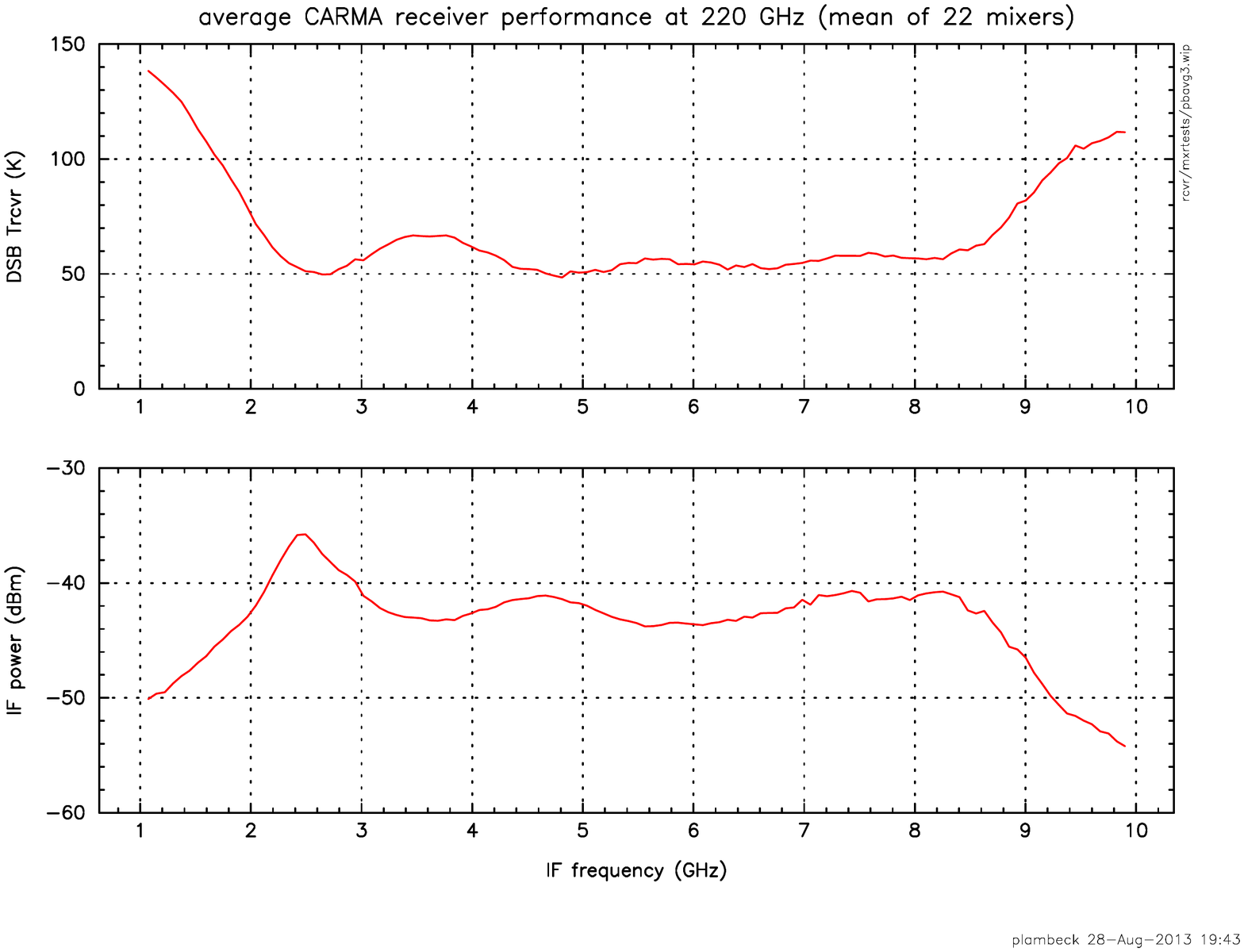}
\caption{\small IF output power and DSB noise temperature as a function of IF frequency, from an
average of 22 mixers.  Noise temperatures were based on measurements of the IF power made
with an Agilent E4407 spectrum analyzer, using a 3\,MHz resolution bandwidth.}
\label{fig:IFperformance} 
\end{figure}

Although at ALMA these devices are used in sideband-separating mixers
\citep{Kerr2013}, the CARMA mixers are a simple double sideband (DSB) design.
An advantage of this design is that the local oscillator can be coupled into
the mixer via a beamsplitter mounted outside the dewar.  Although a DSB mixer
folds together signals arriving in the sidebands above and below the local
oscillator frequency so that they appear at the same IF frequency, a
phase-switching pattern applied to the local oscillators allows the correlator
to separate the upper and lower sideband signals in cross correlation spectra.
Unfortunately, this is possible only for signals that are common to a pair of
telescopes, not for atmospheric noise that is independent for each telescope.
Thus, DSB mixers allow each correlator segment to process twice its bandwidth
in sky frequencies, but noise from both sidebands appears in the spectrum of
each sideband. 


A magnet mounted behind the SIS junction is used to tune out Josephson
tunneling through the insulating barrier, which otherwise adds noise and makes
it difficult to tune the mixers stably.  On ALMA an electromagnet is used for
this purpose, so the field can be adjusted to cancel the Josephson tunneling
very precisely; for simplicity, the CARMA mixers use just a tiny permanent
magnet for this purpose.  The magnet clings to a steel 4-40 set screw in the
aluminum block attached to the mixer.  The set screw may be adjusted (at room
temperature) to position the magnet to get about the right field.


One of the steps in optimizing the tuning of an SIS mixer is to adjust the LO
power level.  In the CARMA system the LO power cannot be adjusted
independently for the RCP and LCP mixers.  The LO is linearly polarized.  It is
coupled in through a beamsplitter.  As shown in Figure~\ref{fig:AllPolTests}, a
linearly polarized signal injected into the module will be coupled about
equally---but not perfectly equally---into the LCP and RCP mixer blocks.
However, mixers sometimes prefer to operate at different power levels.  It
would be possible to install a rotatable quarter-wave plate between the LO
waveguide horn and the beamsplitter to produce any elliptical polarization,
which would offer the flexibility of adjusting the power levels independently
on the two mixers, but to keep the system as simple as possible this is not
done.  Instead, the procedure is to optimize the LO power for the LCP mixer
only; the RCP mixer must live with whatever it gets.  Thus the
RCP noise temperatures at CARMA tend to be slightly higher than the LCP noise
temperatures.

Figure~\ref{fig:mixerscompare} displays the DSB receiver temperatures as a
function of LO frequency for the mixers installed on the telescopes.  These
noise temperatures were measured in the lab using 295\,K ambient and 77\,K cold
loads.  They are based on total power measurements of the entire 1--9\,GHz IF
band using a broadband power meter, with no bandpass limiting filters.  No
corrections were made for optics losses in the dewar or for noise contributions
from warm amplifiers.  The best devices have DSB receiver temperatures of
approximately 40\,K over much of the 210--270 GHz band.  The median receiver
temperature, shown by the thick black curve in the plot, is slightly greater
than 50\,K from 210--240\,GHz.

The average power and DSB noise temperature across the IF passband, at an LO
frequency of 220\,GHz, are shown in Figure~\ref{fig:IFperformance}.  The usable
part of the IF band is between about 2 and 8.5\,GHz.

\section{Calibrating polarization observations}

One approach to calibrating polarization data is through the use of Mueller
matrices.  A Mueller matrix is a transfer function between the observed and the
actual Stokes parameters.  This is the approach discussed, for example, by
\citet{Sault1995} and by \citet{Heiles2001b}.  We \textit{do not} use this
approach because CARMA data typically are analyzed with the 
\miriad{}\footnote{\,This font signifies the data reduction package \miriad{} itself; a
task, procedure, or keyword within \miriad{}; or a Python task.  \miriad{}
\citep{Sault1995} is one of the standard data-reduction software packages for
(sub)millimeter-wave interferometry.} software package, which breaks the
calibration into separate passband, gain, $R$--$L$ phase,\footnote{\,While we are actually 
measuring the $R$--$L$ phases at CARMA, the quantity is known in \miriad{} as \XYphase{}.}
and leakage steps.  

The passband and gain calibrations are handled by \miriad{} tasks \mir{MFCAL} and
\mir{SELFCAL}.  The passbands generally are stable for the duration of each 6--8 hour
observation, and the gain phases of the $R$ and $L$ channels track each other very
closely as well.  The $R$ vs. $L$ gain amplitudes vary substantially only if one
receiver is mistuned; generally amplitude variations are due to pointing or
focus errors, which apply equally to the $R$ and $L$ channels.

The $R$--$L$ phase calibration corrects for the phase difference between the $R$
and $L$ channels on each telescope (i.e., the $R$--$L$ phase) caused by delay
differences in the receiver, underground cables, and correlator cabling.  The
$R$--$L$ phase is not a single number, but is a function of frequency, due to
fiber- and cable-length differences.  That is, the same piece of the IF in
different correlator sections can have completely different $R$--$L$ phases.
See Section \ref{sec:xyphase} for a detailed discussion of $R$--$L$ phase and position-angle calibration.

The leakage corrections compensate for cross-coupling between the $R$ and $L$
channels, caused by imperfections in the polarizers and OMTs, reflections
inside the dewars, and cross-coupling of IF signals in the analog electronics
both preceding and within the correlator. 
See Section \ref{sec:leakage} for a detailed discussion of leakage calibration.

\section{$R$--$L$ phase calibration}
\label{sec:xyphase}
For circularly polarized feeds, the phase difference between the $R$ and $L$
channels is what measures the position angle of an incoming linearly polarized
signal.  An $R$--$L$ phase difference of $2\chi$ corresponds to a
linearly polarized position angle $\chi$.\footnote{\,For example, 
a $180^{\circ}$ rotation of the $R$--$L$ phase leads to a
$90^{\circ}$ rotation of $\chi$.  This is because of the circular-to-linear conversion: an
$R$--$L$ phase change of 90$^{\circ}$ will transform the radiation from linear
($R$ and $L$ in phase) to circular (90$^{\circ}$ out of phase); another 90$^{\circ}$ change
will transform the radiation from circular to linear (180$^{\circ}$ out of
phase, with a position angle perpendicular to the original linear
orientation).}
The ordinary passband correction done with \mir{MFCAL} analyzes only on the
parallel-hand signals $LL$ and $RR$; it does not solve for the $R$--$L$
phase difference.  In order to fit the $R$--$L$ phase it is necessary to observe a
linearly polarized source with a known polarization position angle.  $R$--$L$ phase
calibration on astronomical sources is difficult at millimeter wavelengths, since most
calibrators are weakly polarized and their polarization position angle varies
on time scales of weeks or months.  At centimeter wavelengths 3C286 is the
usual polarization calibrator.  It is approximately 15\% polarized, and its
position angle has been stable for decades \citep{Perley2013}.
Unfortunately, its flux density at 230\,GHz is only about 0.4\,Jy.

\subsection{Grid calibration}
\label{sec:grids}

\begin{figure*} [bt!]
\begin{minipage}[b]{0.4\linewidth}
\centering
\includegraphics[scale=0.12]{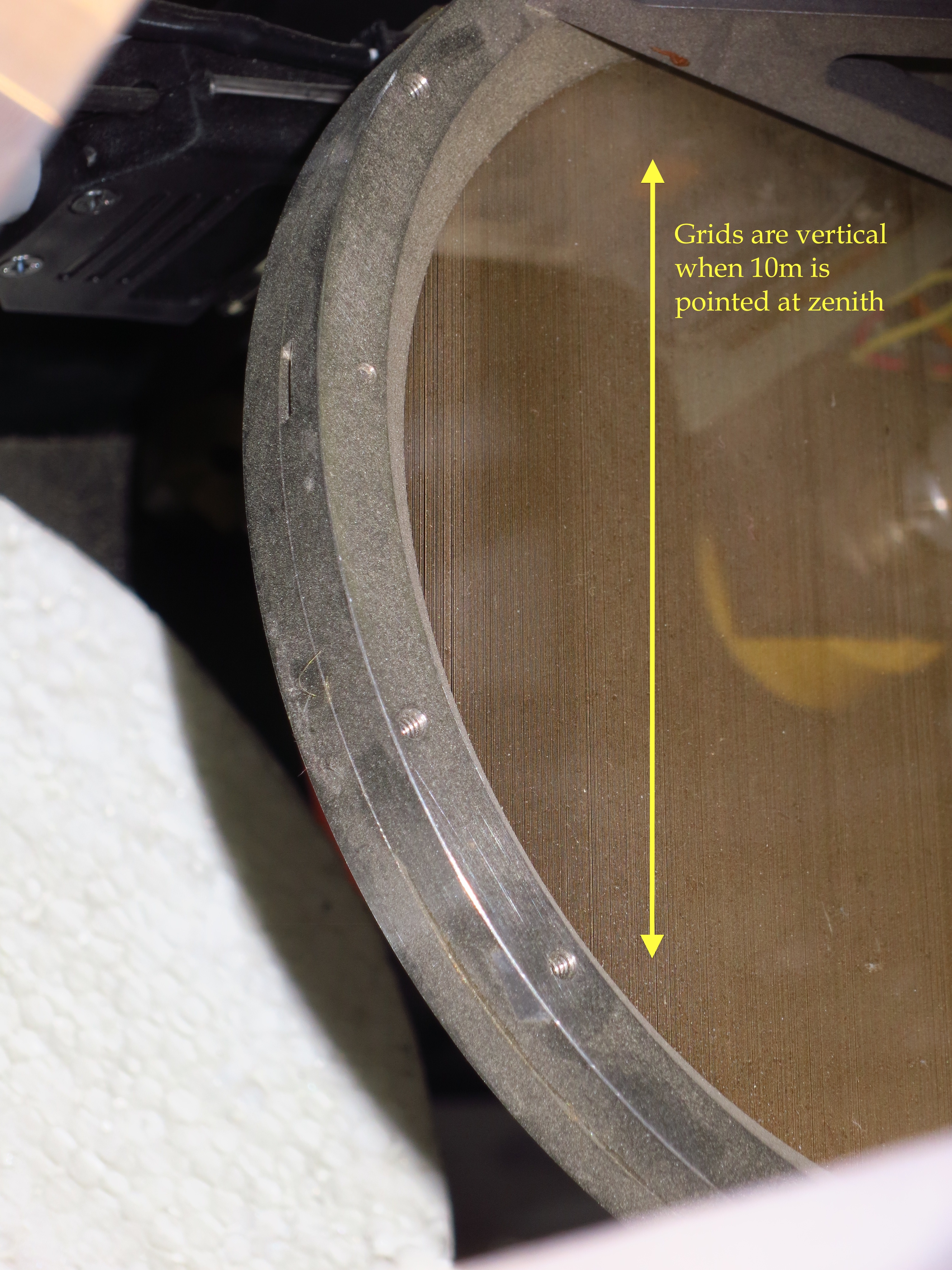}
\end{minipage}
\hspace{0.05\linewidth}
\begin{minipage}[b]{0.5\linewidth}
\centering
\includegraphics[scale=0.15]{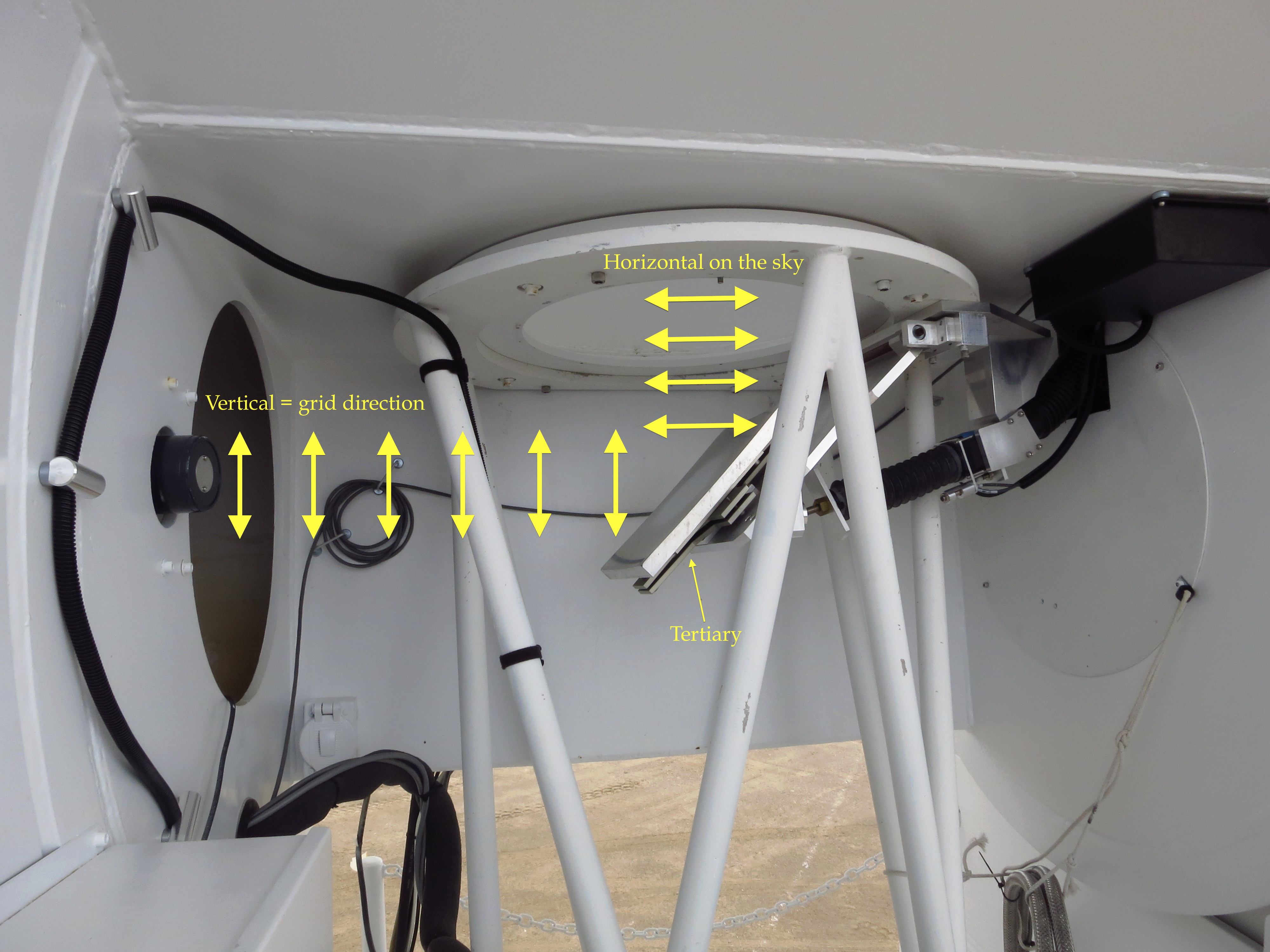}
\end{minipage}
\caption{\small \textit{Left:} Photo of a wire grid polarizer in the receiver
cabin of a 10\,m telescope, with the telescope pointed at zenith.  The wires in
the grid are vertical.  \textit{Right:} Photo of the tertiary mirror with the
10\,m telescope pointed at zenith.  Horizontally polarized radiation from the sky
reflects off the tertiary such that it is vertically polarized at the receiver.
The receiver with its wire grid rotates in elevation along with the telescope,
so this correspondence is preserved at all elevations.
}
\label{fig:gridPA}
\end{figure*}

\begin{figure} [bt!]
\begin{center}
\includegraphics[scale=0.355,clip,trim=2.3cm 1cm 0cm 3.5cm]{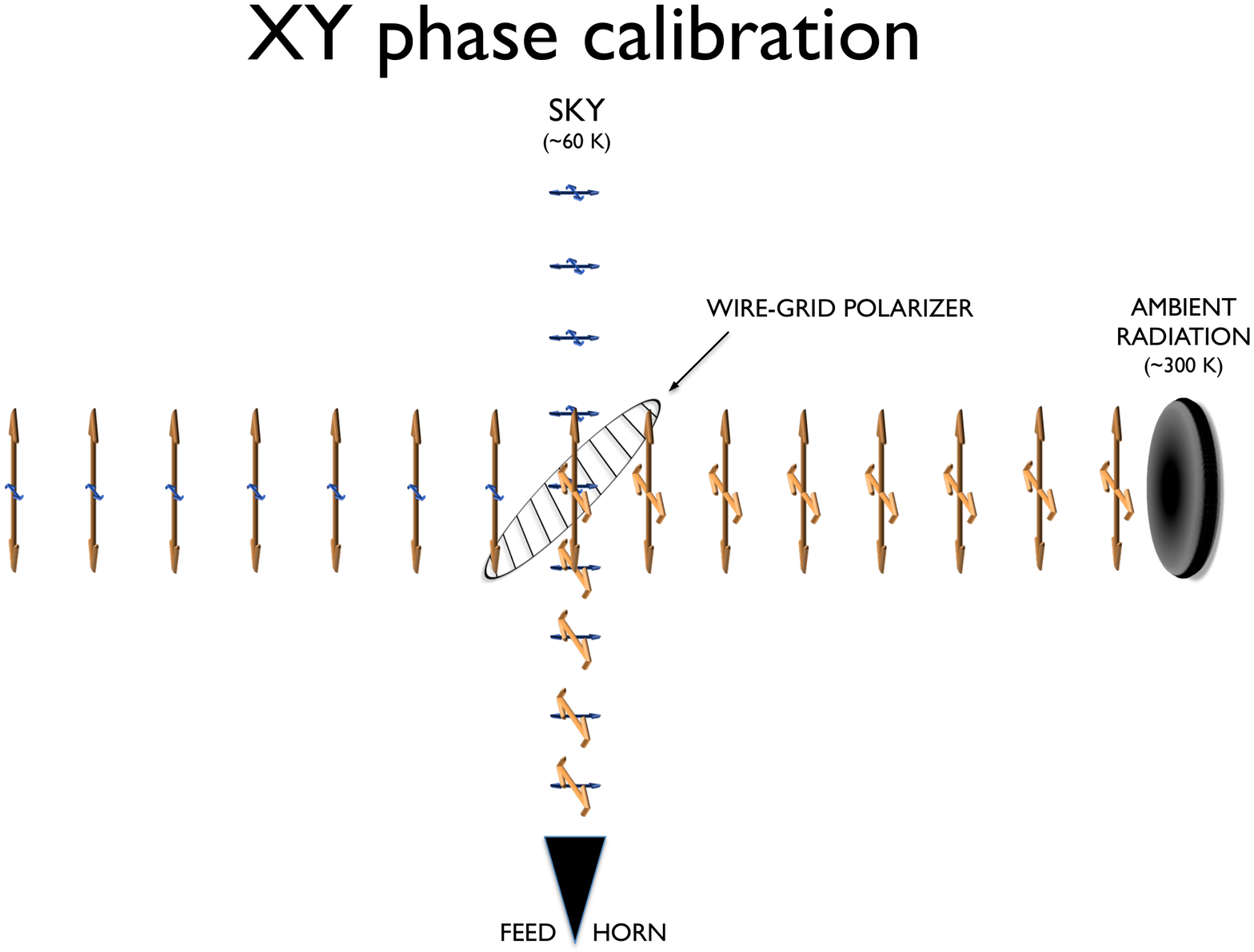}
\caption[$R$--$L$ phase and wire grid schematic]{
\small A schematic showing how the wire grid polarizers produce a highly polarized noise
source by reflecting ambient radiation into the receiver.
}
\label{fig:XYphase_cartoon}
\end{center}
\end{figure}

Not only is it impractical to calibrate the $R$--$L$ phase using astronomical sources,
but matters are further complicated because in the CARMA system 
there is not a single $R$--$L$ phase difference for each telescope.  Rather, there
is a different $R$--$L$ \textit{delay}---hence, a phase slope as a function of
frequency---for each correlator band. 

These band-dependent delays arise in a number of places in the signal path:
(1) on the front end, the path lengths from the input port of the
OMT to the $R$ and $L$ outputs differ by 0.1\,inch.\footnote{\,A distance of 0.1\,in is a
couple of wavelengths at 230\,GHz.  However, both the incoming radiation and the
LO travel through this same length, and it is only the difference in phase
shift at these two frequencies that matters.  For an IF frequency of 5\,GHz,
this differential phase shift is $\sim$\,20$\degree$.}  (2) There are 
chromatic effects in the polarizer and OMT that cause slight differences
in the phase shifts for signals in the upper and lower sidebands---see Section
\ref{sec:XYphase_limits} for further discussion.  (3) There are differences 
in the $R$ and $L$ fiber lengths from the receivers to the correlator.
(4) In the block downconverters, there are cable-length differences between the upper
(>\,5\,GHz) path (the path with the mixer) and the
lower (<\,5\,GHz) path (without the mixer).  

To overcome these various difficulties, we use
observations of artificial linearly polarized noise sources to calibrate the
$R$--$L$ phase.  The noise sources are created by inserting wire grid polarizers
(Figure~\ref{fig:gridPA}) into the beams of the 10\,m telescopes.  With the grid
in place, one linear polarization reaching the receiver originates from the
sky, while the other originates from a room temperature load.  Since the room
temperature load is much hotter than the sky, the receiver sees thermal noise
that is strongly polarized (Figure~\ref{fig:XYphase_cartoon}).  

By default, the master observing script performs grid observations
approximately every 45 minutes during full polarization observing tracks.  The
calibration requires between 1 and 2 minutes: 45 seconds for the grids to move
in and out of the beams, and 30 seconds of integration.  The grid data are
identified by setting the \mir{purpose} keyword to \mir{P} in the data header.
Grid observations are done at the end of phase calibration observations.
Although the sky signal includes flux from the calibration source, this is
negligible compared with the huge signal from the ambient load, and thus does
not affect the measured $R$--$L$ phase.\footnote{\,For example, 3C279 has a
brightness of $\sim$\,10\,Jy, and is about 10\% polarized.  The antenna gains
of the 10\,m telescopes are about 65 Jy/K, so in terms of brightness
temperature, the flux from the 3C279 is roughly 15\,mK.  This is two to three orders of
magnitude smaller than the (polarized) temperature difference between sky and
ambient load, which is of order 100\,K even at low elevation in bad weather.} 

Since the polarized noise is local to each telescope, it does not show up in
cross-correlations with other telescopes.  However, it leads to a strong $LR$
\textit{autocorrelation} (or ``cross-auto'') signal, which is the cross-correlation of the $L$ and
$R$ channels from a single antenna.  The \miriad{} task \mir{XYAUTO} averages
together all the $LR$ autocorrelation data for the grid observations in each
dataset to create a channel-by-channel passband correction for the 10\,m
telescopes.  Rewriting the data with \mir{UVCAT} or \mir{UVCAL} applies these
corrections.  In the new dataset, the phases of all the $R$ and $L$ channels
are equal on the 10\,m telescopes.  This means that a linearly polarized signal
reaching the receiver with the same PA as the noise source will produce $LR$
and $RL$ correlations with phases of zero.  One of these 10\,m telescopes
\textit{must} then be used as the reference antenna for the regular passband
correction performed with \mir{MFCAL}.  The passband correction synchronizes
the $R$ phases on all telescopes with the $R$ phase of the reference antenna,
and the $L$ phases on all telescopes with the $L$ phase for the reference
antenna.  Since the $R$ and $L$ phases of the reference antenna are equal, then
the $R$ phases of all antennas equal the $L$ phases on all antennas.

\smallskip
\noindent
\textbf{Absolute position angle of the grids.} 
\miriad{} computes the polarization position angle of a source as 
$\chi = q + \mir{evector}$,
where $q$ is the source's parallactic angle\footnote{\,The parallactic angle is
defined as the angle between two great circles, one through a celestial object
and the zenith, the other (the ``hour circle'') through the object and the
celestial poles.  The hour circle is analogous to a longitude line on a globe.
In the zenith--object--celestial pole triangle, the parallactic angle is the
angle at the celestial object's corner.  The parallactic angle is zero when the object crosses the
meridian.} and the $uv$-variable \mir{evector} is defined, for linear feeds, as the position
angle of the $X$-feed relative to the local vertical.  For circular feeds,
\mir{evector} is interpreted as the position angle $\chi$ for which RCP and LCP
radiation are in phase with one another, which is determined by
the angle of the wire grid noise source.

With the grids in place and the telescopes pointed at zenith, noise reaching
the receivers is vertically polarized.  However, because of the reflection off
the tertiary mirror, a vertically polarized signal in the receiver cabin
corresponds to a horizontally polarized ($\chi = 90\degree$) source on the sky
(Figure~\ref{fig:gridPA}).  
Consequently, in datasets from CARMA the \miriad{} $uv$-variable \mir{evector} is set to
90$\degree$. 


\subsection{Systematic limits to $R$--$L$ phase accuracy}
\label{sec:XYphase_limits}

The grid calibration of $R$--$L$ phase is susceptible to a number of limitations
that an astronomical calibration would not be.

\smallskip
\noindent
\textbf{USB and LSB are averaged together.} The autocorrelation averages
together the $R$--$L$ phase differences in the upper and lower sidebands.  In
normal cross-correlation spectra between antennas $m$ and $n$, the LSB and USB
are separated by demodulating the phase-switching pattern between the $m$ and
$n$ local oscillators.  This is not possible for the $LR$ autocorrelation
spectrum for a single antenna, since both the RCP and LCP channels use a common
LO.

The delays that contribute to the $R$--$L$ phase difference are mostly at the IF frequency, and
thus produce precisely the same phase shifts for signals in the LSB and USB.
The only exception is the difference in delay from the beamsplitter
to the RCP and LCP mixers. There is a 0.1\,in length difference through the OMT
for the two polarizations, and the waveguide is slightly dispersive, so the
delay is slightly different for the USB and LSB.  A quick calculation
shows that this effect is small: the phase shift of the LSB relative to the LO
is at most 2$^{\circ}$ greater than the phase shift of the USB relative to the
LO.\footnote{\,The guide wavelength is 
$\lambda_g = \lambda_0/\sqrt{1 - (\lambda_0/2a)^2}$,
where $\lambda_0$ is the wavelength of the radiation in free space and 
$a$ is the broad dimension of the waveguide.  The dispersion is greatest at the low end of the band.
Taking an LO frequency of 210\,GHz and an IF of 8\,GHz, the phase shift is
36.2$^{\circ}$ for the LSB and 34.8$^\circ$ for the USB.} The waveguide polarizer
also is slightly chromatic, so the phase shifts through it will differ slightly
for the upper and lower sidebands.

\smallskip
\noindent
\textbf{Common noise.} Noise that is common to both the $R$ and $L$ channels
will generate an $LR$ autocorrelation signal even when the wire grid is out of the beam.
For example, noise radiated out the input of one mixer could be transmitted out
through the OMT, polarizer, and feed horn, reflect back into the module from
the dewar window (as the opposite circular polarization), and be coupled into
the opposite mixer.    
If the $R$ and $L$ mixers had independent local oscillators with different
phase switch patterns, these signals could be rejected in the $LR$
autocorrelations; however the local oscillator is shared between the two
polarizations in the CARMA receivers, so this is not possible.  

In order to derive the $R$--$L$ phase with high accuracy, the signal
from the polarized noise source
must be much larger than the background level.  To test this, we compared the
$LR$ amplitudes in typical 1\,mm weather with and without the grids in place.
The amplitudes are $\sim$\,20 times
higher with the grids in.  This corresponds to an rms error of
$\sim$1/20~radian, about 3$^\circ$, which leads to an uncertainy of
$1.5^{\circ}$ in position angle.  Of course, in poor weather or at low
elevation the uncertainty can be greater because the contrast between sky and
ambient is less, so the polarized noise level is lower.

\smallskip
\noindent
\textbf{Leakage.}
There is always some crosstalk between the $R$ and $L$ channels due to
polarization leakage.  As described in Section~\ref{sec:leakage}, the leakages
exhibit considerable frequency structure. 
Ideally one would solve for $R$--$L$ phase and leakage in an iterative way, but
this is not easily accomplished with \miriad{}, since there is no simple way of
applying leakage corrections on a channel-by-channel basis.  This means
that the leakages can introduce ripples into the $R$--$L$ phase
calibration.

\begin{figure} [hbt!]
\begin{center}
\includegraphics[scale=0.33,clip, trim=0.75cm 1cm 0cm 1cm]{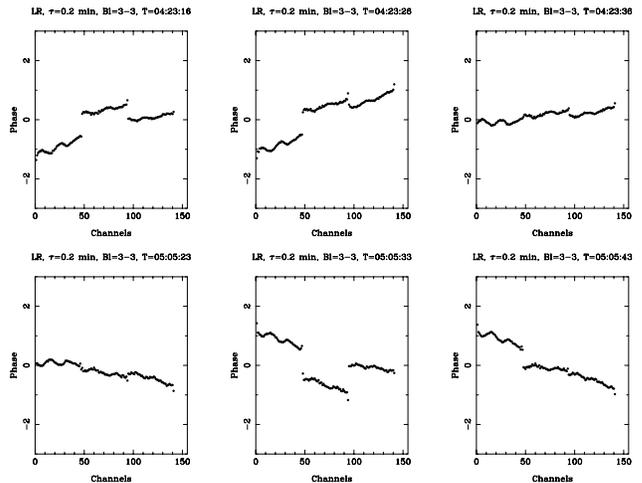}
\caption[$R$--$L$ phase jitter from PLL noise]{ \small
A sequence of cross-auto ($LR$) grid calibration data on antenna C3, taken at 10 second
intervals, after the $R$--$L$ phase correction has been applied.  The 
1--2$^{\circ}$ residuals are caused by phase jitter 
in the digitizer clocks.
}
\label{fig:XYphase_jitter}
\end{center}
\end{figure}

\smallskip
\noindent
\textbf{Phase-lock loop (PLL) jitter.}
Each correlator band uses separate digitizers for $R$ and $L$. The digitizers are not 
run off of a single 1\,GHz clock, but instead are run off of independent clocks that are phase-locked to a common reference signal.
One systematic effect that limits our ability to measure absolute position angle is rapid
variation in $R$--$L$ phase solutions caused by jitter (phase noise) in the 1\,GHz PLL outputs.
See Figure \ref{fig:XYphase_jitter} for plots of the $R$--$L$ phase residuals (after correction),
which show 1--2$^{\circ}$ variations on very short (10\,s) timescales.
This effect should average out for most astronomical measurements.

\section{Leakage Calibration}
\label{sec:leakage}

Leakage corrections compensate for cross-coupling between the $R$ and $L$
channels, caused by imperfections in the receivers or crosstalk in the analog
electronics that precede the correlator. Leakages 
are measured in terms of
voltages (1\% [or 0.01] leakage in voltage corresponds to $10^{-4}$ in power),
and are defined in the following way \citep[][Equation 4.42]{TMS}:
\begin{align}
v^\prime_R &= v_R + D_R\,v_L \\
v^\prime_L &= v_L + D_L\,v_R \,\, ,
\end{align}
\noindent
where $v^\prime_R$ and $v^\prime_L$ are the observed signals, 
$v_R$ and $v_L$ are the true signals, 
$D_R$ is the leakage from $L$ into $R$, and
$D_L$ is the leakage from $R$ into $L$.\footnote{\,In \mir{MIRIAD}, $D_x \rightarrow D_R$, and $D_y \rightarrow D_L$.}
Note that the leakages are complex numbers.

Leakages are calibrated by observing a strong source (usually the gain
calibrator) over a range of parallactic angle.  
The calibrator may be polarized or unpolarized.  Since the telescopes have
altitude-azimuth (alt-az) mounts, the source's polarization will appear to vary, in the frame of
the receivers, as the telescopes track it across the sky, modulating the $LR$
and $RL$ cross correlation amplitudes in a predictable way.
The component of the cross correlation amplitude that does \textit{not}
vary with parallactic angle must then be due to instrumental leakage.

\mir{MIRIAD} task \mir{GPCAL} fits observations of the calibrator in order to
solve simultaneously for the antenna gains vs. time, the polarization leakages,
and the source polarization.  Given good weather and normal antenna
performance, a 4--6\,hr observation for which the source parallactic angle
varies by more than about 60$^\circ$ yields a reliable leakage solution.

Unfortunately, \mir{MIRIAD} allows for only a single pair of leakage
corrections $D_x$ and $D_y$ for each telescope.  To solve for
channel-by-channel leakages requires one to run \mir{GPCAL} many times,
specifying each channel range in turn and saving the results.  We have written
a library of Python routines to handle this chore, and have accumulated a
library of leakage solutions covering many frequency ranges.

\begin{figure*} [hbt!]
\begin{center}
\includegraphics[scale=.66, clip, trim=1cm 0cm 1cm 1cm]{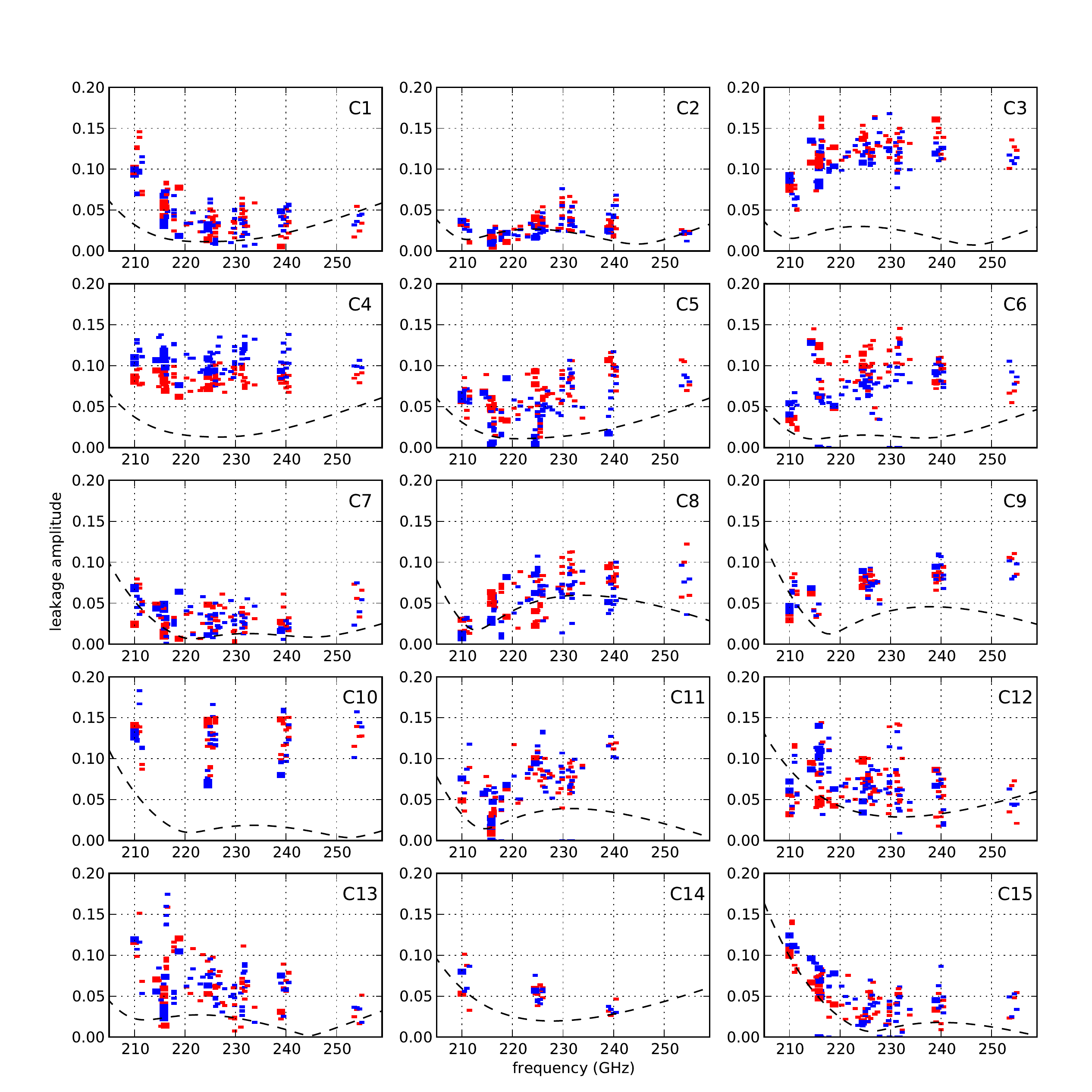}

\caption{ \small Leakage amplitudes vs. frequency for all telescopes, derived
from 12 datasets obtained over a 3.5 year period.  Each point shows the leakage
amplitude for a single 0.5\,GHz wide correlator window.  Red points indicate
$D_R$; blue points, $D_L$.  The dashed curve in each plot shows the leakage
expected from the circular polarizer on that antenna, based on the polarizer
measurements shown in Figure~\ref{fig:AllPolTests}.  Fewer data are plotted for some
antennas because of receiver swaps.  }

\label{fig:LeakAll}
\end{center}
\end{figure*}

\smallskip
\noindent
\textbf{Expected leakage amplitudes.} Ideally, the polarization
leakages would be due exclusively to imperfections in the waveguide polarizers.
Figure~\ref{fig:PolLeak} shows that we expect leakage amplitudes of 
$\sim$\,2--3\% for polarizers that are within the expected dimensional
tolerances.  From the polarizer test data in Figure~\ref{fig:AllPolTests}, we inferred
that many of the polarizers were \textit{not} within these tolerances; however, the
leakage vs. frequency expected from each polarizer can be computed if its dimensions are
known, using the software described in \citet{Plambeck2010}.

Figure \ref{fig:LeakAll} shows the band-averaged leakages for all telescopes
based on observations obtained from 2011--2015.  In some cases dewars were
swapped, so fewer datasets were used.  
The dashed curve on each plot shows the
theoretical leakage expected for the particular circular polarizer on that
telescope, based on the fits to the polarizer dimensions shown in
Figure~\ref{fig:AllPolTests}.  In most cases the leakage amplitudes are
substantially larger than expected from the polarizer alone, although for a few
telescopes (C1, C2, C8, C15) the polarizer curve appears to form a lower bound
to the measured leakages.  

\begin{figure*} [hbt!]
\begin{center}
\includegraphics[scale=.49, clip, trim=0cm 0cm 0cm 0cm]{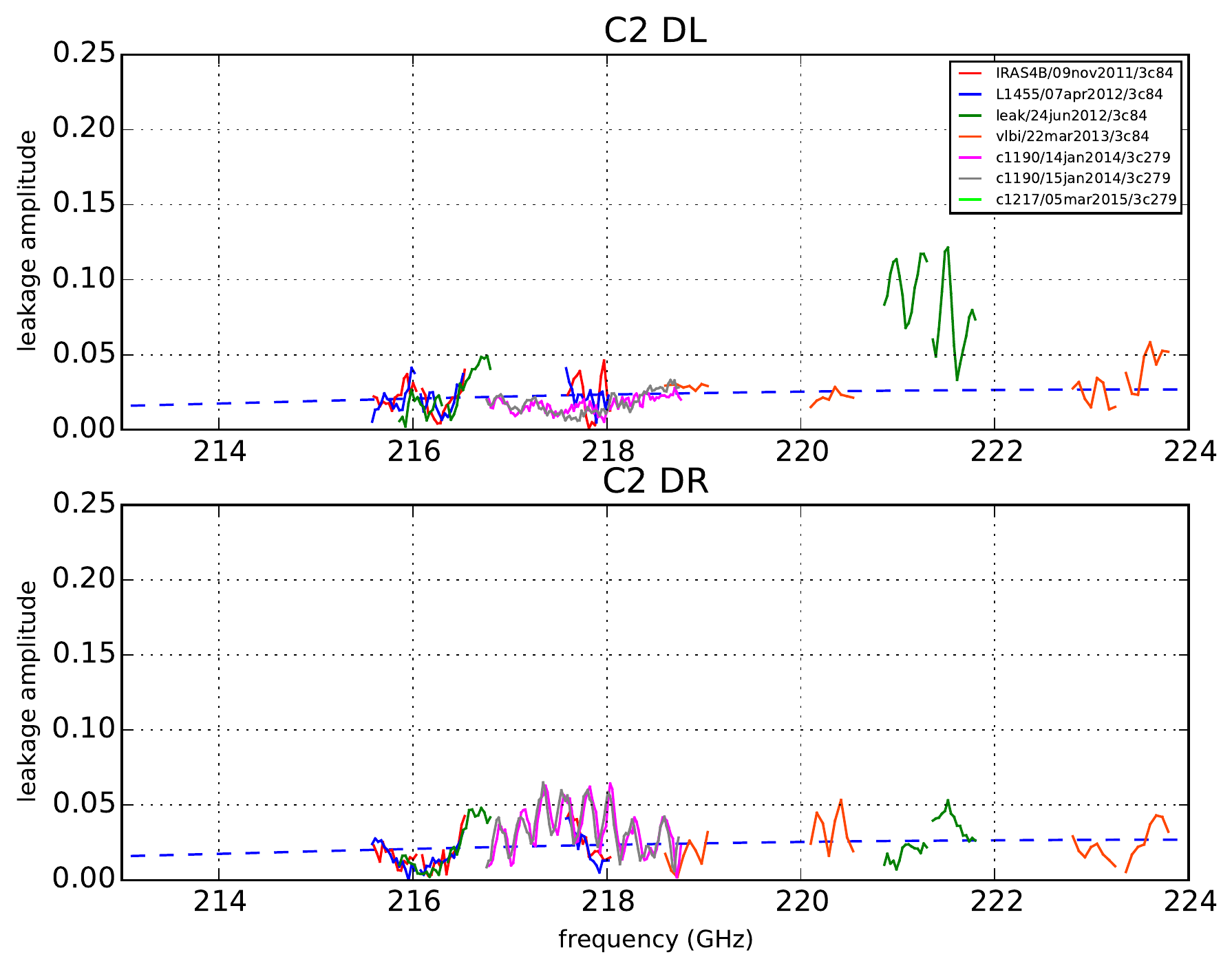}
\vspace{0.1in}
\includegraphics[scale=.49, clip, trim=0cm 0cm 0cm 0cm]{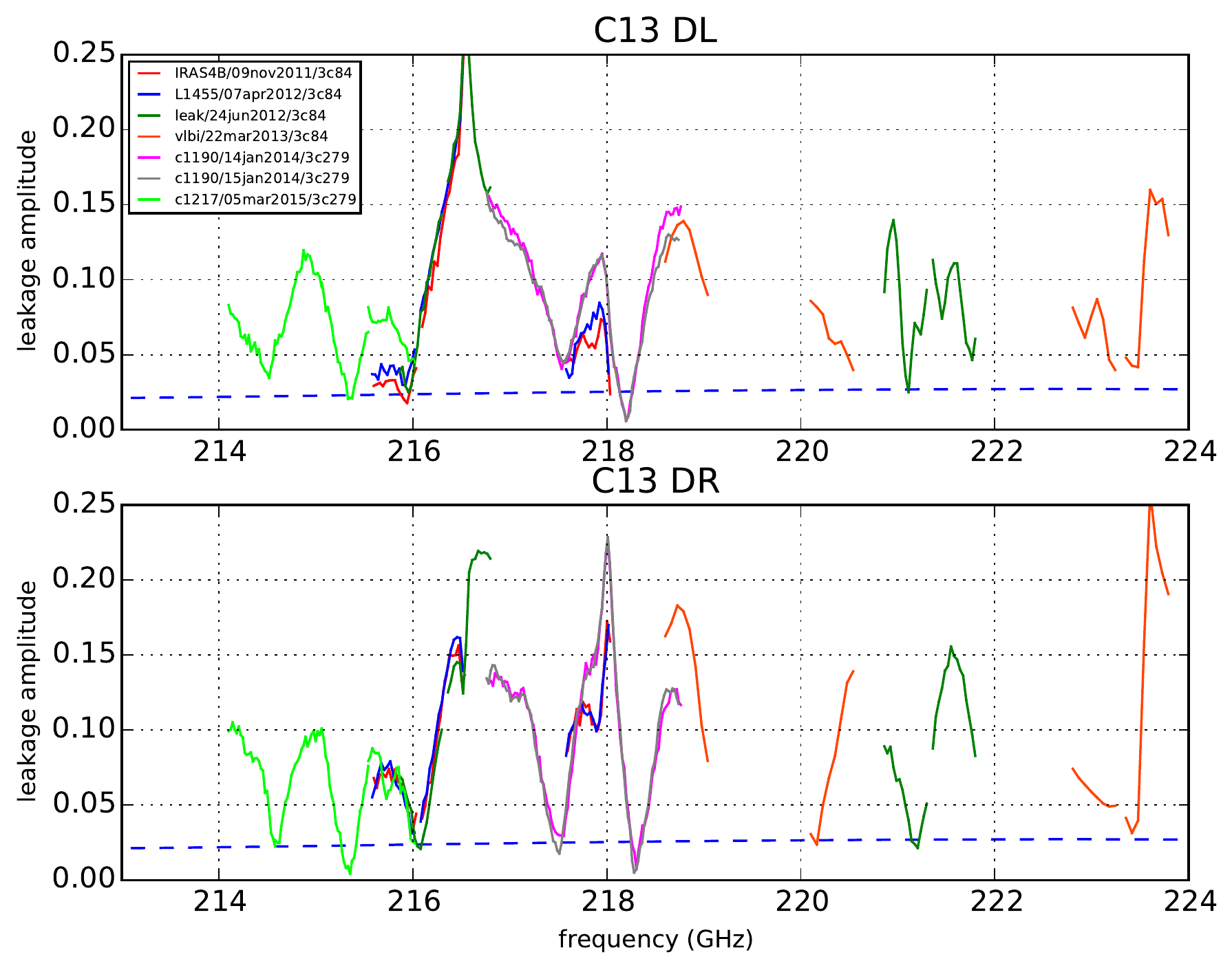}
\caption[Ripples in leakage terms]{ \small Leakage amplitudes vs. frequency
for antennas C2 and C13 over the frequency range 214--224\,GHz.  The
frequency resolution is typically 50\,MHz.  In most cases there
is excellent reproducibility in the leakages obtained months or years apart.
Dashed curves
show the leakage amplitude expected from the circular polarizers on these
telescopes.}
\label{fig:C2andC13leaks}
\end{center}
\end{figure*}

With higher frequency resolution, one finds that in most cases the leakages have
substantial frequency structure, 
with periods of a few $\times$\,100\,MHz to a few GHz.  Figure~\ref{fig:C2andC13leaks}
shows examples of this structure for C2, one of the
best antennas, and for C13, one of the worst.  The ripples in the leakages
suggest that cross coupling of the unwanted polarization takes place via
multiple paths with different delays.  We discuss possible sources of
cross-coupling in the sections below.

\smallskip
\noindent
\textbf{Cross-coupling in the block downconverter.} One source of leakage
ripples is cross-coupling of IF signals in the correlator room, probably in the
block downconverters.  Each of the 8 analog downconverters assigned to a
telescope obtains its input from a 4-way switch on that telescope's block
downconverter.  The 4 inputs to the switch are the low-band RCP, low-band LCP,
high-band RCP, and high-band LCP, where ``low-band'' is the 1--5\,GHz piece of
the IF band, and ``high-band'' is the 5--9\,GHz piece (which has been downconverted to
1--5\,GHz).  Although nominally the switch provides 45\,dB isolation between ports,
this is not adequate if the RCP and LCP power levels differ substantially.  For
example, if the RCP power is 15\,dB greater than the LCP power, then the
relative level of RCP coupled into the LCP IF is --30\,dB, which means that
$|V_R| = 0.03\,|V_L|$.
This RCP signal beats with the RCP voltage coupled via the polarizer, producing
a 6\% peak-peak ripple in $D_L$.  Meanwhile, the relative level of LCP in the
RCP IF is --60\,dB, which produces just 0.2\% ripple in $D_R$.  

\begin{figure*} [bt!]
\begin{center}
\includegraphics[scale=.7, clip, trim=1.5cm 6cm 1.5cm 5.5cm]{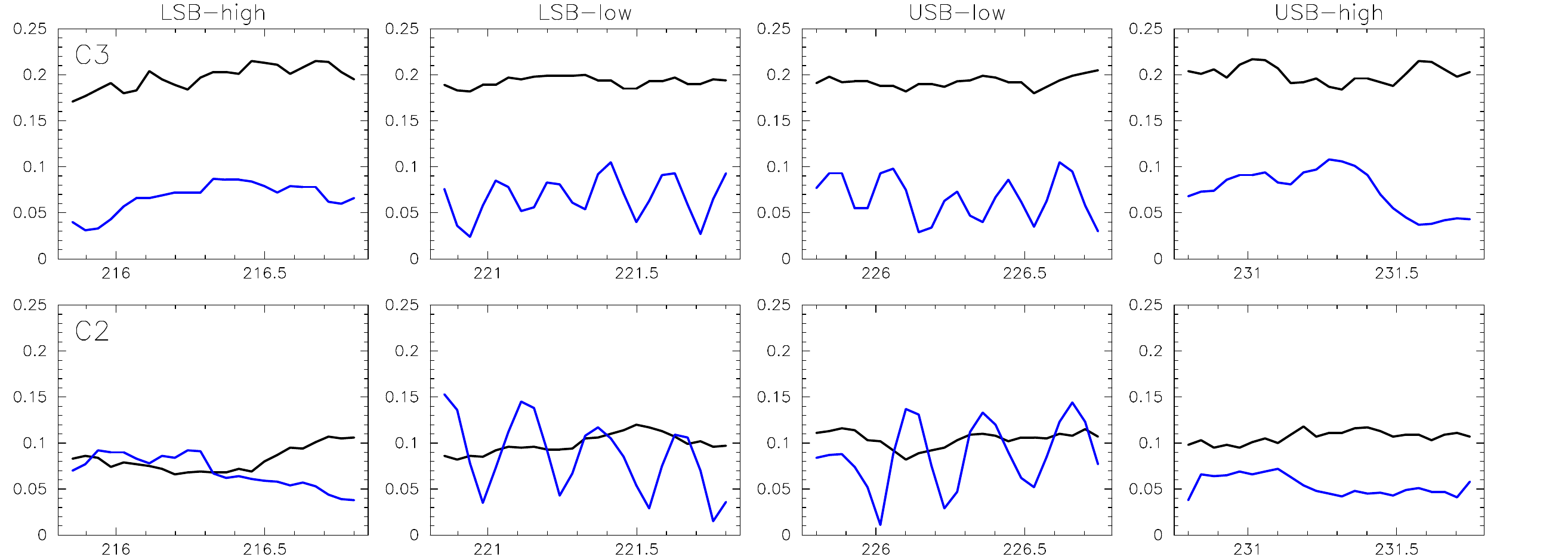}
\caption[Ripples caused by
cross-coupling in block downconverters]{ \small Leakage amplitudes for antennas
C2 and C3 plotted as a function of sky frequency, derived from 
a dataset where two
correlator sections were positioned above 5\,GHz in the IF (outer two columns,
``LSB-high'' and ``USB-high''), and two below 5\,GHz (middle two
columns, ``LSB-low'' and ``USB-low''); note that the lowest sky frequency corresponds
to the highest IF frequency in the LSB.    Black curves are $D_R$ and blue curves
are $D_L$.  Large ripples are evident only in $D_L$, and only below IF frequencies of 5~GHz (the ``low'' bands),
indicative of cross-coupling in the block downconverters.}
\label{fig:ripples_blockdcon} \end{center} \end{figure*}

Thus, cross-coupling in the block downconverter manifests itself as leakage
ripples in one polarization, but not the other.  An example of this behavior is
shown in Figure \ref{fig:ripples_blockdcon}.  In this observation 2 correlator
sections were centered at 2.5\,GHz in the IF, while the other 2 were centered at
7.5\,GHz.  Prominent leakage ripples are seen only in $D_L$, and only in the 1--5\,GHz
section of the IF band.  Evidently the RCP and LCP power levels are more
closely balanced above 5\,GHz.

\begin{figure*} [hbt!]
\begin{center}
\includegraphics[scale=.73, clip, trim=2cm 13cm 0cm 2.5cm]{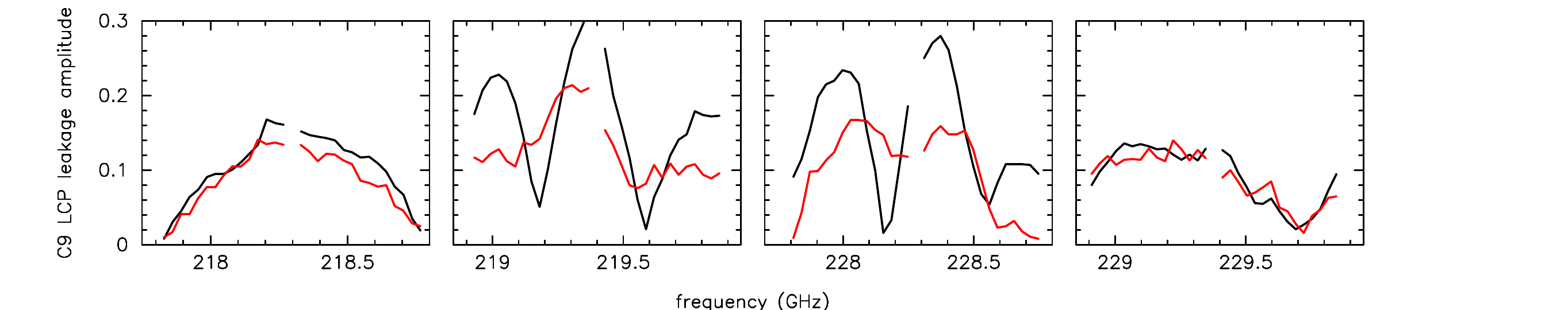}
\caption[Ripples before and after attenuation]{ \small
$D_L$ leakage amplitude vs. frequency for antenna C9 before (black) and after (red)
attenuating the RCP input to the block downconverter by 10\,dB.  The frequency
setup is similar to that in
Figure~\ref{fig:ripples_blockdcon}.  Lowering the RCP power level in the downconverter
reduced coupling of RCP into LCP, hence reduced the magnitudes of the leakage ripples
in $D_L$ for the two ``low-band'' correlator sections.  
The $D_R$ leakages were unchanged, as were the leakages on all other telescopes. \\
}
\label{fig:ripples_blockdcon_corrected}
\end{center}
\end{figure*}

Figure~\ref{fig:ripples_blockdcon_corrected} shows that the ripples can be
reduced by equalizing the RCP and LCP power levels.  For these observations the
RCP power into the block downconverter on telescope C9 was initially 
9\,dB greater than the LCP power.  Installing a 10\,dB attenuator on the RCP
input to the block downconverter substantially reduced the ripple in $D_L$ for the
the low-band sections, and caused no deterioration in performance above 5\,GHz or
in $D_R$.

\smallskip
\noindent
\textbf{Reflections in the receiver.}
The ripples that we attribute to cross-coupling in the block downconverters
have periods of about 250\,MHz.  Many antennas, particularly the 6\,m antennas (C7--C15),
also have leakage ripples with a period of about 1\,GHz (see the results for
C13 plotted in Figure~\ref{fig:C2andC13leaks}), which corresponds to
cross-coupling with a 1\,nsec delay.  We hypothesize that this is caused by
reflections inside the dewar.  In the 6\,m dewars the front of the feed horn is
approximately 5\,cm behind the dewar window, and the path lengths through the
components are roughly 5\,cm for the horn, 2\,cm for the waveguide polarizer,
and 3\,cm for the OMT.  Thus, an RCP signal that reflected off the RCP mixer
would travel back through a 15\,cm path to the dewar window (as RCP).
Reflection off the window would convert it to LCP.  It then would travel
through another 15\,cm path to the LCP mixer, accounting for a total delay of
30\,cm/$c$, or
1\,nsec.

\begin{figure} [bt!]
\begin{center}
\includegraphics[scale=.48, clip, trim=0cm 0cm 0cm 0cm]{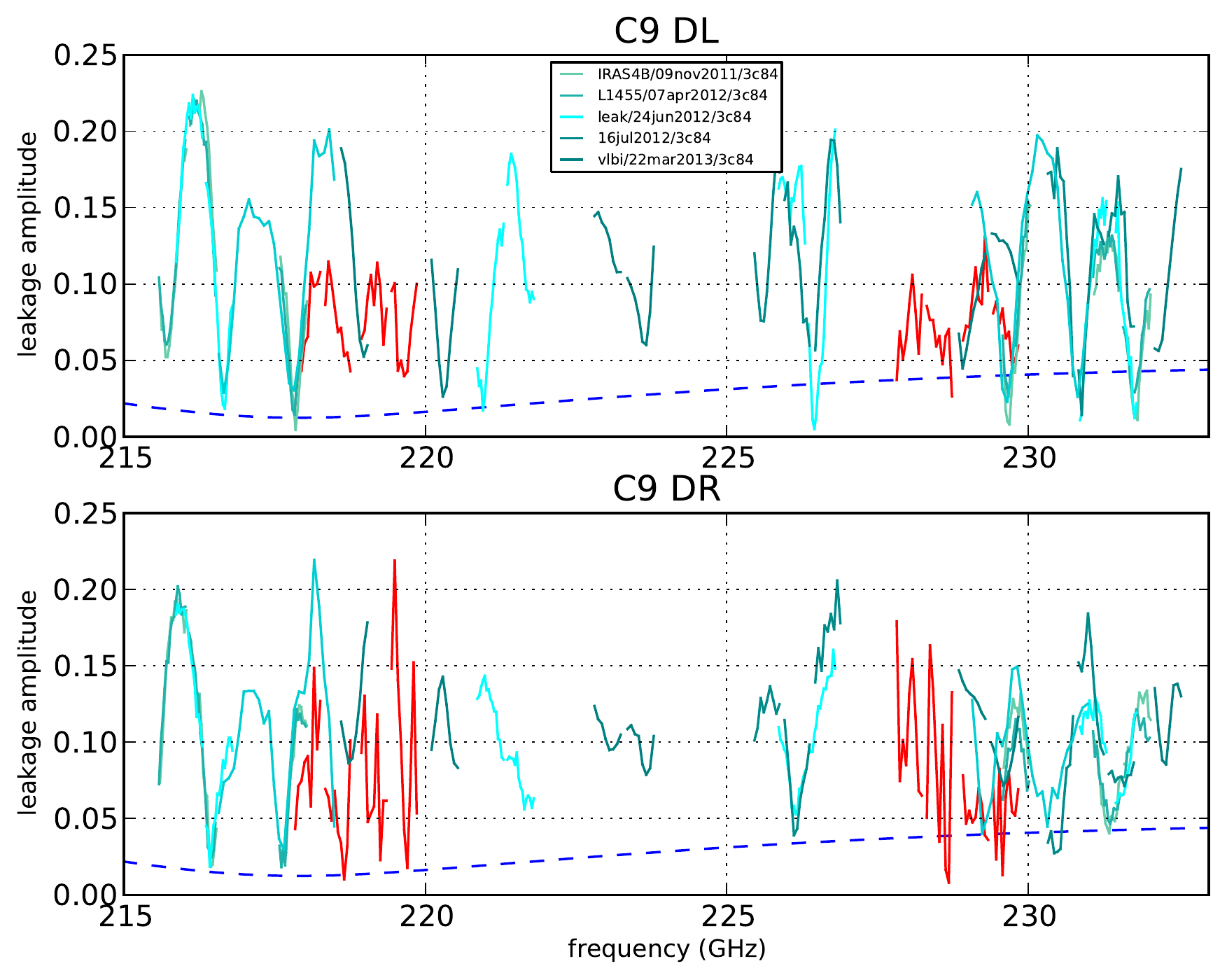}
\caption[Ripples after changing mixer bias]{ \small Leakage amplitudes as a
function of frequency for antenna C9 derived from 9 different datasets.  For the data
shown in red the voltage bias on the RCP mixer was set above the 
SIS junction's superconducting
energy gap so that the junction behaved as a resistive load.  This reduced
reflections of RCP signals from this mixer, which lowered the magnitude of the
LCP leakage ripples.
The RCP leakages for this dataset were
then just noise, since there was no signal coming from the RCP receiver.  }  
\label{fig:ripples_mixerbias}
\end{center}
\end{figure}

To test this  hypothesis, we attempted to reduce the reflection from the RCP
mixer in antenna C9 by changing the mixer bias (see Figure
\ref{fig:ripples_mixerbias}).  Reducing the reflection off of one mixer should
reduce the leakage ripples for the opposite polarization.  For the data shown
in red in Figure \ref{fig:ripples_mixerbias}, the RCP mixer was biased to
13\,mV.  This is above the superconducting energy gap, so no RCP astronomical
signals would have been downconverted to the IF: and indeed, the RCP
leakages (red curves, lower panel) are just noise.  However, at this bias there
should be a better impedance match between the SIS mixer and waveguide,
reducing the magnitude of the signal reflecting back from the mixer.  And, 
indeed, the LCP leakage ripples (red curves, upper panel) appear to be
reduced, consistent with the reflection hypothesis.

We then tried to reduce the reflection from the dewar window (a double convex
Teflon lens) by tilting it with a special clamp ring.  It was physically
possible to tilt the lens by only about 5$\degree$, and this had no apparent
effect on the leakages.  In another experiment, we attempted to \textit{worsen}
the reflection from the dewar window on C7---an antenna with relatively
small leakage ripples---by installing a flat 0.010\,in thick piece of Mylar just
in front of the window.  Again, however, there was no discernable effect on the
leakages.  Finally, to confirm that the ripples originate in the dewar and
not elsewhere in the telescope or correlator room, we physically swapped the
dewars between antennas C9 and C10.  This swapped the leakage pattern between
these antennas, confirming that the problem does originate in the dewar.

\begin{figure} [hbt!]
\includegraphics[scale=0.36, clip, trim=2.5cm 1cm 1cm 1cm] {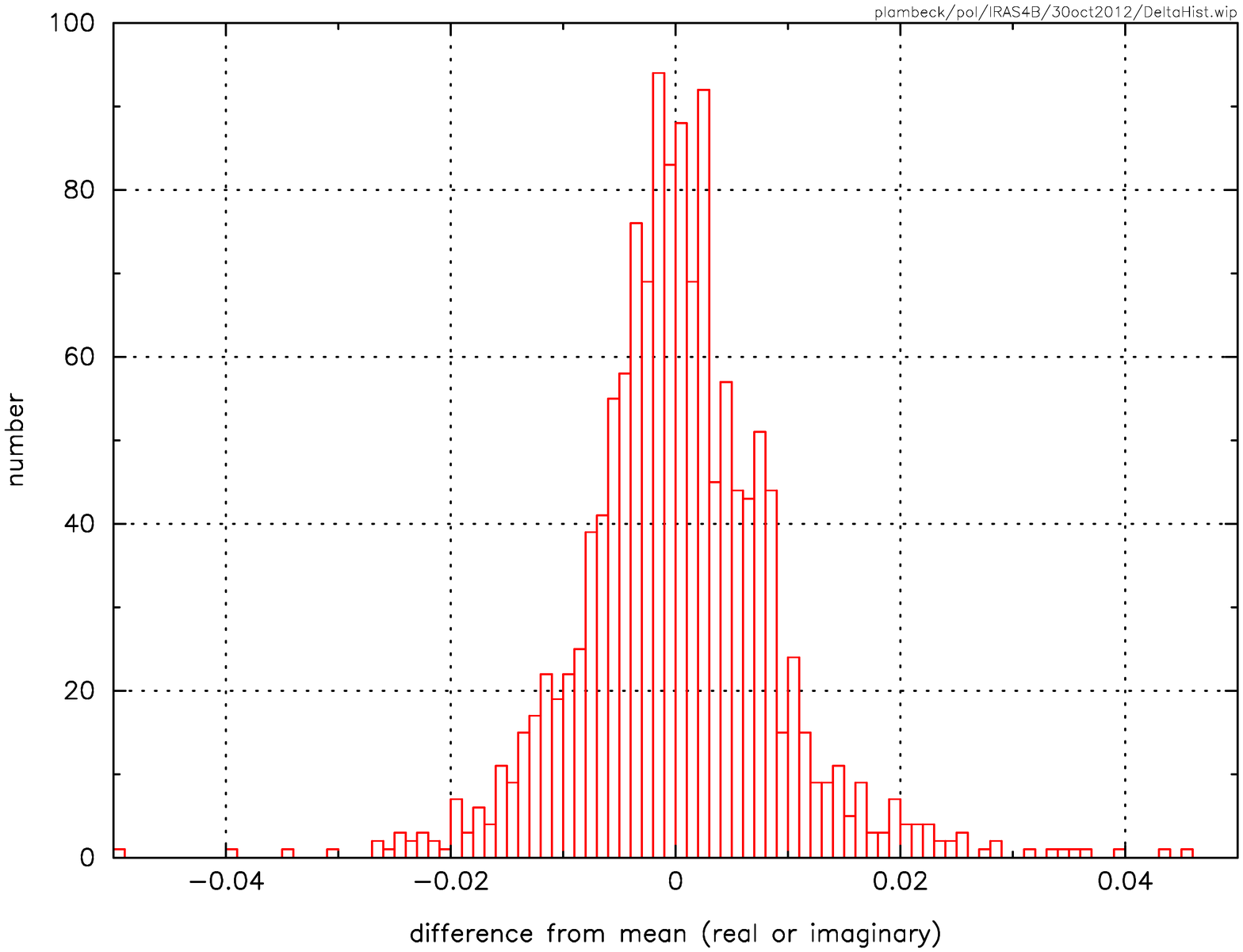}
\caption[]{ \small
A histogram of the differences of 4 leakage solutions from their mean.  The values encompass the real and imaginary 
components for $D_R$ (leakage from LCP into RCP) and $D_L$ (leakage from RCP into LCP), for 6 correlator windows and all 15 telescopes.  
The leakage solutions are from 02 September 2012, 25 September 2012, 25 October 2012, and 30 October 2012.  
The standard deviation is 0.0089.
\vspace{0.3in}
}
\label{fig:leakage_scatter}
\end{figure}

\begin{figure*} [bt!]
\begin{center}
\includegraphics[scale=.9, clip, trim=2cm 1cm 2cm 1cm]{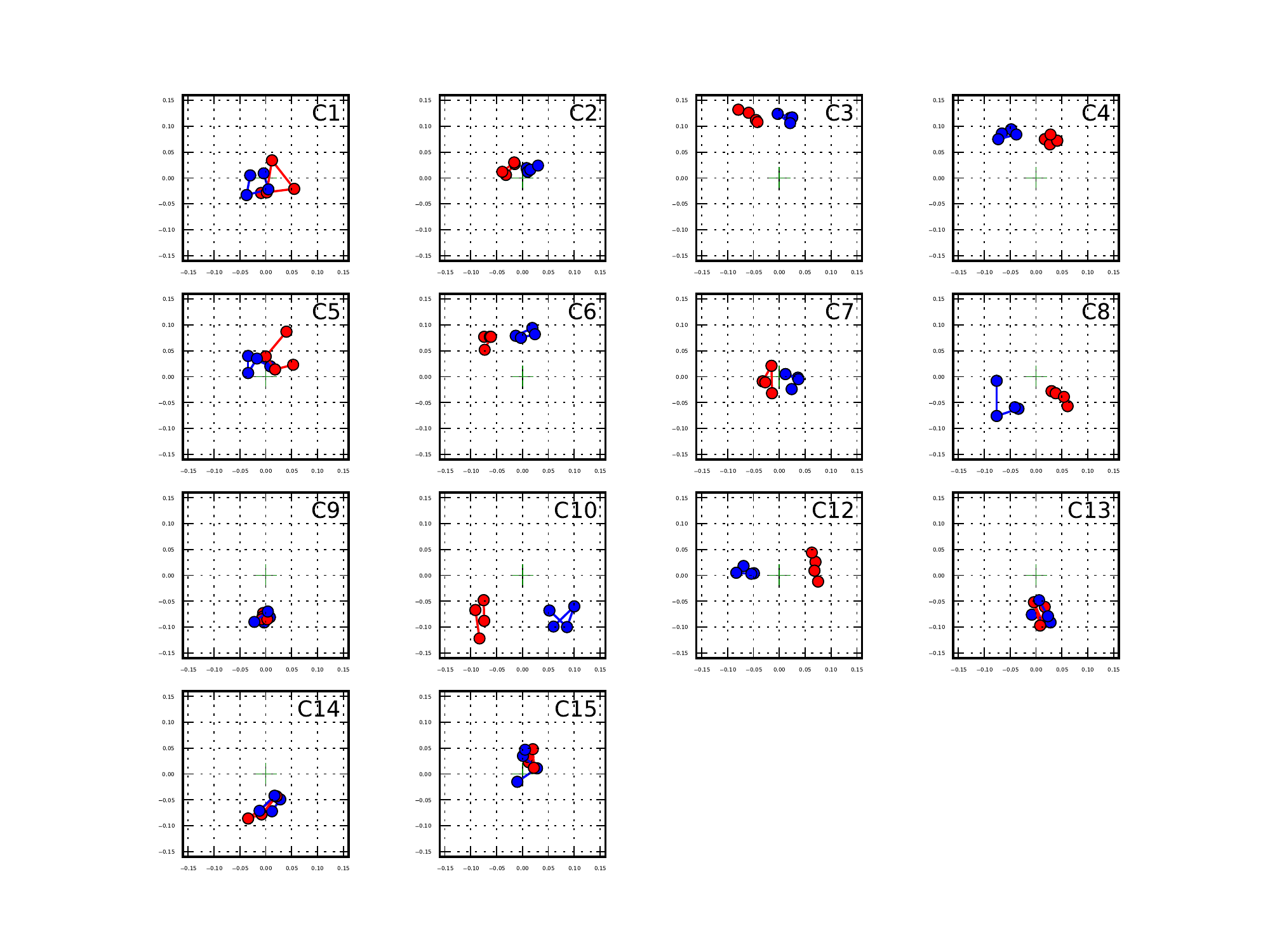}
\caption[complex leaks]{ \small Leakages plotted in the complex plane.
$D_R$ (red) and $D_L$ (blue) for four 500-MHz wide bands centered
at 224.5, 225.0, 225.5, and 226.0\,GHz, from data set \texttt{c1217.2D\_2303c279.miriad.2}.
Typically $D_R$ and $D_L$ have mirror symmetry about the imaginary axis.}
\label{fig:ComplexLeaks}
\end{center}
\end{figure*}

\smallskip
\noindent
\textbf{Reproducibility of the leakage solutions.}
Since the CARMA receivers have no moving parts, the leakages are expected to be
stable and reproducible.  Figure \ref{fig:leakage_scatter} presents a histogram
of the scatter in leakage solutions from several different datasets obtained
over a two-month interval.  In this case the standard deviation of the Re and
Im parts of the leakage terms was $\sim$\,0.009; however, in data taken a few days apart the
standard deviation can be as small as $\sim$\,0.002.

Factors limiting the reproducibility are reflections from the mixers---which
will depend on their physical temperature and on their voltage and current bias---and 
the relative RCP and LCP power levels in the block downconverters, which
will depend on the tuning and correlator setup.  

There is an ambiguity in the leakage terms because they always appear in
pairs in expressions for the $LR$ and $RL$ cross-correlations \citep{Sault1995}.\footnote
{\,The ambiguity does not occur for VLBI observations where sources are
observed at different parallactic angles from different observatories.}
For example, the observed $RL^*$ cross correlation between antennas $m$ and $n$ is given by:
\begin{align}
\langle v^\prime_{Rm} v^{\prime*}_{Ln} \rangle &= \langle\ \ (v_{Rm} + D_{Rm} v_{Lm}) (v_{Ln} + D_{Ln} v_{Rn})^*\ \ \rangle \\
 &= \langle v_{Rm} v_{Ln}^* \rangle + D_{Rm} \langle v_{Lm} v_{Ln}^* \rangle + D_{Ln}^* \langle v_{Rm} v_{Rn}^* \rangle \\
 &= \langle v_{Rm} v_{Ln}^* \rangle + \frac{1}{2} I_{mn} (D_{Rm} + D_{Ln}^*) 
\end{align}
This expression is unchanged if one adds an arbitrary complex number $c = a + jb$
to all the $D_R$ terms, and subracts its conjugate $c^* = a - jb$ from all the
$D_L$ terms: \begin{equation} (D_{Rm} + a + jb) + (D_{Ln} - a + jb)^* = D_{Rm}
+ D_{Ln}^*.  \end{equation}
By convention, \miriad{} chooses this complex offset $c$ such that $\sum
\textrm{Re}(D_{Rm}) = \sum \textrm{Re}(D_{Lm})$ and $ \sum \textrm{Im}(D_{Rm})
= - \sum \textrm{Im}(D_{Lm})$.  As a consequence, the absolute leakages may
change if some antennas are missing from the data set.  When plotted in the
complex plane,  $D_R$ and $D_L$ have a reflection symmetry across the imaginary
axis for most antennas (see Figure~\ref{fig:ComplexLeaks}).  That is,
$\textrm{Re}(D_R) = -\textrm{Re}(D_L)$ and $\textrm{Im}(D_R) =
\textrm{Im}(D_L)$.  Thus, the \miriad{} convention causes the average of the
imaginary parts of the leakages to be close to 0.  If an antenna like C3, for which the
leakages have a large imaginary offset, is omitted from the solution, then the
imaginary parts of all the other antennas will increase by about 0.01.  One
must therefore be cautious when comparing leakages derived from different data
sets.

\section{Systematic limitations to polarization measurements}

There are several effects that limit the accuracy of polarization measurements made with CARMA.  
(1) The accuracy is limited by the signal-to-noise ratio;
noise bias leads to an overestimate of the polarization fraction in the low-SNR case.  
(2) There is a dynamic range limit: for a bright
source that is weakly polarized, errors in the leakage calibrations can give
false detections of polarization.  (3) The absolute position angle accuracy
depends on the grid calibrations.  (4) For extended (non-point-like) sources, the polarization
varies across the beam.  These effects are discussed in the sections below.

\subsection{Signal-to-noise limitations: debiasing polarimetric images}

Polarization measurements have a positive bias because the polarization $P =
\sqrt{Q^2 + U^2}$ is always positive, even though the Stokes parameters $Q$ and
$U$ from which $P$ is derived can be either positive or negative.  This bias
has a significant effect in low-SNR measurements, i.e., when
$P \lesssim 3\,\sigma_P$, where $\sigma_P$ is the rms noise in the polarization
maps.  (The rms noise values in the $Q$ and $U$ maps are generally comparable,
such that we set $\sigma_P \approx \sigma_Q \approx \sigma_U$).  The bias can
be taken into account by calculating the bias-corrected polarized intensity
$P_c$ (\eg{} \citealt{Vaillancourt2006}; see also \citealt{Naghizadeh1993} for
a discussion of the statistics of position angles in low SNR measurements).

The probability density function (PDF) for the observed polarization $P$ of a signal
with true polarization $P_c$ is given by the Rice distribution
(\citealt[][Equation B1]{Killeen1986}, and \citealt[][Equation
6]{Vaillancourt2006}):

\begin{equation}
\textrm{PDF}(P | P_c, \sigma_P) = \frac{P}{\sigma_P^2} \, I_0\left(\frac{PP_c}{\sigma_P^2}\right) \exp{\left[-(P^2 + P_c^2) / 2\sigma_P^2\right]}\,\,.
\label{eqn:PDF_obs}
\end{equation} 

\noindent
However, to calculate the debiased intensity of the true polarization $P_c$ 
one needs the opposite: $\textrm{PDF}(P_c | P, \sigma_P)$. 
If one assumes a uniform prior for the true polarization $P_c$, then by Bayes's theorem 
$\textrm{PDF}(P_c | P, \sigma_P)$ is the same as Equation \ref{eqn:PDF_obs}:

\begin{equation}
\textrm{PDF}(P_c | P, \sigma_P) = \frac{P}{\sigma_P^2} \, I_0\left(\frac{PP_c}{\sigma_P^2}\right) \exp{\left[-(P^2 + P_c^2) / 2\sigma_P^2\right]}\,\,.
\label{eqn:PDF_true}
\end{equation} 

\noindent
Thus, given the observed polarization $P$, one can calculate the true polarization $P_c$ 
by finding the maximum (i.e., the most probable value) of the PDF in Equation \ref{eqn:PDF_true}.

For very significant polarization detections ($P \gtrsim 5\,\sigma_P$), the simple high-SNR limit is valid
(see \citealt{Vaillancourt2006}, Equation 12):

\begin{equation}
P_c \approx \sqrt{Q^2 + U^2 - \sigma_P^2}\,.
\end{equation}

\noindent
However, for low-SNR detections ($P \lesssim 5\,\sigma_P$), things are not so simple, and one must use 
Equation \ref{eqn:PDF_true} to calculate the debiased polarization intensity.

The position angle $\chi$ and uncertainty $\delta\chi$ (calculated using standard error propagation) of the incoming
radiation are
\begin{equation}
\chi = \frac{1}{2} \arctan{\left(\frac{U}{Q}\right)}\,,
\end{equation}
\begin{equation}
\delta \chi = \frac{1}{2} \frac{\sigma_P}{P_c}\,.
\end{equation}
\noindent
I.e., for a detection $P_c = 2\,\sigma_P$, the uncertainty in the position angle is 1/4 of a radian,
or $\pm\,14^{\circ}$.

\subsection{Limitations due to leakage uncertainties.}

The accuracy of polarization measurements also is limited by uncertainties in
the polarization leakage corrections.  These uncertainties affect bright
sources as well as weak ones; thus they provide a dynamic range limit
for polarization measurements with CARMA.

To test the effects of leakage variations on the solutions of polarization
position angle and fraction, we performed simulations using \mir{UVGEN}.  As
discussed above in Section~\ref{sec:leakage}, the Re and Im parts of the
leakages vary by up to a few $\times$ 0.01 from track to track.  Thus, we set up
the simulations to probe the effects of Gaussian random errors in the leakages with standard
deviations between 0.01 and 0.03.  The simulated $uv$ data included thermal noise
consistent with typical 1\,mm weather at CARMA: system temperatures of 300\,K
and opacity $\tau \approx 0.3$.  The data comprised 8\,GHz of continuum
bandwidth, and included a point source with a flux of 1\,Jy and at an elevation
of 30$^{\circ}$.  We used actual leakage terms from CARMA polarization data as
the initial set of values.

We performed two tests, examining how variations in leakage terms affect the PA and the polarization fraction (\%) of
(1) a snapshot observation (for example, when measuring the polarization of the passband calibrator, which is only observed for 10\,min),  
and (2) a longer, $\sim$\,6\,hr observation of the calibrator or the science target. 

The simulations proceeded as follows:

\begin{itemize}
\item Used \mir{UVGEN} to simulate leakage-free data: either a 10\,min snapshot, or a long track covering $\pm 3$\,hr around transit
\item Copied ``real'' leakages (from a typical 1\,mm CARMA polarization track) into the data file
\item Rewrote the data to apply the leakages, thus corrupting the data
\item Copied a negated version of the above leakages into the dataset after varying the leakages' Re and Im parts 
using Gaussian random errors with standard deviations between 0.01 and 0.03
\item Used \mir{UVFLUX} to find the \% and PA of the source
\item Ran a Monte-Carlo test for point sources with 0\%, 1\%, and 10\% polarization, and calculated the standard deviation of the 
polarization fraction and PA for each value of the leakage error
\end{itemize}

Copying negated leakages into the dataset corrects the corrupted data exactly.  
However, copying a set of leakages whose Re and Im parts were modified slightly result in 
slight differences in the calculated PA and \%.  Performing these simulations gave us a handle on how robust
our calculated PA and \% values were to typical leakage variations. 

\begin{figure*} [hbt!]
\begin{center}
\includegraphics[scale=0.4]{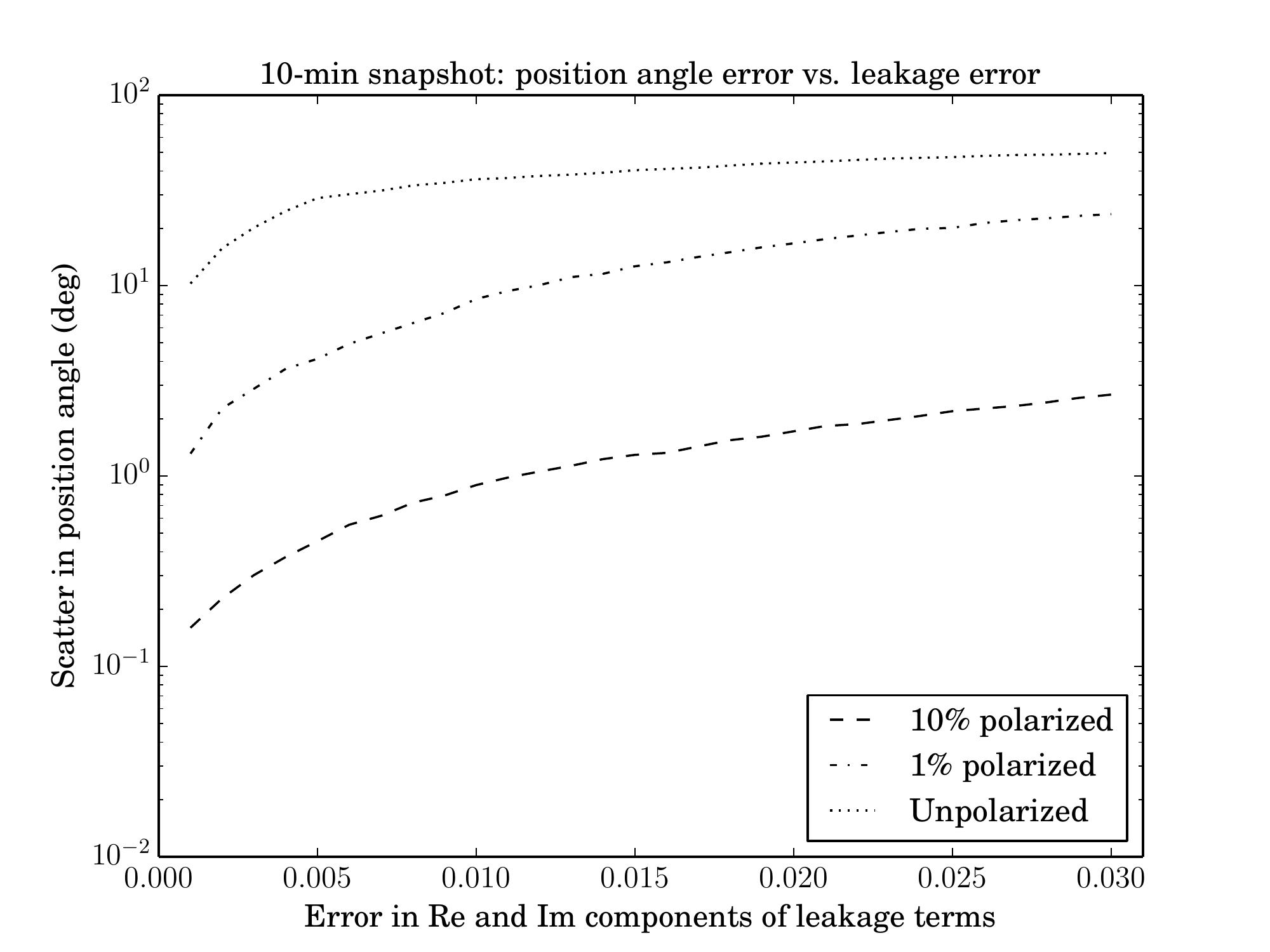}
\includegraphics[scale=0.4]{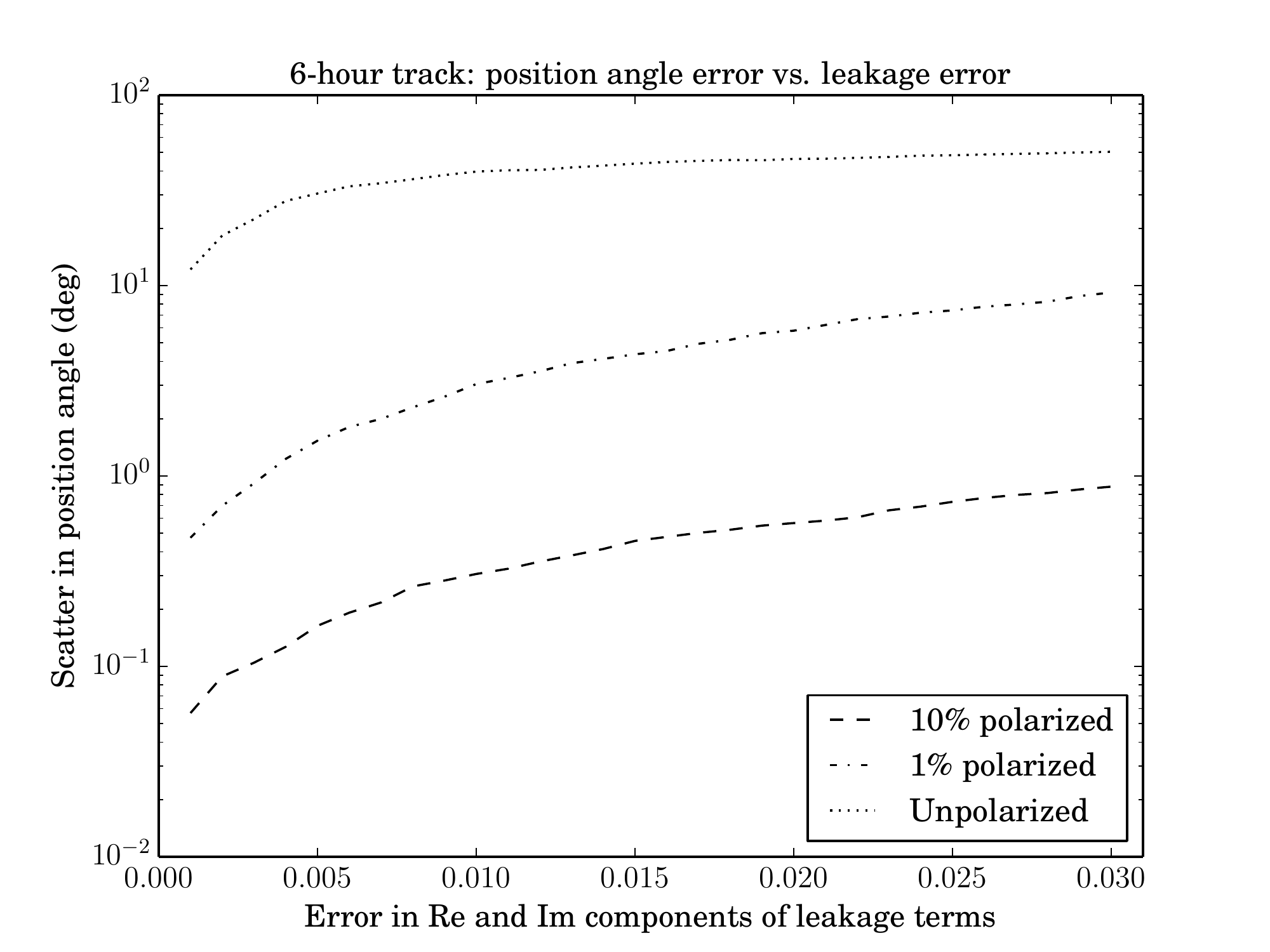}
\includegraphics[scale=0.4]{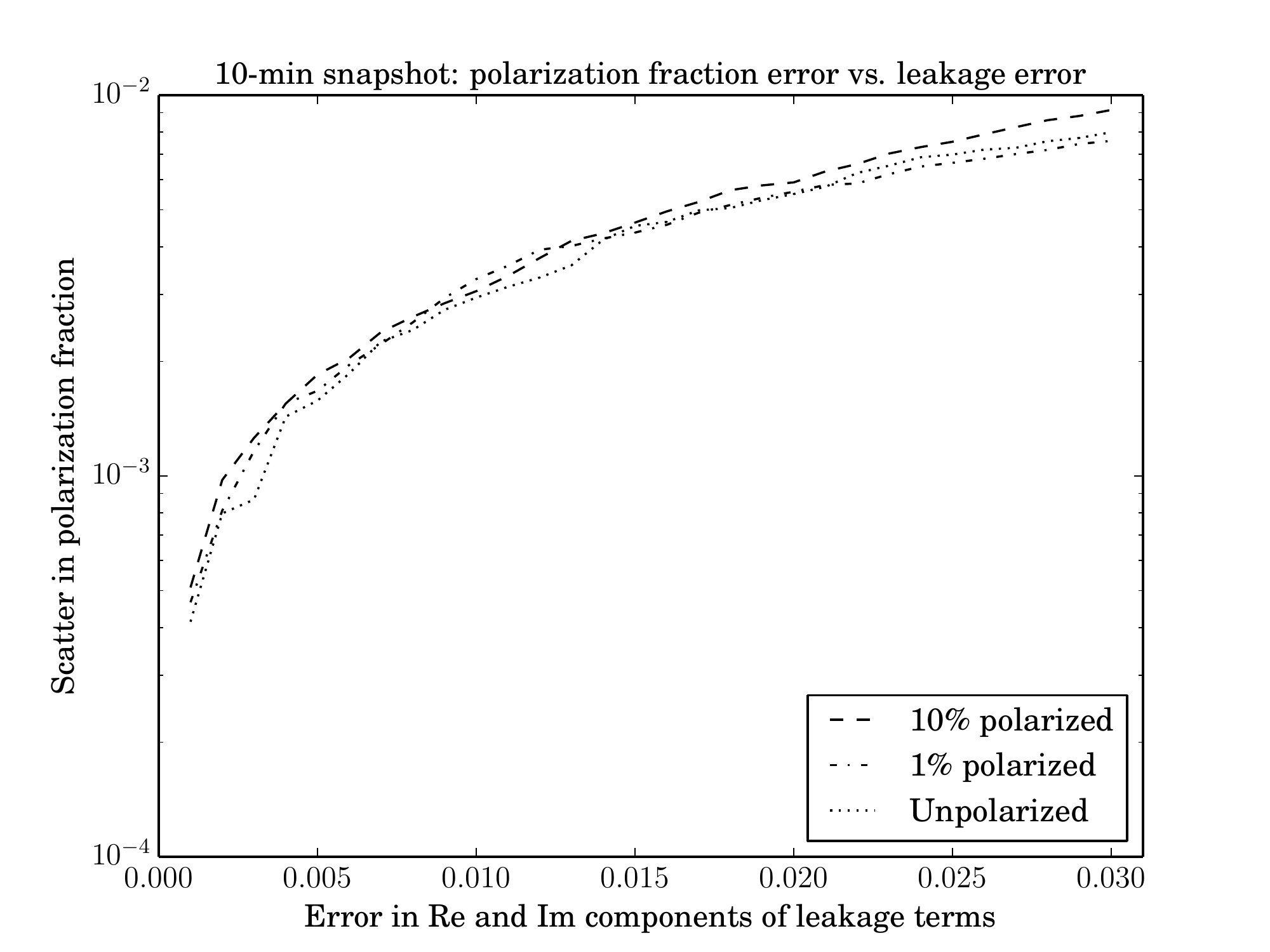}
\includegraphics[scale=0.4]{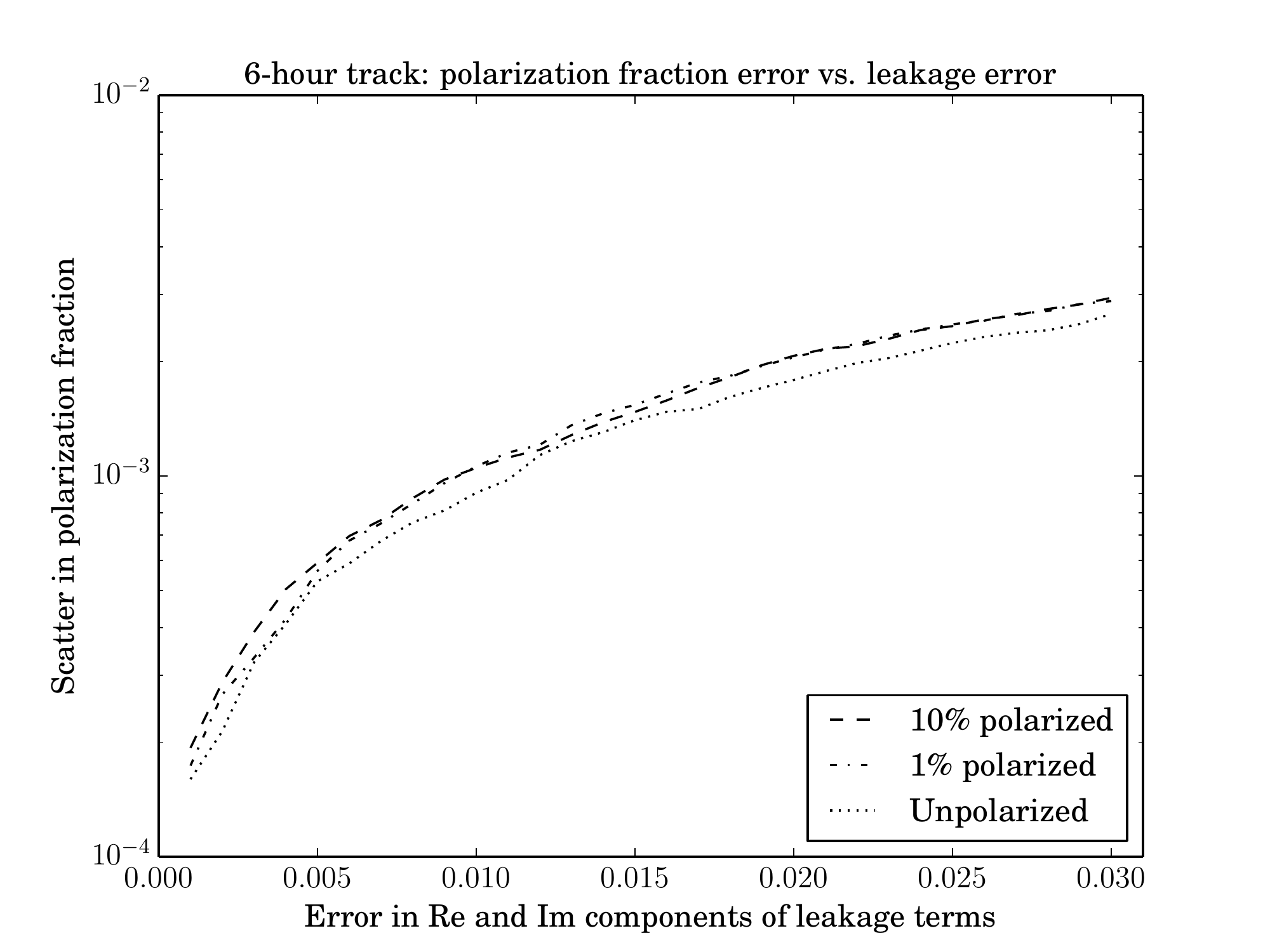}
\caption[Errors in PA and fractional polarization as a function of leakage errors]
{ \small
Errors in PA and fractional polarization as a function of leakage errors.
}
\label{fig:PA_pct_vs_leakage_errors}
\end{center}
\end{figure*}

See Figure \ref{fig:PA_pct_vs_leakage_errors} for the results.  Unsurprisingly,
the scatter in PA and \% is larger for snapshot observations than for
full tracks.  Additionally, the scatter in PA is much larger for weakly
polarized sources.  Looking at the plots of the scatter in \%,
we conclude that the instrumental polarization caused by the typical
leakage errors of $\sim$\,0.01 is $\sim$\,0.3\% for a snapshot observation and
$\sim$\,0.1\% for a longer track.  The corresponding variations in position
angle are $\sim$\,8$^{\circ}$ and $\sim$\,3$^{\circ}$ for a 1\% polarized
source.

Note that very small variations in the leakages can cause drastic changes to
the measured polarization position angles if a source is very weakly polarized
($P_c/I \lesssim 0.5\%$).  For example, when analyzing CARMA data toward the
protoplanetary nebula CRL~618 (see \citealt{Sabin2014}) the position angles
varied by up to 90$^{\circ}$ from night to night depending on how exactly we
reduced the data.  
The scatter in the Re and Im parts of the leakage solutions on the two nights
was $\sim$\,0.01, which is standard; however, in this case these slight
differences led to significantly different $Q$ and $U$ maps.  This was not a
problem for the TADPOL sources \citep[see][]{Hull2013, Hull2014}, which were on
average at least a few percent polarized, and which tended to be much fainter
sources where the ability to detect polarization was limited by the SNR
instead of by dynamic range.

We therefore urge caution when interpreting the position angles of sources with polarization fractions of $< 0.5\%$.

\subsection{Absolute position-angle accuracy}
\label{sec:mars}

Even for a strongly polarized source, there is an uncertainty in PA due to the
absolute accuracy of the $R$--$L$ phase calibration.  As discussed in
Section~\ref{sec:xyphase}, polarized noise sources in the cabins of the 10\,m
telescopes are used to measure the $R$--$L$ phase.  One of the 10\,m telescopes
(usually C1, since it has relatively small leakages) is used as the reference
antenna for the passband calibration.  This transfers its $R$--$L$ phase
calibration to all the other telescopes, including the 6\,m telescopes, which
are not equipped with polarized noise sources.  

Observations of an astronomical source with a stable, well defined PA are
required in order to check the accuracy of the PA measurements.  At centimeter
wavelengths, 3C286 is the usual polarization calibrator; its position angle
has been stable for decades \citep{Perley2013}.  However, the PA of 3C286 increases
slowly with frequency, from $\chi = 33^{\circ}$ at $\lambda \gtrsim 3.7$\,cm to
$\chi = 36^{\circ}$ at $\lambda$\,0.7\,cm.  At millimeter wavelengths,
\citet{2012A&A...541A.111A} measured $\chi = 37.3 \pm 0.8^{\circ}$ at
$\lambda\,$3\,mm and $\chi = 33.1 \pm 5.7^{\circ}$ at $\lambda\,$1.3\,mm.  ALMA
commissioning results at $\lambda\,$1.3\,mm gave $\chi = 39^{\circ}$ 
(Stuartt Corder, priv. comm., 2013).

\begin{figure} [bt!]
\includegraphics[width=0.52\textwidth, clip, trim=5.0cm 3.18cm 5.0cm 1.0cm] {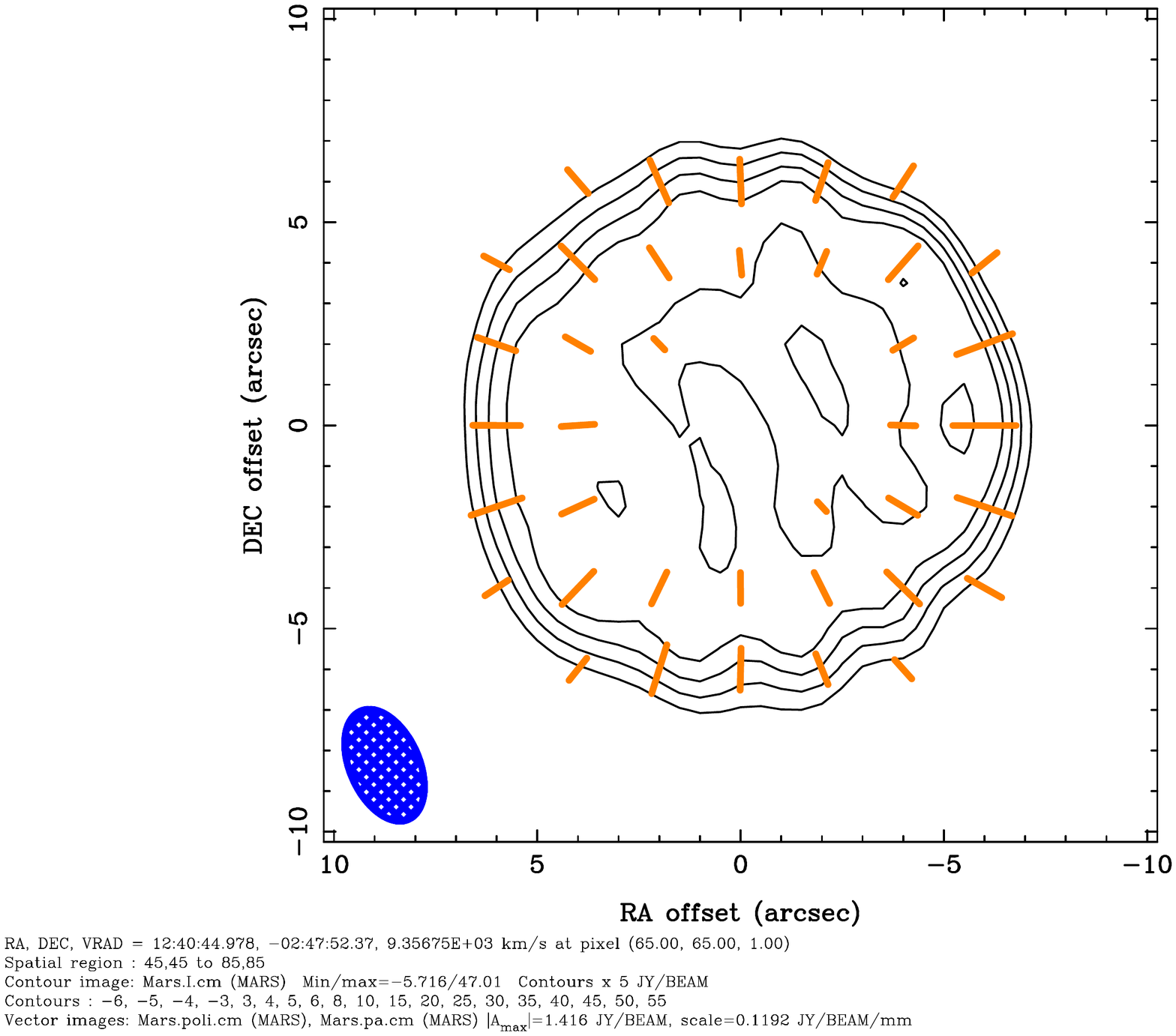}
\caption[]{ \small
Mars polarization map generated using C1 as the passband (and hence $R$--$L$ phase)
reference antenna, based on data set \texttt{c1217.2D\_2303c279.miriad.2}.  
Mars is $14.2\arcsec$ in diameter and the synthesized beam (blue ellipse) is
$2.9\arcsec \times 1.7\arcsec$.  The Stokes $I$ intensity is shown by black contours at 15,
20, 25, 30, and 40\,\jybm{}.  The peak Stokes I flux density is 47\,\jybm{} and
the maximum polarized intensity is 1.4\,\jybm{}.
Line segments indicate polarization orientation from the CARMA data; 
segment lengths are proportional to polarization intensity.  
The polarization of spherical, solid bodies like Mars should be radial
(see Section \ref{sec:mars} and Figure \ref{fig:mars_dev}).
\vspace{0.1in}
}
\label{fig:mars_pol}
\end{figure}

Fortunately, observations of the polarized thermal emission from a rocky planet
or satellite provide an absolute standard by which the PA of 3C286 can me
measured.  The planet must be well resolved by the synthesized beam.  The
polarization is expected to be \textit{radial} with respect to the center of
the planet's disk.\footnote{
\,The radiation that we receive is a mixture of thermal emission from the
planet that is transmitted through its surface, and 3\,K background radiation
that is reflected off this surface.  If the surface is tilted with respect to
our line of sight, as it is near the limb of the planet, the transmission and
reflection coefficients are functions of the radiation's polarization
direction.  The transmission coefficient is greater, and the reflection
coefficient smaller, for radiation that is polarized parallel to the plane of
incidence, which is the plane defined by the incident and reflected rays.  (In fact, 
at Brewster's angle the reflection coefficient for this polarization drops to
zero).  Thus, emission from the planet is preferentially polarized
parallel to this plane, i.e., radially.} 

The radial polarization pattern has been seen before in, for example,
observations of the Moon \citep{Heiles1963, Davies1966, White1973};
\citet{Heiles1963} found that the maximum polarization of the Moon was
$\sim$\,20\% when observed at $\lambda$\,21\,cm.  The Moon is much too large to
fit into the primary beams of millimeter wave telescopes like CARMA or ALMA,
however.  Mars is a better choice for these telescopes.

\begin{figure*} [bt!]
\begin{center}
\includegraphics[scale=0.8]{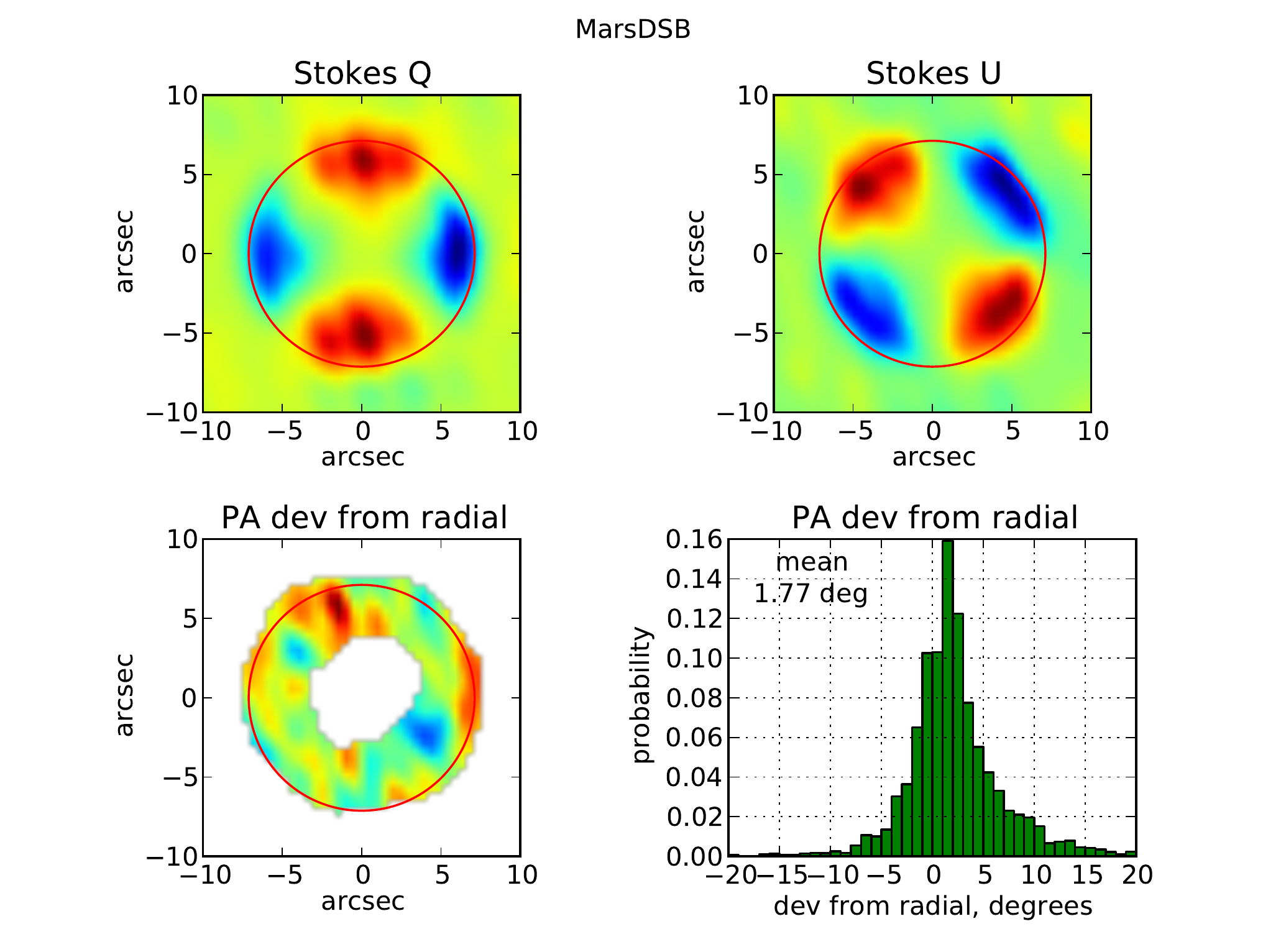}
\caption[Polarization of Mars (deviation analysis)]{ \small
Analysis of deviation from radial of the Mars polarization data shown in Figure~\ref{fig:mars_pol}. The
red circle indicates the diameter of Mars.  Radial deviations are computed only at
pixels where the polarized intensity is >\,0.3\,\jybm{}. The color scale on the deviation map runs
from --10 to 10$^{\circ}$. Values are weighted by polarized intensity when computing the mean deviation.
}
\label{fig:mars_dev}
\end{center}
\end{figure*}
  
Figure~\ref{fig:mars_pol} shows 225\,GHz polarization observations of Mars made
with CARMA on 04 May 2014.  The planet was $14.2''$ in diameter and the
synthesized beam was $2.9'' \times 1.7''$.  Observations of Mars, 3C279, and
3C286 were interleaved over a 3 hour long track.  Leakages were derived from
the 3C279 data.  A careful analysis of the polarization vectors (Figure~\ref{fig:mars_dev}) 
shows that they are skewed 
by about 1.8$^\circ$ from the radial direction.   
  
\smallskip
\noindent
\textbf{PA depends on the choice of reference antenna.}
Unfortunately, a different choice of reference antenna yields a slightly
different absolute position angle.  As shown in column 5 of Table
\ref{table:PA_deviations}, for the 04 May 2014 data the offset ranges from
--0.56$^{\circ}$ for C5 up to 8.7$^{\circ}$ for C6.  These discrepancies are
unlikely to originate from leakages; C3 has the worst leakages (magnitudes of
order 10\%), but its PA deviation is no larger than average.  The deviations
may originate in the optics between the wire grids and the primary mirrors of
the 10\,m telescopes, but further tests should be made.

For each choice of reference antenna we also computed the PA of 3C286.  Column
10 of Table \ref{table:PA_deviations} shows the 3C286 position angles after
subtracting the corresponding Mars radial deviation.  The answers are
surprisingly consistent, with an average PA value of $39.9^{\circ}$.  Using
another leakage solution derived from 3C279 observations two days earlier, on
02 May 2014, gave a similar result: 40.0$^{\circ}$ for the PA of 3C286.

Table \ref{table:PA_deviations_avg} summarizes  Mars polarization observations
on two different days: 04 May 2014 (see Table \ref{table:PA_deviations}) and 17
Jan 2014.  The difference in the final corrected PA for 3C286 between the two
days gives some indication of the uncertainty in the calibration.  For the
January observation Mars was observed over a narrow range in parallactic angle,
and the planet's disk was less well-resolved by the synthesized beam, so it is
tempting to assume that those data are less reliable.  On the other hand,
primary beam polarization (Section~\ref{sec:beampol}) might have skewed the
results of the May observations, when the planet's disk was larger.  Taking the
average of the two measurements, we conclude that the polarization position
angle of 3C286 at 225~GHz is approximately $39.2^\circ \pm 1^\circ$. 

\smallskip
\noindent
\textbf{USB -- LSB position angles.} For Faraday rotation measurements, one
measures the difference in a source's polarization position angle in the LSB
and USB.   Slight differences in the $R$--$L$ phase difference are expected at
these two frequencies because of chromatic effects in the waveguide polarizers.
It is not possible to derive the USB and LSB $R$--$L$ phases from the wire grid
noise source, since the sidebands cannot be separated in noise that is local to
each antenna.  

Column 4 in Table \ref{table:PA_deviations_avg} shows the difference in the Mars radial deviation for the
USB and LSB.  Depending on the choice of antenna used for the $R$--$L$ phase
calibration, this value ranges from 0.12 to 2.81$^{\circ}$.  Column 8 shows the
difference in PA measured on 3C286 for the USB and LSB.  For reasons that are still unclear, 
there is a systematic difference of about 1.2$^{\circ}$ between
the Mars and 3C286 offsets.  We have higher confidence in the 3C286 values,
since 3C286 is a point source.  


\vspace{0.1in}
\begin{table*}
\begin{center}
{\normalsize
\begin{tabular}{|c|cccc|cccc|c|}
\hline 
   (1)   & (2) & (3) & (4) & (5)               & (6) & (7) & (8) & (9) & (10) \\
 $R$--$L$ cal & \multicolumn{4}{c|}{Mars radial dev (deg)} & \multicolumn{4}{c|}{3C286 PA (deg)}  & 3C286 PA (deg) \\
 antenna & LSB & USB & USB$-$LSB & DSB           & LSB & USB & USB$-$LSB & DSB    & corrected \\
\hline
   C1    & 0.69 & 3.50 & 2.81 & 1.77           & 40.71 & 42.25 & 1.54 & 41.62 & 39.85 \\
   C2    & 4.96 & 6.20 & 1.24 & 5.29           & 44.92 & 45.07 & 0.15  & 45.13 & 39.84   \\
   C3    & 4.21 & 4.86 & --0.53\phantom{1} & 3.97          & 44.15 & 43.51 & --0.64\phantom{1} & 43.95 & 39.98  \\
   C4    & 5.73 & 6.87 & 1.14 & 5.74           & 45.63 & 45.51 & --0.12\phantom{1} & 45.69 & 39.95  \\
   C5    & --0.14\phantom{1} & --0.02\phantom{1} & 0.12 & --0.56\phantom{1}        & 39.71 & 38.62 & --1.09\phantom{1} & 39.29 & 39.85  \\
   C6    & 8.23 & 9.74 & 1.51 & 8.70           & 48.19 & 48.80 & 0.61  & 48.63 & 39.93  \\
\hline
\end{tabular}

\caption[]{ \small
Deviation from radial of the Mars polarization orientations, and 3C286 position angles, as
a function of the $R$--$L$ phase calibration antenna.
Lower sideband (LSB; 217.75~GHz), upper sideband (USB; 232.25~GHz),
and double sideband (DSB; 225~GHz) values are given.  For reasons that aren't clear, the DSB values are not necessarily
the mean of the LSB and USB values.  The last column shows the 3C286 DSB position angles 
corrected by the Mars DSB radial deviations.  Leakages were derived from data taken on 04 May 2014, during an observation with
wide parallactic-angle coverage.
\label{table:PA_deviations}
}  
}
\end{center}
\end{table*}

\begin{table*}
\begin{center}
{\normalsize
\begin{tabular}{|c|c|c|cccccc|c|}
\hline 
 Date & Mars diameter & Beam size & \multicolumn{6}{c|}{Mars PA dev, DSB (deg)}  & 3C286 PA (deg) \\
         &                          &                   & C1 & C2 & C3 & C4 & C5 & C6 &   corrected \\
\hline
17 Jan 2014  &  7.8$\arcsec$  &  2.49 $\times$ 2.15$\arcsec$  &  0.54 & 4.96 & 2.54 & 4.39 & --1.83 & 4.54  &  38.5 \\
04 May 2014  &  14.2$\arcsec$  &  2.92 $\times$ 1.71$\arcsec$  &  1.77 & 5.29 & 3.97 & 5.74 & --0.56 & 8.70  &  39.9 \\
\hline
\end{tabular}

\caption[]{ \small Average PA deviations in measurements of Mars, and 3C286 position angles.
Same as Table \ref{table:PA_deviations}, but for two separate observations on 17 Jan 2014 and 04 May 2014.
The difference in the final corrected PA for 3C286 is unexplained, but may be because the 17 Jan track was a short, snapshot
observation and the 04 May track had wide parallactic-angle coverage.
}  
\label{table:PA_deviations_avg}
}
\end{center}
\end{table*}


\subsection{Primary-beam Polarization}
\label{sec:beampol}

If a source is not perfectly on-axis, then additional distortions are introduced across
the primary beams of the telescopes.
To check for these variations in the instrumental polarization, we observed BL Lac (a bright, highly polarized quasar)
at 16 offset positions, eight of which were $12\arcsec$ and eight of which were $24\arcsec$ from the 
field center.\footnote{\,The analysis presented below only uses the data taken $12\arcsec$ from the field center.}
Our aim was to characterize the ``beam squint'' and ``beam squash.''

Beam squint arises when the feeds of the telescope are tilted with respect to the optical axis of the primary reflector,
causing the LCP and RCP responses to be slightly offset from the symmetry axis.
Beam squint is discussed in \citet{Chu1973, Adatia1975, Rudge1978}.  Squint manifests itself as a 
double-lobed pattern in Stokes $V$, which is the difference of RCP and LCP (see Equation \ref{eqn:stokes_rl4}).  

Beam squash is caused by differences in the beam widths of the orthogonal polarizations.
This is caused largely by the slight difference in reflectivity of the two polarizations off the
surface of the curved parabolic reflector.
Squash manifests itself as a four-lobed ``cloverleaf'' pattern in the linearly polarized Stokes $Q$ and $U$ maps.
This phenomenon is discussed in \citet{Napier1994, Napier1999}; the term ``beam squash'' was coined in \citet{Heiles2001c}.

Maps of both squint and squash have been made
by \citet{RobishawThesis} for the Green Bank Telescope (GBT) and \citet{Heiles2001c} for Arecibo.

Additionally, there is a large body of work on wide-field polarimetric calibration and imaging with the 
Karl G. Jansky Very Large Array (VLA). \citet{Uson2008} primarily discuss Stokes $V$ (squint) calibration.
\citet{Cotton2010} discuss both off-axis circular (squint) and linear instrumental polarization, and present a model for 
correcting for both.  They measure the off-axis instrumental polarization 
for various frequencies using the method we employ at CARMA, where all antennas move to various offset
positions simultaneously.

A more robust method of removing polarization artifacts is 
beam holography, where one or more ``reference antennas'' remain pointed at a strong source while the
remaining antennas move in a raster pattern, thus measuring the primary beam of each dish 
in all four Stokes parameters $IQUV$
(see, e.g., \citealt{Corder2006, Lamb2008}, who map the total power [Stokes $I$] beam at CARMA).   
After mapping all four beams for all antennas, one can then use those beam models 
to correct the full-Stokes data.

\smallskip
\noindent
\textbf{Beam squint.}
While the CARMA system is almost always used for measuring linear polarization, we are still able to
make maps of Stokes $V$, and are thus able to measure squint.   Beam squint is normally characterized
using the squint angle $\Psi_s$ (\citealt{Rudge1978}, Equation 25; see also \citealt{Napier1994}):

\begin{align}
\Psi_s &= \arcsin{\frac{\lambda\sin{\theta_0}}{4\pi F}} \\
&\approx \frac{\lambda\theta_0}{4\pi F} \,\, ,
\label{eqn:squint_theo}
\end{align}

\noindent
where $\lambda$ is the observing wavelength, $\theta_0$ is the angular offset between the feed and the telescope's optical axis,
and $F$ is the focal length of the primary reflector.  The approximation in Equation \ref{eqn:squint_theo}
is good when the angular offset $\theta_0$ is small, which is usually the case.

\begin{figure*} [bt!]
\begin{center}
\includegraphics[scale=0.4]{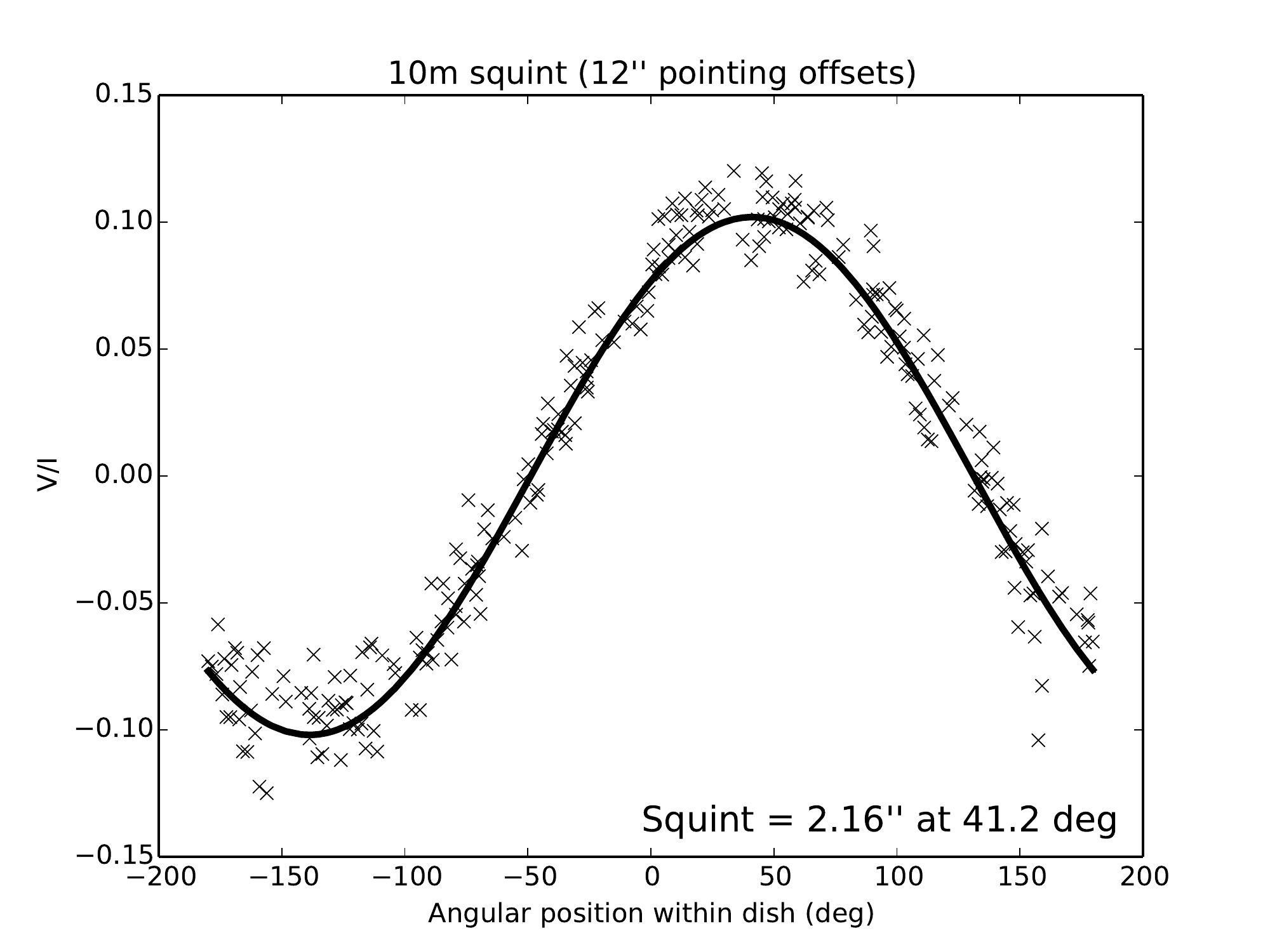}
\includegraphics[scale=0.4]{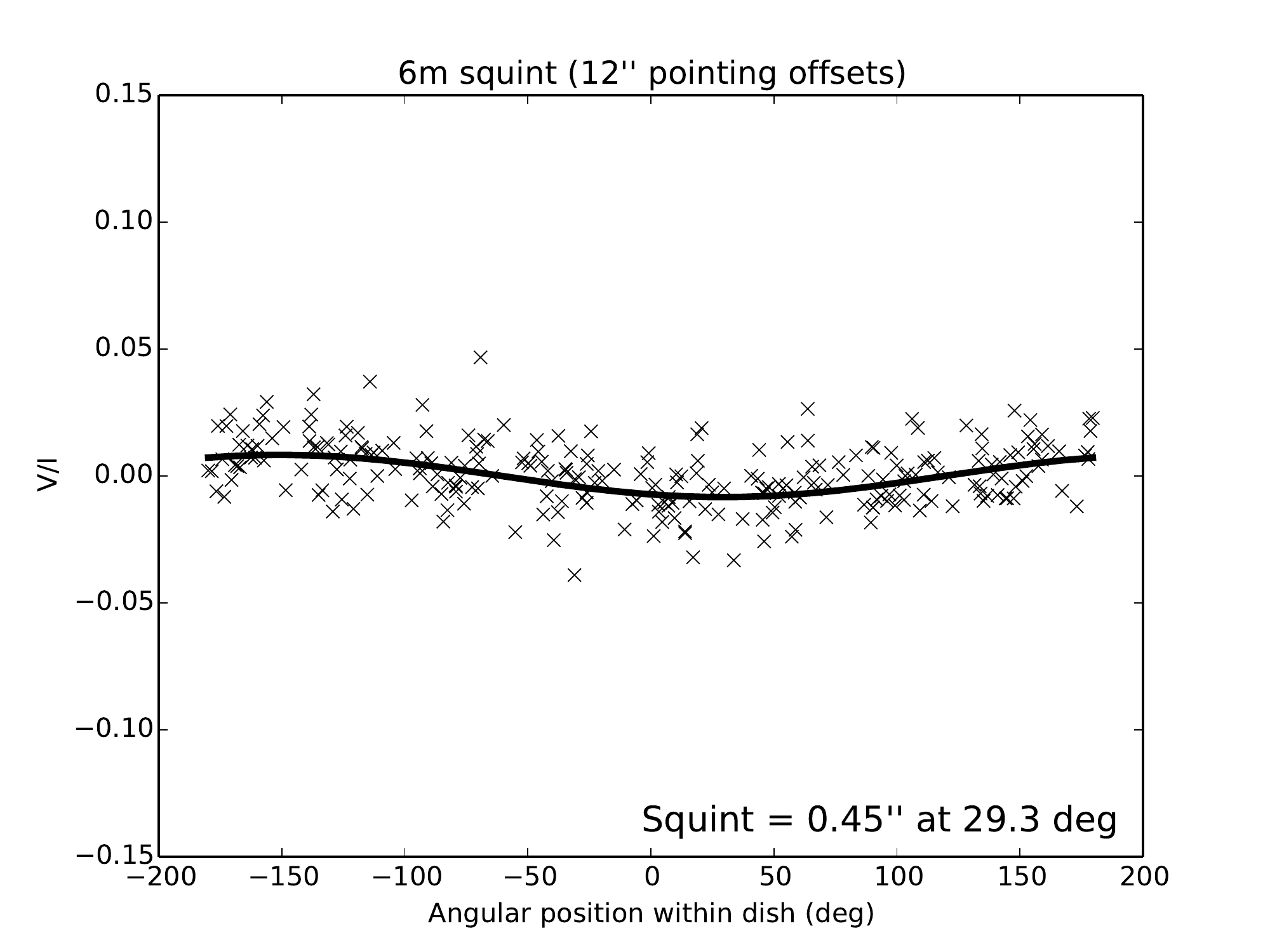}
\caption[6\,m and 10\,m beam squint]{ \small
6\,m and 10\,m beam squint, fit using data positions offset by 12$\arcsec$ from the pointing center.  
The solid curve is the the best-fit squint model (see Equations \ref{eqn:squint1} and \ref{eqn:squint2}).
}
\label{fig:squint}
\end{center}
\end{figure*}

The CARMA 6\,m antennas have a Cassegrain design, with feed horn offset by
$\Delta \textrm{az} = 7.126\arcmin$ and $\Delta \textrm{el} = 1.91\arcmin$ \citep{Plambeck2000}.
Using Equation \ref{eqn:squint_theo} with an effective focal length of $F = 1186$\,inches, 
the theoretical squint produced by these offsets should be 
incredibly small: $\approx 0.002\arcsec$.  The measured squint is significantly larger ($\approx 0.45\arcsec$;
see Figure \ref{fig:squint}, right panel).
The feeds in the 10\,m antennas are on-axis (and should thus be squint-free); however, 
there are multiple reflections in the optical path, including three curved 
mirrors with $\sim$\,90$^{\circ}$ bends that are all in different planes.  The odd number of off-axis reflections 
may be the cause of the 10\,m squint, which is even worse than the 6\,m squint 
($\approx 2.16\arcsec$; see Figure \ref{fig:squint}, left panel).  

In order to fit for the squint we solved for Stokes $V$ at each (12$\arcsec$) offset position and for each 30\,s integration on BL~Lac.  
Note that we applied the gain and leakage corrections solved at the center position only.
We calculated the offsets in the frame of the dish (in azimuth and elevation) by de-rotating the 
offset position (in RA and DEC) by the parallactic angle $\chi$ associated with the given integration.
In order to derive the squint,
we fit the data to the difference of two circular Gaussians $G_2 - G_1$:

\small
\begin{align}
\label{eqn:squint1}
G_1 &= \exp(-\left[ (\textrm{az}-\Delta \textrm{az}/2)^2 + (\textrm{el}-\Delta \textrm{el}/2)^2 \right] / (2\sigma^2)) \\
\label{eqn:squint2}
G_2 &= \exp(-\left[ (\textrm{az}+\Delta \textrm{az}/2)^2 + (\textrm{el}+\Delta \textrm{el}/2)^2 \right] / (2\sigma^2)) \,\,,
\end{align}
\normalsize

\noindent
where $\sigma = \textrm{FWHM} / (2\sqrt{2\log2})$, and FWHM is the full-width at half-maximum of the primary beam at $\lambda$\,1.3\,mm
(30$\arcsec$ for the 10\,m antennas and 56$\arcsec$ for the 6\,m antennas); $\Delta \textrm{az}$ and 
$\Delta \textrm{el}$ are the vertical and horizontal components of the offset between the RCP and LCP beams, and the angle
of the beam offset in the frame of the dish $\theta = \arctan{(\Delta \textrm{el} / \Delta \textrm{az})} - 90^{\circ}$, measured east from north
(or counterclockwise from vertical, in the reference frame of the antenna).  We assume that the amplitudes of the Gaussians
are identical, and that the two beams are offset by equal and opposite amounts ($\pm \Delta \textrm{az}/2$ and
$\pm \Delta \textrm{el}/2$) from the pointing center.

\smallskip
\noindent
\textbf{Beam squash.}
Telescopes with native circular feeds like CARMA should see squash in both Stokes $Q$ and $U$,
which are both combinations of the cross-polarizations $RL$ and $LR$ (see Equations \ref{eqn:stokes_rl2} and \ref{eqn:stokes_rl3}).  
For telescopes like the GBT and Arecibo with native linear feeds,
it should be easier to see squash in Stokes $U$, which is calculated using the cross-polarizations $XY$ and $YX$
(see Equation \ref{eqn:stokes_xy3}) and is thus unaffected by $XX$ and $YY$ gain variations
that can plague Stokes $Q$ (see Equation \ref{eqn:stokes_xy2}).
Maps of beam squash have been made in Stokes $U$
for the GBT \citep{RobishawThesis} and in both Stokes $Q$ and $U$ for Arecibo (\citealt{Heiles2001c}; note that
for Arecibo the gain variations had a minimal effect, allowing excellent maps of both Stokes parameters).

We see no evidence in the $Q$ or $U$ CARMA maps for a quadrupolar squash pattern, 
which should have twice the frequency of the squint as one moves around the dish.
However, variations in $Q$ and $U$ do lead to squint-like variations (i.e., a single crest and trough) in the polarization 
fraction and position angle.

\begin{figure*} [bt!]
\begin{center}
\includegraphics[scale=0.40]{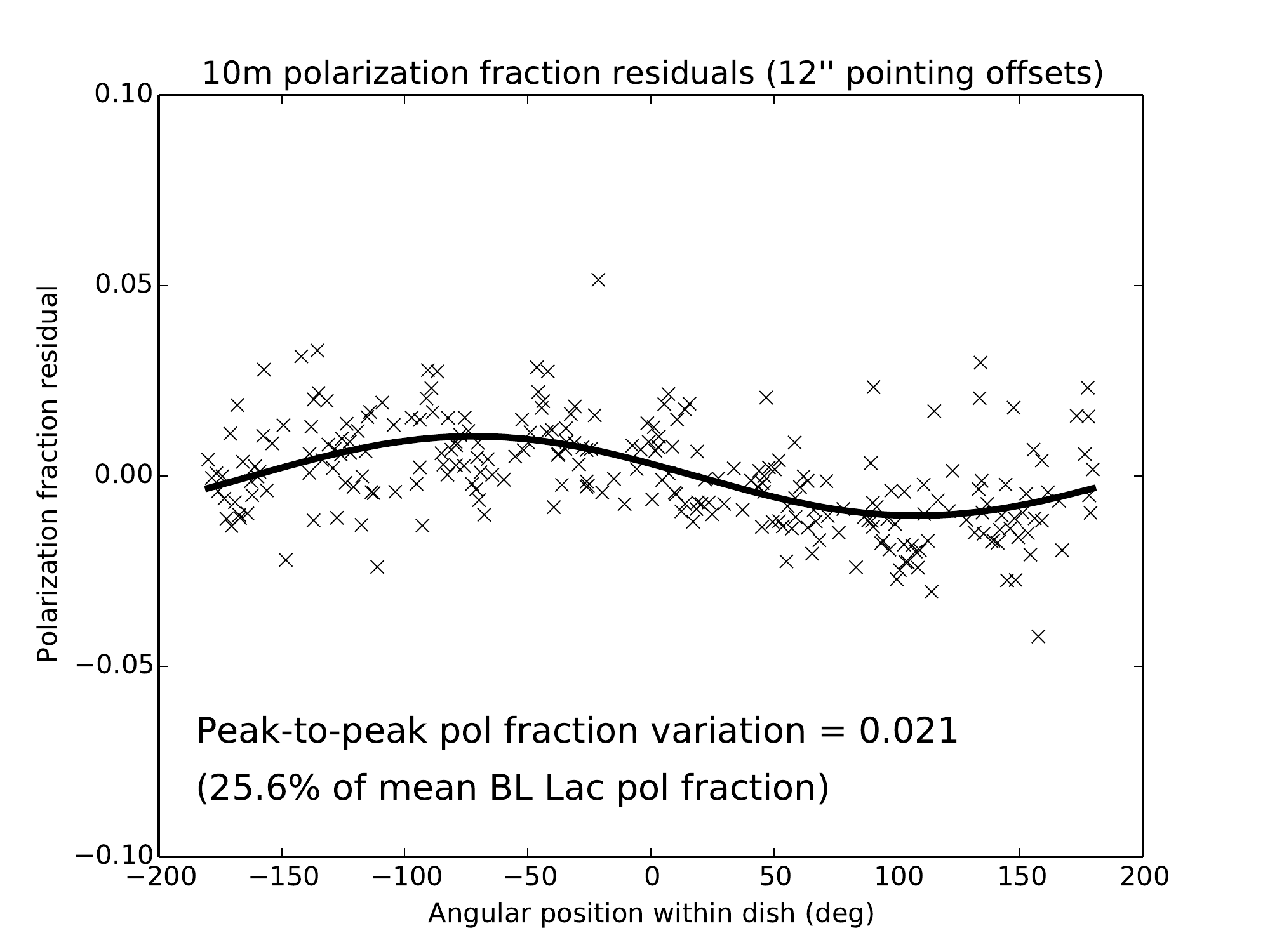}
\includegraphics[scale=0.40]{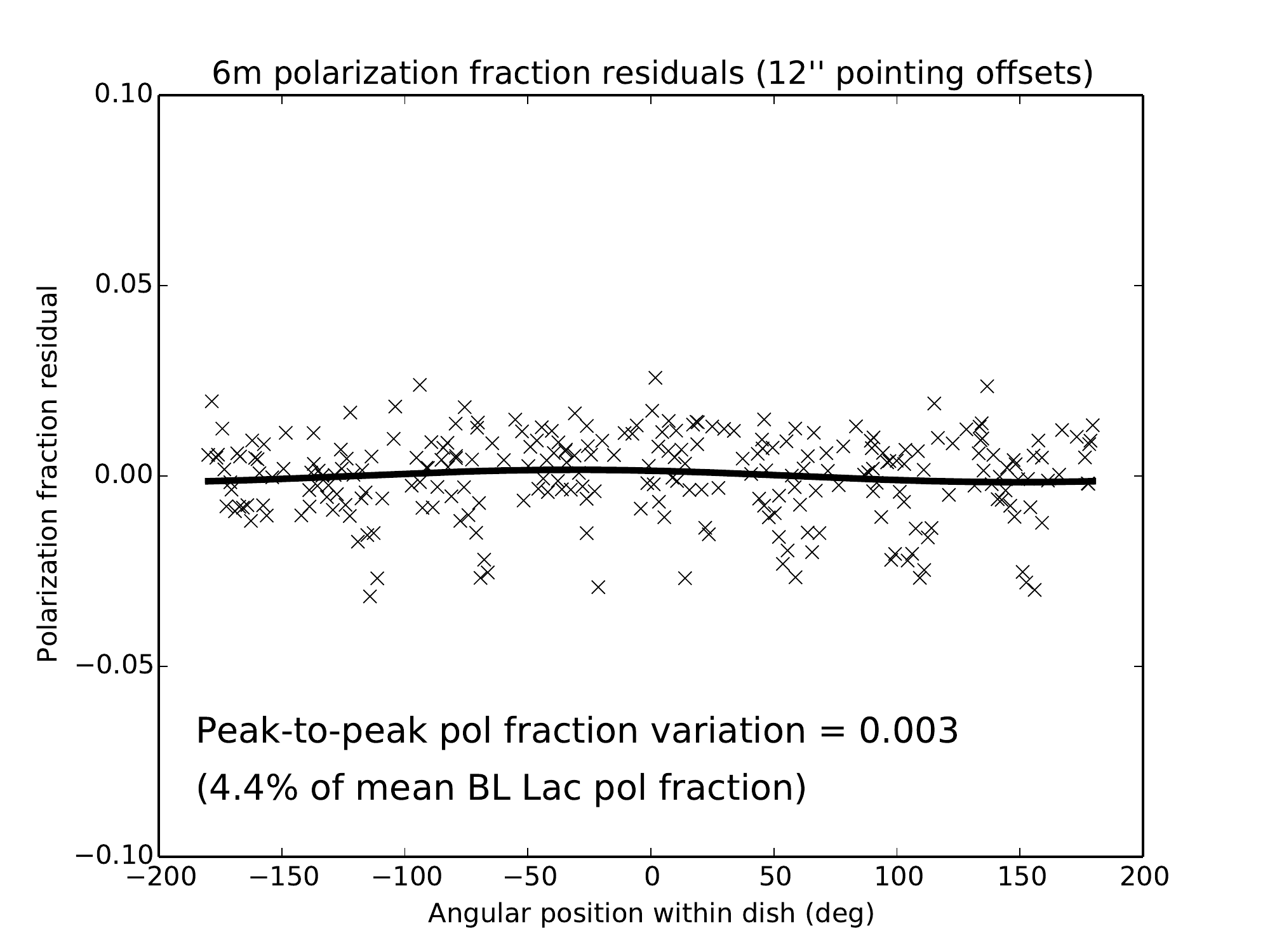}
\includegraphics[scale=0.40]{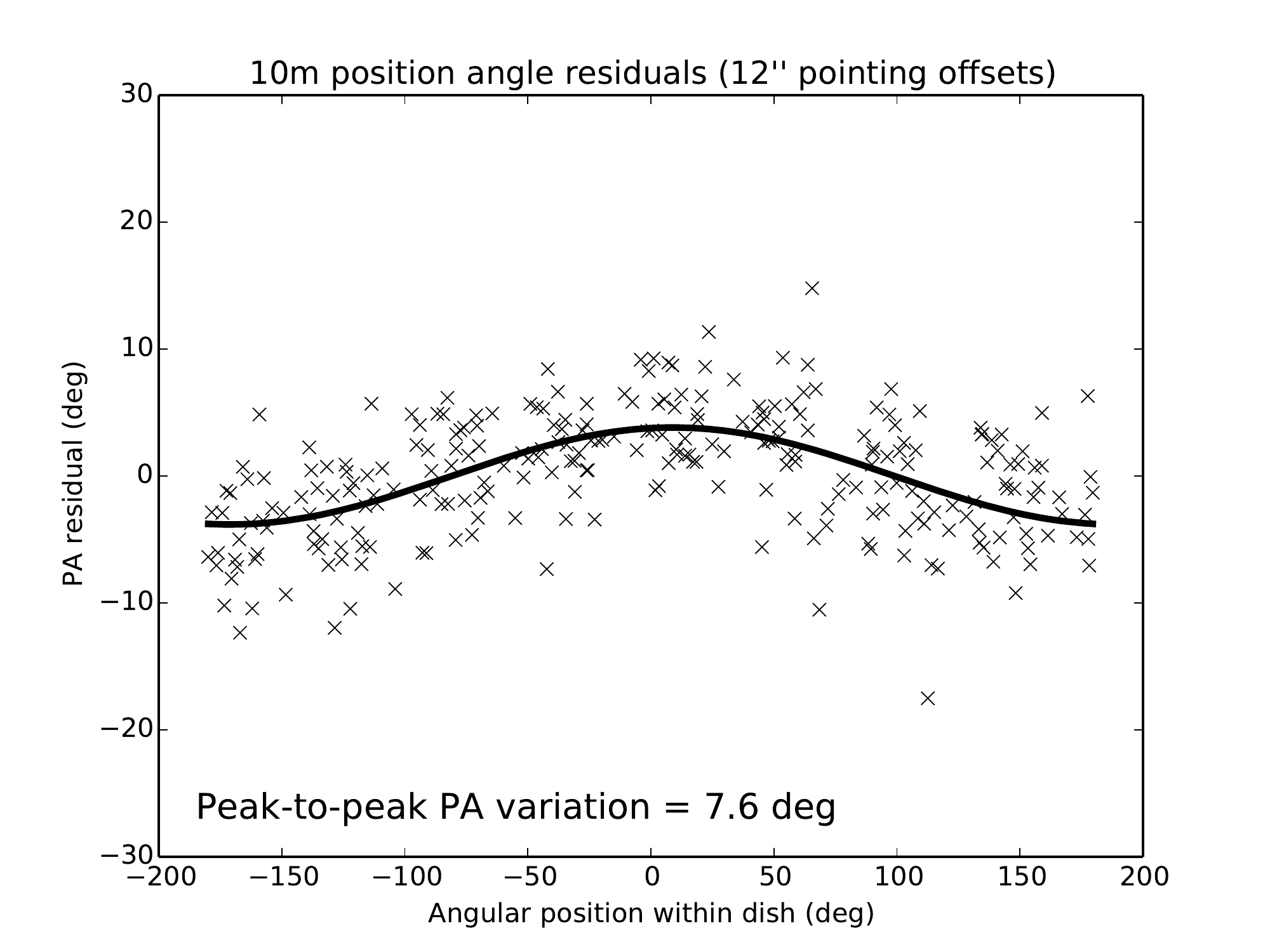}
\includegraphics[scale=0.40]{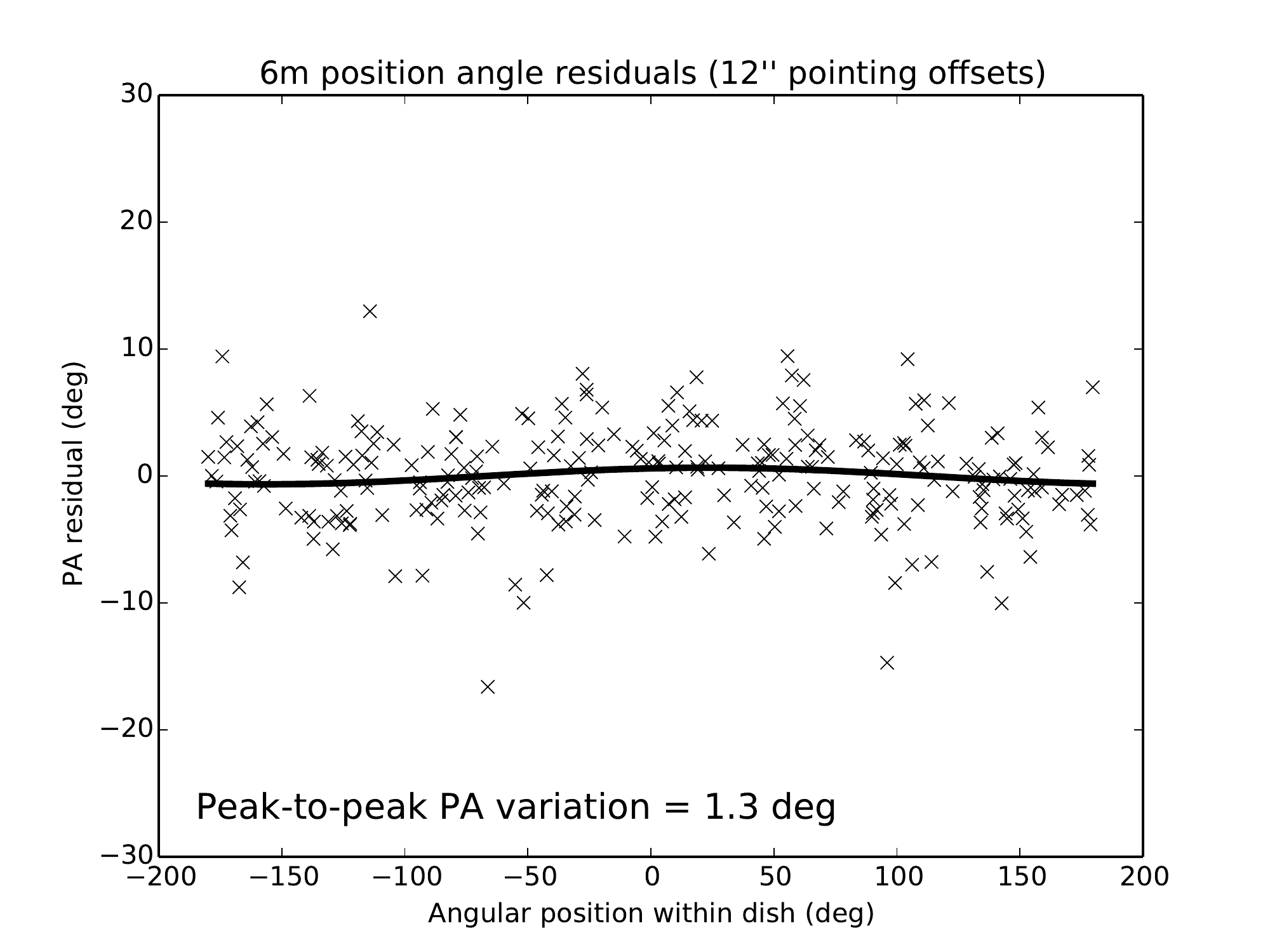}
\caption[Variations of PA and polarization fraction across 6\,m and 10\,m primary beams]{ \small
Variations of PA and polarization fraction across the 6\,m and 10\,m primary beams,
fit using data from positions offset by 12$\arcsec$ from the pointing center.  The solid curves are sinusoids
fit to the data to find the peak-to-peak variations.
}
\label{fig:squash}
\end{center}
\end{figure*}

\smallskip
\noindent
\textbf{Variation in polarization fraction and position angle across the primary beam.}
While we saw no evidence of squash in the data, we did see squint-like behavior in $Q$, $U$, and the PA and polarization
fraction derived from them.  This behavior is not expected theoretically, but is seen at Arecibo and discussed in various
publications and technical memos including \citet{Heiles1999, Heiles2001c, Heiles2003, Heiles2005b}.

We fit a sinusoid to find the peak-to-peak variation in the calculated PA and polarization 
fraction of both the 10\,m and 6\,m dishes.
The 6\,m dishes showed very little variation in PA and polarization fraction,
whereas the 10\,m dishes showed more
(possibly because observations were further out into the primary beam).  The 10\,m antennas had
peak-to-peak variations of $\sim$\,8$^{\circ}$ in PA and $\sim$\,0.02 in polarization fraction (0.02 represents a 
$\sim$\,25\% variation in the mean of the BL~Lac polarization fraction).  See Figure \ref{fig:squash}.

\section{Summary}
\label{sec:summary}

In this overview of the 1.3\,mm dual polarization receiver system built for
CARMA, we described the design and performance of the key hardware components---circular
polarizers, orthomode transducers, and mixers---and discussed the
calibration of the system for polarization observations, focusing particularly
on the calibration of the $R$--$L$ phase offsets and the polarization leakage
terms.  The leakages were found to exhibit considerable frequency structure,
which we attribute to reflections inside the dewars and to cross-coupling in
the analog block downconverter, which precedes the correlator.

We discussed limitations on the accuracy of polarization measurements made with
this system, including the effects of signal-to-noise, leakage uncertainties,
and primary beam polarization. The absolute accuracy of polarization position
angle measurements was checked by mapping the radial polarization pattern
across the disk of Mars.  Transferring the Mars calibration to 3C826, we measured
a polarization position angle for 3C286 at 225\,GHz of
$39.2^\circ \pm 1^\circ$.

\smallskip
\noindent
\textbf{Lessons learned.}  In retrospect, we are not sure that the choice of
circularly polarized feeds was a good one. In a dual polarization system with
circularly polarized feeds, signals that reflect off a mixer can bounce back
from the dewar window with the opposite polarization, introducing
frequency-dependent cross-coupling between the two polarization channels.
Accurately characterizing these ripples in the polarization leakages is
extremely time-consuming, since it requires observations of a strong calibrator
over a wide parallactic angle range.  
Although the leakage ripples in the CARMA system tended to average out over the
full passband, and were not a severe limitation for continuum polarization
observations of dust or extragalactic synchrotron sources, these ripples did
hamper our ability to search for polarization in spectral lines.

In contrast, in a dual polarization system with linearly polarized feeds,\footnote{\,We note 
that, if desired, a beamsplitter can still be used
to couple the local oscillator into the mixers in such a system, simply by
rotating the plane of polarization of the LO with a waveguide twist or half
wave plate so that it is oriented at 45$\degree$ to the OMT axes.}  
reflections would not cause cross-coupling between the two polarizations ($XY$
and $YX$), but instead would create ripples in the receiver bandpass ($XX$ and
$YY$), which are easily calibrated with a brief observation of a bright
continuum source. Of course, as
discussed in Section~\ref{sec:intro}, crossed-linear feeds have their own
problems.  Specifically, in order to detect weak linear polarization with
crossed linear feeds, the receiver gains must be extremely stable.  The newest
interferometer with crossed-linear receivers is ALMA.  The promising continuum
polarization results obtained thus far with ALMA suggest that gain stability is
not a significant limitation; further tests of the ALMA system will reveal whether
there are other issues unique to crossed-linear systems that were not
encountered at CARMA.

\section*{Acknowledgments}
C.L.H.H. would like to acknowledge the advice and guidance of the members of
the Berkeley Radio Astronomy Laboratory, the Berkeley Astronomy Department,
the Owens Valley Radio Observatory, and the CARMA consortium.
He would also like to thank Tim Robishaw for his clear and concise
thesis introduction, which provided an excellent review of polarization basics.
Finally, he thanks Michael Johnson for his help during the stimulating foray into
the conversion from linear to circular Stokes parameters.

C.L.H.H. acknowledges support from an NRAO Jansky Fellowship, an NSF Graduate Fellowship, 
and a Ford Foundation Dissertation Fellowship. 

R.L.P. and C.L.H.H. gratefully acknowledge Suren Singh (Agilent, Inc., Petaluma, CA) 
for making possible the tests of the orthomode transducers.

Support for CARMA construction was derived from the states of California,
Illinois, and Maryland, the James S. McDonnell Foundation, the Gordon and
Betty Moore Foundation, the Kenneth T. and Eileen L. Norris Foundation, the
University of Chicago, the Associates of the California Institute of
Technology, and the National Science Foundation. Ongoing CARMA development and
operations are supported by the National Science Foundation under a
cooperative agreement, and by the CARMA partner universities.

\section*{Appendix: Converting from linear to circular Stokes parameters}

Here we will work through the details of converting the Stokes parameters from their commonly 
seen linear forms to their circular forms.  As we described in Section \ref{sec:intro}, 
in an $X$--$Y$ coordinate system
where $+x$ points North, $+y$ points East, and radiation propagates toward us
along the $+z$ axis, the Stokes parameters for crossed-linear feeds are:

\begin{align}
\label{app_eqn:stokes_xy1}
I &= \langle E_X E_X^* \rangle + \langle E_Y E_Y^* \rangle \\
\label{app_eqn:stokes_xy2}
Q &= \langle E_X E_X^* \rangle - \langle E_Y E_Y^* \rangle \\
\label{app_eqn:stokes_xy3}
U &= \langle E_X E_Y^* \rangle + \langle E_X^* E_Y \rangle \\
\label{app_eqn:stokes_xy4}
V &= -i \left( \langle E_X E_Y^* \rangle - \langle E_X^* E_Y \rangle \right) \,\,,
\end{align}

\noindent
where $E_X^*$ denotes the complex conjugate of $E_X$.  These equations are consistent with
any number of textbooks and publications, including 
Equations 2.47a--2.47d in \citealt{RybickiLightman1979}, 
Equations 1 in \citealt{Hamaker1996b},
Equations 4.19 in \citealt{TMS}, and others.

In radio astronomy $E_R$ is defined
using the IEEE (Institute of Electrical and Electronics Engineers) convention:
the phase $\phi$ of the sinusoid $E_Y$ lags the sinusoid $E_X$ by 90$\degree$.
Thus, a right-circularly polarized wave (one with pure $E_R$) can be represented in the following way:

\begin{align}
\label{app_eqn:phaselag_1}
E_Y &= E_X e^{-i\frac{\pi}{2}} \\
\label{app_eqn:phaselag_2}
&= -i E_X \,\,.
\end{align}

\noindent
As defined and discussed in \citet{IAU1974,Hamaker1996b,IEEE1997}, this lag
results in right-circularly polarized (RCP) radiation whose phasor rotates
\textit{counterclockwise} as viewed by the \textit{receiver}.  

Additionally, in \citet{IAU1974} the IAU deemed Stokes $V$ to be \textit{positive} if the signal has net RCP.  
To check this, we plug $E_X = e^{i\phi}$ and a lagging $E_Y = e^{i\left(\phi - \frac{\pi}{2}\right)}$ into
the crossed-linear equation for Stokes $V$ (Equation \ref{app_eqn:stokes_xy4}),
and we find that indeed, Stokes $V$ is positive for RCP radiation.

To convert from linear to circular we use the linear-to-circular coordinate transform reported in Equation 17 of \citet{Hamaker1996}, and further
elucidated in Section 3 of \citet{Hamaker1996b} (note that we have chosen the upper sign convention in
\citealt{Hamaker1996b}, which is the convention used in \citealt{Hamaker1996} as well as in the references they quote):

\begin{align}
\textbf{C$_A$} = 
\frac{1}{\sqrt{2}} \left[ \begin{array}{rr}
1 & i \\
1 & -i 
\end{array} \right] \,\,.
\label{app_eqn:convert_XY_to_RL}
\end{align}

\noindent
\citet{Hamaker1996b} explicitly refer to the above matrix as \textit{circular-rl,} clearly defining 
the order of their matrix (top row $\rightarrow R$, bottom row $\rightarrow L$).
We can then convert from [$E_X, E_Y$] to [$E_R, E_L$]:

\begin{align}
\left[ \begin{array}{r}
E_R \\
E_L 
\end{array} \right] = 
\frac{1}{\sqrt{2}} \left[ \begin{array}{rr}
1 & i \\
1 & -i 
\end{array} \right] 
\left[ \begin{array}{r}
E_X \\
E_Y
\end{array} \right] \,\,.
\end{align}

\noindent
$E_R$ and $E_L$ are thus:

\begin{align}
\label{app_eqn:E_R}
E_R &= \frac{1}{\sqrt{2}}\left(E_X + iE_Y\right) \\
\label{app_eqn:E_L}
E_L &= \frac{1}{\sqrt{2}}\left(E_X - iE_Y\right)
\end{align}

As a check, we substitute $E_Y = -iE_X$ (Equations \ref{app_eqn:phaselag_1} and \ref{app_eqn:phaselag_2},
which describe how $E_Y$ lags $E_X$ in RCP radiation) 
into the above expression for $E_L$.  We find that $E_L = 0$, as expected.

One can also invert matrix \textbf{C$_A$} to obtain the matrix required to convert from circular back to linear:

\begin{align}
\left[ \begin{array}{r}
E_X \\
E_Y 
\end{array} \right] = 
\frac{1}{\sqrt{2}} \left[ \begin{array}{rr}
1 & 1 \\
-i & i 
\end{array} \right] 
\left[ \begin{array}{r}
E_R \\
E_L
\end{array} \right] \,\,;
\end{align}

\noindent
this yields expressions for $E_X$ and $E_Y$:

\begin{align}
\label{app_eqn:E_X}
E_X &= \frac{1}{\sqrt{2}}\left(E_R + E_L\right) \\
\label{app_eqn:E_Y}
E_Y &= \frac{1}{\sqrt{2}}\left(-iE_R + iE_L\right)
\end{align}

We can substitute these expressions for $E_X$ and $E_Y$ (Equations \ref{app_eqn:E_X}
and \ref{app_eqn:E_Y}) into the crossed-linear Stokes parameters 
(Equations \ref{app_eqn:stokes_xy1}--\ref{app_eqn:stokes_xy4}).  
We find the crossed-circular Stokes parameters to be:

\begin{align}
\label{app_eqn:stokes_rl1}
I &= \langle E_R E_R^* \rangle + \langle E_L E_L^* \rangle  \\
\label{app_eqn:stokes_rl2}
Q &= \langle E_R E_L^* \rangle + \langle E_R^* E_L \rangle  \\
\label{app_eqn:stokes_rl3}
U &= -i\, \left( \langle E_R E_L^* \rangle - \langle E_R^* E_L \rangle \right)  \\
\label{app_eqn:stokes_rl4}
V &= \langle E_R E_R^* \rangle - \langle E_L E_L^* \rangle \,\,.
\end{align}

Note that occasionally the Stokes parameters are defined as $\frac{1}{2}$ $\times$ the expressions
listed in Equations \ref{app_eqn:stokes_rl1}--\ref{app_eqn:stokes_rl4} \citep[e.g., Equations 1--4 in][]{Roberts1994}.
When this is the case, one can express the four cross-correlations of $R$ and $L$ in terms of the Stokes parameters:

\vspace{-.1in}
\begin{align}
E_R E_R^* &= I + V \\
E_L E_L^*  &= I - V \\
E_R E_L^* &= Q + iU \\
E_R^* E_L &= Q - iU \,\,.
\end{align}

\bibliographystyle{apj}
\bibliography{carma_pol_arxiv}

\end{document}